\documentclass[12pt]{article}
\usepackage{amsmath}
\usepackage{pstricks,pst-plot,pst-grad,pstcol,pst-char}

\makeatletter

\@addtoreset{equation}{section}

\def\draftdate{\relax}
\def\mda{\relax}
\def\mua{\relax}
\def\mla{\relax}
\def\draft{
\def\thtystars{******************************}
\def\sixtystars{\thtystars\thtystars}
\typeout{}
\typeout{\sixtystars**}
\typeout{* Draft mode!
         For final version remove \protect\draft\space in source file *}
\typeout{\sixtystars**}
\typeout{}
\def\draftdate{\today}
\def\mua{\marginpar[\boldmath\hfil$\uparrow$]%
                   {\boldmath$\uparrow$\hfil}%
                    \typeout{marginpar: $\uparrow$}\ignorespaces}
\def\mda{\marginpar[\boldmath\hfil$\downarrow$]%
                   {\boldmath$\downarrow$\hfil}%
                    \typeout{marginpar: $\downarrow$}\ignorespaces}
\def\mla{\marginpar[\boldmath\hfil$\rightarrow$]%
                   {\boldmath$\leftarrow $\hfil}%
                    \typeout{marginpar: $\leftrightarrow$}\ignorespaces}
\overfullrule 5pt
\oddsidemargin -15mm
\marginparwidth 29mm
}

\def\stars{\strut\leaders\hbox{*}\hfill\strut}
\def\starline{\hfil\strut\hfil\hbox to \textwidth {\stars}\hfil}

\newcommand\Ref[1]     {Ref.\,\cite{#1}}
\newcommand\Refs[1]    {Refs.\,\cite{#1}}
\newcommand\eqn[1]     {Eq.\,(\ref{#1})}
\newcommand\eqns[2]    {Eqs.\,(\ref{#1}) and~(\ref{#2})}
\newcommand\eqnss[2]   {Eqs.\,(\ref{#1})--(\ref{#2})}
\newcommand\fig[1]     {Fig.\,{\ref{#1}}}

\newcommand\sect[1]    {Sect.\,{\ref{#1}}}

\newcommand\tab[1]     {Table~\ref{#1}}

\newcommand\subtitle[1] {\noindent{\bf #1}}

\def\beq{\begin{equation}}
\def\eeq{\end{equation}}
\def\beeq{\begin{eqnarray}}
\def\eeeq{\end{eqnarray}}
\def\aand{\!\!\!\!\!\!\!\!&&}
\def\arrowlimit#1{\mathrel{\mathop{\longrightarrow}\limits_{#1}}}
\newcommand\bom[1]     {{\mbox{\boldmath $#1$}}}
\newcommand\nn         {\nonumber}
\newcommand\res[3]     {$(#1\pm #2)\cdot 10^{#3}$}

\newcommand\as         {\ensuremath{\alpha_{\mathrm{s}}}}

\newcommand{\CF}       {C_{\mathrm{F}}}
\newcommand{\CA}       {C_{\mathrm{A}}}
\newcommand{\TR}       {T_{\mathrm{R}}}

\newcommand{\bT}       {\bom{T}}

\newcommand\qb         {{\bar q}}
\newcommand\Qb         {{\bar Q}}

\newcommand{\eps}      {\varepsilon}

\newcommand{\PS}[1]    {\rd\phi_{#1}}
\newcommand{\rd}{{\mathrm{d}}}
\newcommand\tsig[1]    {\sigma^{\mathrm{#1}}}
\newcommand\dsig[1]    {\rd\sigma^{{\rm #1}}}
\newcommand\dsiga[2]   {\rd\sigma^{{\rm #1,A}_{\scriptscriptstyle #2}}}
\newcommand{\Jac}[2]   {{\cal J}^{(#1)}_{#2}}

\newcommand\la         {\langle}
\newcommand\ra         {\rangle}

\newcommand\SME[3]     {|{\cal M}_{#1}^{(#2)}{(#3)}|^2}
\newcommand\M[2]       {\ensuremath{|{\cal{M}}_{#1}^{#2}|^2}}
\newcommand\bra[3]     {\la {\cal M}_{#1}^{#2}#3|}
\newcommand\ket[3]     {|{\cal M}_{#1}^{#2}#3\ra}

\newcommand{\mom}[1]   {\{p\}^{#1}}
\newcommand{\momt}[2]   {\{\ti{p}\}^{#1}_{m#2}}
\newcommand{\momh}[2]   {\{\ha{p}\}^{#1}_{m#2}}

\newcommand{\cmap}[1]   {\stackrel{{\mathsf C}_{#1}}{\longrightarrow}}
\newcommand{\smap}[1]   {\stackrel{{\mathsf S}_{#1}}{\longrightarrow}}

\newcommand{\bA}[1]    {\bom{\mathrm A}_{#1}}

\newcommand{\bSCS}[1]  {\bom{\mathrm C}\kern-2pt\bom{\mathrm S}_{#1}}

\def\hP{\hat{P}}
\newcommand{\calS}       {{\cal S}}
\newcommand{\bcA}[2]   {{\bom{\cal A}}_{#1}}
\newcommand{\cC}[2]    {{\cal C}_{#1}^{#2}}
\newcommand{\cS}[2]    {{\cal S}_{#1}^{#2}}
\newcommand{\cCS}[3]   {{\cal C}_{#1}^{~}{\cal S}_{#2}^{#3}}

\newcommand{\cSCS}[2]  {{\cal C}\kern-2pt{\cal S}_{#1}^{#2}}

\newcommand{\ti}[1]    {\tilde{#1}}
\newcommand{\wti}[1]   {\widetilde{\,#1\,}}
\newcommand{\ha}[1]    {\hat{#1}}
\newcommand{\wha}[1]   {\widehat{\,#1\,}}

\newcommand\tzz[2]     {z_{#1,#2}}

\newcommand\kT[1]      {k_{\perp,#1}}
\newcommand\kTt[1]     {k_{\perp,#1}}
\newcommand\kTtm[2]    {k_{\perp,#1}^{#2}}
\newcommand\kappatkt   {\ti{k}_{\perp,k,t}}
\newcommand\kappatir   {\ti{k}_{\perp,i,r}}

\begin{document}


\begin{titlepage}
\renewcommand{\thefootnote}{\fnsymbol{footnote}}
\begin{flushright}
hep-ph/0609042 \\
DFTT 15/2006
\end{flushright}
\par \vspace{5mm}
\begin{center}
{\Large \bf A subtraction scheme for computing QCD jet cross sections
at NNLO: \\[.5em] 
regularization of doubly-real emissions}
\end{center}

\par \vspace{2mm}
\begin{center}
{\bf G\'abor Somogyi}, {\bf Zolt\'an Tr\'ocs\'anyi},\\[.5em]
{University of Debrecen and \\Institute of Nuclear Research of
the Hungarian Academy of Sciences\\ H-4001 Debrecen, PO Box 51, Hungary}
\vspace{3mm}

and
\vspace{3mm}
                                                                                
{\bf Vittorio Del Duca},\\[.5em]
{Istituto Nazionale di Fisica Nucleare, Sez. di Torino\\
via P. Giuria, 1 - 10125 Torino, Italy\\
E-mail: delduca@to.infn.it}
\end{center}

\par \vspace{2mm}
\begin{center} {\large \bf Abstract} \end{center}
\begin{quote}
\pretolerance 10000
We present a subtraction scheme for computing jet cross sections
in electron-positron annihilation at next-to-next-to-leading
order accuracy in perturbative QCD. In this first part we deal with the
regularization of the doubly-real contribution to the NNLO correction.
\end{quote}

\vspace*{\fill}
\begin{flushleft}
September 2006
\end{flushleft}
\end{titlepage}
\clearpage

\tableofcontents

\renewcommand{\thefootnote}{\fnsymbol{footnote}}


%
%

\section{Introduction}
\label{sec:intro}

QCD, the theory of strong interactions, is an important component of
the Standard Model of elementary particle interactions. It is
asymptotically free, which allows us to compute cross sections of 
elementary particle interactions at high energies as a perturbative
expansion in the running strong coupling $\as(\mu_R)$. However,
the running coupling $\as(\mu_R)$ remains rather large at energies
relevant at recent and future colliders. In addition, to leading order
in the perturbative expansion, the coupling varies sizeably with the
choice of the (unphysical) renormalization scale $\mu_R$.  In
hadron-initiated processes, the situation is worsened by the
dependence of the cross section on the (also unphysical) factorization 
scale $\mu_F$, which separates the long-distance from the short-distance
part of the strong interaction. Thus, a leading-order evaluation of the
cross section yields rather unreliable predictions for most processes in
the theory. To improve this situation, in the past 25 years the
radiative corrections at the next-to-leading order (NLO) accuracy have
been computed. These efforts have culminated, when process-independent
methods were presented for computing QCD cross sections to NLO
accuracy, namely the slicing~\cite{Giele:1991vf,Giele:1993dj},
subtraction~\cite{Frixione:1995ms,Nagy:1996bz,Frixione:1997np}
and dipole subtraction~\cite{Catani:1996vz} methods.  In some cases,
though, the NLO corrections were found to be disturbingly large,
and/or the dependence on $\mu_R$ (and eventually $\mu_F$) was found to
be still sizeable, thus casting doubts on the applicability of the
perturbative predictions. When the NLO corrections are found to
be of the same order as the leading-order prediction, the only way to 
assess the reliability of QCD perturbation theory is the computation
of the next-to-next-to-leading order (NNLO) corrections.

In recent years severe efforts have been made to compute the NNLO
corrections to the parton distribution functions \cite{Moch:2004pa}
and important basic processes, such as vector boson production 
\cite{Hamberg:1990np,Harlander:2002wh,Anastasiou:2003yy,%
Anastasiou:2003ds,Melnikov:2006di} and Higgs production 
\cite{Harlander:2002wh,Anastasiou:2002yz,Ravindran:2003um,%
Anastasiou:2004xq,Anastasiou:2005qj} in hadron collisions and jet
production in electron-positron annihilation \cite{Anastasiou:2004qd,%
Gehrmann-DeRidder:2004tv}. These computations evaluate also the phase
space integrals in $d$ dimensions, thus, do not follow the
process-independent methods used to compute the NLO corrections. 

The more traditional approach relies on defining approximate
cross sections which match the singular behaviour of the QCD cross
sections in all the relevant unresolved limits. Various attempts 
were made in this direction in \Refs{Kosower:1997zr,Campbell:1998nn,%
Weinzierl:2003fx,Weinzierl:2003ra, Gehrmann-DeRidder:2003bm,
Gehrmann-DeRidder:2004xe,Frixione:2004is,Gehrmann-DeRidder:2005hi,%
Ridder:2005aw,Gehrmann-DeRidder:2005cm,Weinzierl:2006ij}. In general,
the definition of the approximate cross sections must rely on the
singly- and doubly-unresolved limits of the QCD squared matrix
elements.  Although the infrared limits of QCD matrix elements have
been extensively studied both at tree-level~\cite{Altarelli:1977zs,%
Berends:1988zn,Mangano:1990by,Gehrmann-DeRidder:1997gf,Campbell:1997hg,%
Catani:1998nv,Kosower:1999xi,Catani:1999ss,DelDuca:1999ha,%
Kosower:2002su,Kosower:2003cz,Kosower:2003bh}, and at one-loop
\cite{Bern:1994zx,Bern:1998sc,Kosower:1999rx,Bern:1999ry,Catani:2000pi},
the formulae presented in the literature do not lend themselves directly
for devising the approximate cross sections for two reasons.  The first
problem is that the various single and double soft and/or collinear
limits overlap in a very complicated way and the infrared factorization
formulae have to be written in such forms that these overlaps can be
disentangled so that double subtraction is avoided.  The second problem
is that even if the factorization formulae are written such that double
subtraction does not happen, the expressions cannot straightforwardly
be used as subtraction formulae, because the momenta of the partons in
the factorized matrix elements are unambiguously defined only
in the strict soft and collinear limits. In order to define the
approximate cross sections one also has to factorize the phase space of
the unresolved partons such that the singular factors can be integrated
and the remaining expressions can be combined with the virtual correction
leading to cross sections which are finite and integrable in four
dimensions.

In \Ref{Somogyi:2005xz} we presented a solution to the first problem,
but did not explicitly define the approximate cross sections, which we
left for later work.  In this paper we turn to the second problem and
define the complete approximate cross sections that regularize the
doubly-real emission.  For factorizing the phase space one has two
options. On the one hand we may decompose the squared matrix elements
into expressions that contain only single singular factors and use the
factorization formulae as subtraction terms. In NLO computations this
method was termed `residuum subtraction'. On the other hand one may
use exact phase-space factorization (keeping momentum conservation and
particles on-shell) to maintain gauge invariance of the factorized matrix
elements. In NLO computations the `dipole subtraction method' represents
an example of this approach.

The single singular factor decomposition cannot be followed in a NNLO
computation because a singular factor in a doubly-unresolved region of
the phase space naturally contains singular factors in
singly-unresolved regions. Therefore, we have to use exact
factorizations of the unresolved phase space measures.
There are only two known ways of exact phase space factorization, one
termed `dipole factorization' \cite{Catani:1996vz}, while the other
called `antennae factorization' \cite{Kosower:1997zr}. Actually, both
belong to the same general class. The important feature of these is that
in order to factorize the unresolved phase space measures, two partons,
called `emitter' and `spectator', are singled out for each subtraction
term. The emitter emits the unresolved parton and the spectator
takes away the momentum recoil to maintain momentum conservation.
In a NLO computation the choice for the emitter-spectator pair is
naturally a colour-connected pair of partons in the soft factorization
formula. The collinear emissions are distributed among the soft ones
using colour conservation. Although this trick provides a fairly elegant
framework for computing NLO corrections, the unnatural distribution of
collinear emissions seems impossible to maintain in a NNLO computation
because it leads to simultaneously spin- and colour-correlated squared
matrix elements for which collinear factorization formulae do not exist
\cite{Somogyi:2005xz}.\footnote{Note a misprint in \Ref{Somogyi:2005xz},
where `soft' factorization is written in this respect instead of `collinear'.} 

In \Ref{Gehrmann-DeRidder:2005cm} the antennae factorization is used for
NNLO subtractions, which is made possible by the use of colour-ordered
subamplitudes, where the emitter-spectator `antenna pairs' can naturally
and unambiguously be selected because singular emission for a given
amplitude can occur only between ordered pairs of momenta (the emitter
and the spectator). Note however, that the effect of quantum
interference in the squared matrix element mixes the colour
subamplitudes in a rather complicated way and a general scheme to
construct the subtraction terms in a process-independent way has not
been given yet.  It seems to us that in order to construct a general
method for computing NNLO corrections we are forced to give up the
antennae (or dipole) factorization of the phase space. 

In \Ref{Somogyi:2006_1}, we defined a new NLO subtraction scheme with
exact phase-space factorization that can be generalized to any order in
perturbation theory. In this paper we extend that scheme to computing
the contribution of the doubly-real emission to the NNLO corrections.
We demonstrate that the regularized cross section is indeed numerically
integrable by computing the corresponding contribution to the cross
section of electron-positron annihilation into three jets from the
$e^+ e^- \to q \qb g g g$ subprocess.

\section{Jet cross sections at NNLO accuracy}
\label{sec:xsec_NNLO}

The jet cross sections in perturbative QCD are represented by an
expansion in the strong coupling $\as$. At NNLO accuracy we keep the
three lowest-order terms,
\beq
\sigma = \tsig{LO} + \tsig{NLO} + \tsig{NNLO}\,.
\eeq
Assuming an $m$-jet quantity, the leading-order contribution is the
integral of the fully differential Born cross section $\dsig{B}_m$ of
$m$ final-state partons over the available $m$-parton phase space
defined by the jet function $J_m$,
\beq
\tsig{LO} = \int_m\!\dsig{B}_m J_m\:.
\eeq
The NLO contribution is a sum of two terms, the real and virtual
corrections,
\beq
\tsig{NLO} =
\int_{m+1}\!\dsig{R}_{m+1} J_{m+1} + \int_m\!\dsig{V}_m J_m\:.
\label{eq:sigmaNLO}
\eeq
Here the notation for the integrals indicates that the real correction
involves $m+1$ final-state partons, one of those being unresolved, while
the virtual correction has $m$-parton kinematics. The NNLO correction is
a sum of three contributions, the doubly-real, the one-loop
singly-unresolved real-virtual and the two-loop doubly-virtual terms,
\beq
\tsig{NNLO} =
\int_{m+2}\!\dsig{RR}_{m+2} J_{m+2}
+ \int_{m+1}\!\dsig{RV}_{m+1} J_{m+1}
+ \int_m\!\dsig{VV}_m J_m\:.
\label{eq:sigmaNNLO}
\eeq
Here the notation for the integrals indicates that the doubly-real
corrections involve $m+2$ final-state partons, the real-virtual
contribution involves $m+1$ final-state partons and the doubly-virtual
term is an integral over the phase space of $m$ partons, and the phase
spaces are restricted by the corresponding jet functions $J_n$ that
define the physical quantity.

In $d=4$ dimensions the two contributions in  \eqn{eq:sigmaNLO} as well
as the three contributions in \eqn{eq:sigmaNNLO} are separately
divergent, but their sum is finite for infrared-safe observables order by
order in the expansion in $\as$. The requirement of infrared-safety puts
constraints on the analytic behaviour of the jet functions that were
spelled out explicitly in \Ref{Somogyi:2005xz}.

Following from kinematical reasons, fully inclusive observables can be
accurately evaluated in QCD perturbation theory relatively simply.
Since these observables are completely inclusive, no phase-space
restriction has to be applied ($J_n = 1$ for any $n$). Real and virtual
contributions can be combined at the integrand level resulting in the
cancellation of soft and collinear singularities before performing the
relevant phase-space integrations.  Owing to these features, general
techniques have been available for some
time~\cite{Tkachov:1981wb,Chetyrkin:1981qh} to carry out NNLO
calculations in analytic form. 

QCD calculations beyond LO for inclusive quantities, such as jet cross
sections or event-shape distributions, are much more
involved. Owing to the complicated phase space for multiparton
configurations, analytic calculations are impossible for most of the
distributions.  Moreover, soft and collinear singularities are separately
present in the real radiation correction (due to integrations over the
phase space of the unresolved parton) and virtual contributions (due to
integrations over the loop momentum) at the intermediate steps. These
singularities have to be first regularized by analytic continuation in
a number of space-time dimensions $d=4-2\eps$ different from four. 
This analytic continuation prevents a straightforward implementation of
numerical integration techniques.%

The traditional approach to finding the finite corrections at NLO
accuracy is to regularize the real radiation contribution by subtracting
a suitably defined approximate cross section $\dsig{R,A}$ such
that (i) $\dsig{R,A}$ matches the pointwise singular behaviour of
$\dsig{R}$ in the one-parton infrared regions of the phase space in any
dimensions (ii) and it can be integrated over the one-parton phase
space of the unresolved parton independently of the jet function,
resulting in a Laurent expansion in $\eps$. After performing this
integration, the approximate cross section can be combined with the
virtual correction $\dsig{V}$ before integration. We then write 
\beq
\tsig{NLO} =
\int_{m+1}\!\left[\dsig{R}_{m+1} J_{m+1} - \dsig{R,A}_{m+1} J_m\right]
+ \int_m\!\left[\dsig{V}_m + \int_1\!  \dsig{R,A}_{m+1}\right] J_m\:,
\label{eq:sigmaNLO2}
\eeq
where both integrals on the right-hand side are finite in $d = 4$
dimensions.  The final result is that one is able to rewrite the two
NLO contributions in \eqn{eq:sigmaNLO} as a sum of two finite integrals,
\beq
\tsig{NLO} =
\int_{m+1}\!\dsig{NLO}_{m+1} + \int_m\!\dsig{NLO}_m\:,
\label{eq:sigmaNLOfin}
\eeq
that are integrable in four dimensions using standard numerical
techniques.

The construction of the suitable approximate cross section $\dsig{R,A}$
is made possible by the universal soft and collinear factorization
properties of QCD matrix elements. Envisaging a similar construction for
computing the NNLO correction, the universal infrared behaviour of the
loop amplitudes and the infrared limits of the real-emission corrections
at NNLO, as well as the singularity structure of the two-loop squared
matrix elements has been computed \cite{Catani:1998bh,Sterman:2002qn}.
However, it is far more complex to disentangle these singularities at
the NNLO accuracy than at NLO \cite{Somogyi:2005xz}, thus up to now,
process independent approximate cross sections for regularizing the
$\dsig{RR}$ and $\dsig{RV}$ terms have not been computed, but for the
relatively simple case of $e^+e^- \to $ 2 and 3 jets, when the
dependence on colour completely factorizes from all matrix elements
\cite{Gehrmann-DeRidder:2004xe,Weinzierl:2006ij}.

In order to avoid this complexity in \Ref{Anastasiou:2003gr} a new
method has been developed for computing the QCD corrections by
combining the real-emission and virtual corrections before integration
for arbitrary jet function. The method is very simple conceptually. It
considers the problem from a purely mathematical point of view:
how to compute a complicated, but finite integral numerically?
The first step is to map the phase spaces onto the unit hypercube of
suitable dimensions. Then the singularities from inside the hypercube are
removed to the edges of the cube by splitting appropriately the
integrations and mapping them back to the $[0,1]$ interval. Next, the
overlapping singularities are disentangled using sector decomposition
\cite{Roth:1996pd,Binoth:2000ps,Heinrich:2002rc,Binoth:2004jv,%
Heinrich:2004jv,Heinrich:2006sw}. At this point the only factors in the
integrand that lead to divergences are of the form $\lambda^{-1+n
\eps}$, therefore, the $\eps$ poles can be extracted in terms of plus
distributions \cite{Anastasiou:2003gr}.  The method is clearly
completely general and its strength has already been demonstrated in
various explicit computations \cite{Anastasiou:2004xq,Anastasiou:2005qj,%
Anastasiou:2004qd,Melnikov:2006di}.
Note however, that the various mappings of the phase space as well as
the sector decompositions are not unique. The particular choices depend
on the analytic structure of the functions one has to integrate,
namely, the squared matrix elements for the given process. In fact, the
different terms in the squared matrix element may prefer different
mappings as in the case of \Ref{Anastasiou:2005qj}. This means that
with this technique the construction of a universal program that
requires only various matrix elements as input for computing NNLO
corrections for arbitrary processes does not seem straightforward.

Such program exists for computing NLO corrections
\cite{Nagy:2003tz,Nagy:2001xb,Nagy:1998bb} based upon the subtraction
method. Therefore, it is of interest whether the subtraction method can
be extended to the computation of NNLO corrections. In the next section
we rewrite \eqn{eq:sigmaNNLO} such that each phase space integral is
finite and thus can be performed numerically in four dimensions using
standard Monte Carlo techniques.  

%
%

\section{Subtraction scheme at NNLO accuracy}
\label{sec:sub_NNLO}

Let us consider first the doubly-real contribution, $\dsig{RR}_{m+2}$.
It is divergent in the doubly-unresolved regions of phase space. In order
to cancel the two-parton singularities we subtract the approximate
cross section $\dsiga{RR}{2}_{m+2}$ that matches the pointwise
singular behaviour of $\dsig{RR}_{m+2}$ in $d$ dimensions in the
two-parton infrared regions. Then we have
\beeq
\tsig{NNLO} \aand= \int_{m+2}\Big[\dsig{RR}_{m+2} J_{m+2} - \dsiga{RR}{2}_{m+2} J_{m}\Big]
\nn \\ &&
+ \int_{m+1}\dsig{RV}_{m+1}  J_{m+1} 
+ \int_m\Big[\dsig{VV}_m + \int_2\dsiga{RR}{2}_{m+2}\Big] J_{m}\,.
\label{eq:sigmaNNLO_v1}
\eeeq
However the first integral is still divergent in the singly-unresolved
regions of the phase space. In order to cancel these remaining
singularities we  subtract the approximate cross sections
$\dsiga{RR}{1}_{m+2}$ and $\dsiga{RR}{12}_{m+2}$ to obtain
\beeq
\tsig{NNLO} \aand=
\int_{m+2}
\Big[\dsig{RR}_{m+2} J_{m+2} - \dsiga{RR}{2}_{m+2} J_{m}
    -\Big(\dsiga{RR}{1}_{m+2} J_{m+1} - \dsiga{RR}{12}_{m+2} J_{m}\Big)\Big]
\nn \\ &&
+ \int_{m+1}\Big[\dsig{RV}_{m+1} + \int_1\dsiga{RR}{1}_{m+2}\Big] J_{m+1} 
\nn\\ &&
+ \int_m
\Big[\dsig{VV}_m + \int_2\dsiga{RR}{2}_{m+2} - \int_2 \dsiga{RR}{12}_{m+2}\Big]
J_{m}\,.
\label{eq:sigmaNNLO_v2}
\eeeq
Here $\dsiga{RR}{1}_{m+2}$ and $\dsiga{RR}{12}_{m+2}$ regularize the
singly-unresolved limits of $\dsig{RR}_{m+2}$ and $\dsiga{RR}{2}_{m+2}$
respectively. For the construction to be consistent, we must also require 
that $\dsiga{RR}{1}_{m+2}-\dsiga{RR}{12}_{m+2}$ be integrable in the
two-parton infrared regions of the phase space,%
\footnote{Formally this means that $\dsiga{RR}{12}_{m+2} =
\dsiga{RR}{21}_{m+2}$, which we have already taken into account in
writing the expressions.}
which restricts the possible forms of $\dsiga{RR}{12}_{m+2}$ severely.
In \eqn{eq:sigmaNNLO_v2} the jet functions, multiplying each
approximate cross section, organize the terms according to in which
integral they should appear. In particular, $\dsiga{RR}{1}_{m+2}$ is
multiplied by $J_{m+1}$, therefore, after integration over the phase
space of the unresolved parton, it is combined with $\dsig{RV}_{m+1}$,
while $\dsiga{RR}{12}_{m+2}$ is multiplied with $J_m$, therefore, it is
added back in the third line. The $m+2$-parton integral above is now
finite by construction. 

Next consider the real-virtual contribution $\dsig{RV}_{m+1}$. It has
two types of singularities: (i) explicit $\eps$ poles in the loop
amplitude and (ii) kinematical singularities in the singly-unresolved
regions of the phase space. In \eqn{eq:sigmaNNLO_v2} the former are
already regularized. Indeed, unitarity guarantees that the second line
of that equation is free of $\eps$ poles if $\dsiga{RR}{1}_{m+2}$ is a
true regulator of $\dsig{RR}_{m+2}$ in the one-parton infrared regions
of the phase space, just as it does in NLO subtraction schemes. To
regularize the kinematical singularities, we subtract the approximate
cross sections $\dsiga{RV}{1}_{m+1}$ and
$\left(\int_1\dsiga{RR}{1}_{m+2}\right)\!\strut^{{\rm A}_{\scriptscriptstyle 1}}$, 
which regularize the real-virtual cross section $\dsig{RV}_{m+1}$ and
$\int_1 \dsiga{RR}{1}_{m+2}$, respectively, when a single parton
becomes unresolved. Thus, the NNLO cross section is written as
\beeq
\tsig{NNLO} \aand=
\int_{m+2}
\Big\{\dsig{RR}_{m+2} J_{m+2} - \dsiga{RR}{2}_{m+2} J_{m}
     -\Big[\dsiga{RR}{1}_{m+2} J_{m+1} - \dsiga{RR}{12}_{m+2} J_{m}\Big]
\Big\}_{\eps = 0}
\label{eq:sigmaNNLO_v3}\\
\aand+
\int_{m+1}
\Big\{\Big(\dsig{RV}_{m+1} + \int_1\dsiga{RR}{1}_{m+2}\Big) J_{m+1} 
     -\Big[\dsiga{RV}{1}_{m+1} + \Big(\int_1\dsiga{RR}{1}_{m+2}\Big)
\strut^{{\rm A}_{\scriptscriptstyle 1}} \Big] J_{m}
\Big\}_{\eps = 0}
\nn\\
\aand+
\int_m
\Big\{\dsig{VV}_m + \int_2\Big[\dsiga{RR}{2}_{m+2} - \dsiga{RR}{12}_{m+2}\Big]
     +\int_1\Big[\dsiga{RV}{1}_{m+1} + \Big(\int_1\dsiga{RR}{1}_{m+2}\Big)
\strut^{{\rm A}_{\scriptscriptstyle 1}} \Big]
\Big\}_{\eps = 0} J_{m}\,.
\nn
\eeeq
Since the first and second integrals on the right hand side of this
equation are finite in $d=4$ dimensions by construction, it follows
from the Kinoshita-Lee-Nauenberg theorem that the combination of
integrals in the last line is finite as well, provided the jet function
defines an infrared-safe observable.

The final result of these manipulations is that we rewrite
\eqn{eq:sigmaNNLO} as
\beq
\tsig{NNLO} =
\int_{m+2}\!\dsig{NNLO}_{m+2}
+ \int_{m+1}\!\dsig{NNLO}_{m+1}
+ \int_m\!\dsig{NNLO}_m\,,
\label{eq:sigmaNNLOfin}
\eeq
that is a sum of three integrals,
\beeq
\dsig{NNLO}_{m+2} \aand=
\Big\{\dsig{RR}_{m+2} J_{m+2} - \dsiga{RR}{2}_{m+2} J_{m}
     -\Big[\dsiga{RR}{1}_{m+2} J_{m+1} - \dsiga{RR}{12}_{m+2} J_{m}\Big]
\Big\}_{\eps=0}\,,
\label{eq:sigmaNNLOm+2}\\
\dsig{NNLO}_{m+1} \aand=
\Big\{\Big[\dsig{RV}_{m+1} + \int_1\dsiga{RR}{1}_{m+2}\Big] J_{m+1} 
     -\Big[\dsiga{RV}{1}_{m+1} + \Big(\int_1\dsiga{RR}{1}_{m+2}\Big)
\strut^{{\rm A}_{\scriptscriptstyle 1}} \Big] J_{m}
\Big\}_{\eps=0}\,, \;\;\;\;\;\;
\label{eq:sigmaNNLOm+1}
\eeeq
and
\beq
\dsig{NNLO}_{m} =
\Big\{\dsig{VV}_m + \int_2\Big[\dsiga{RR}{2}_{m+2} - \dsiga{RR}{12}_{m+2}\Big]
     +\int_1\Big[\dsiga{RV}{1}_{m+1} + \Big(\int_1\dsiga{RR}{1}_{m+2}\Big)
\strut^{{\rm A}_{\scriptscriptstyle 1}} \Big]
\Big\}_{\eps=0} J_{m}\,,
\label{eq:sigmaNNLOm}
\eeq
each integrable in four dimensions using standard numerical techniques.

The subtraction scheme presented here differs somewhat from the one
outlined in \Ref{Somogyi:2005xz}, where we assumed that
$\dsiga{RR}{12}_{m+2}$ can be defined such that (in the
present notation) 
\beq
\int_{m+1}\left[
  \int_1\dsiga{RR}{12}_{m+2}
- \Big(\int_1\dsiga{RR}{1}_{m+2}\Big)
\strut^{{\rm A}_{\scriptscriptstyle 1}}\right] J_m = {\rm finite}
\label{eq:RRA1A1_RRA12}
\eeq
in $d = 4$ dimensions.
However, as already emphasized in \Ref{Somogyi:2005xz},
\eqn{eq:RRA1A1_RRA12} does not follow from unitarity, rather it is 
a constraint on the definitions of $\dsiga{RR}{1}_{m+2}$ and
$\dsiga{RR}{12}_{m+2}$. It turns out more convenient to drop this extra
condition and rearrange the subtraction scheme as presented here.  

In this paper we present all formulae relevant for constructing
$\dsig{NNLO}_{m+2}$ explicitly. The terms needed for defining
$\dsig{NNLO}_{m+1}$ and $\dsig{NNLO}_m$ will be given in separate
papers.  
We use the colour- and spin-state notation introduced in
\Ref{Catani:1996vz}. The complete description of our notation can be
found in \Ref{Somogyi:2005xz}.  

%
%

\section{Subtraction terms for doubly-real emission}
\label{sec:sub_RR}

The cross section $\dsig{RR}_{m+2}$ is the integral of the tree-level
squared matrix element for $m+2$ parton production over the $m+2$
parton phase space
\beq
\dsig{RR}_{m+2} = \PS{m+2}{}\M{m+2}{(0)}\,,
\label{eq:tsigRRm+2}
\eeq
where the phase-space measure is defined as 
\beq
\PS{n}{(p_1,\dots,p_n;Q)} = \prod_{i=1}^n \frac{\rd^d p_i}{(2 \pi)^{d-1}}
\delta_+(p_i^2)\:(2 \pi)^d\,\delta^{(d)}\left(Q-\sum_{i=1}^n p_i\right)\,.
\label{eq:PSn}
\eeq
In \eqn{eq:tsigRRm+2} (and all subsequent formulae) the superscript
$(0)$ refers to tree-level expressions.  We disentangled the overlap
structure of the singularities of $\M{m+2}{(0)}$ into the pieces
$\bA{2}\M{m+2}{(0)}$, $\bA{1}\M{m+2}{(0)}$ and $\bA{12}\M{m+2}{(0)}$ in
\Ref{Somogyi:2005xz}.  These expressions are only defined in the strict
soft and/or collinear limits. To define true counterterms, they need to
be extended over the full phase space. This extension requires a
phase-space factorization that maintains momentum conservation exactly,
but such that in addition it respects the delicate structure of
cancellations among the various subtraction terms.  

The true (extended) counterterms may symbolically be written as
\beeq
\dsiga{RR}2_{m+2} \aand=
\PS{m}{}\:[\rd p_{2}]\:{\bom{\cal A}}_2^{(0)} \M{m+2}{(0)}\,,
\label{eq:dsigRRA2}
\\
\dsiga{RR}1_{m+2} \aand=
\PS{m+1}{}\:[\rd p_{1}]\:{\bom{\cal A}}_1^{(0)} \M{m+2}{(0)}\,,
\label{eq:dsigRRA1}
\eeeq
and
\beq
\dsiga{RR}{12}_{m+2} =
\PS{m}{}\:[\rd p_{1}]\:[\rd p_{1}]\:{\bom{\cal A}}_{12}^{(0)} 
\M{m+2}{(0)}\,,
\label{eq:dsigRRA12}
\eeq
where in \eqnss{eq:dsigRRA2}{eq:dsigRRA12} we used a formal, calligraphic 
notation (to be defined explicitly below) to indicate the extension of the 
terms $\bA{2}\M{m+2}{(0)}$, $\bA{1}\M{m+2}{(0)}$ and $\bA{12}\M{m+2}{(0)}$ 
over the whole phase space that was written in exactly factorized forms,
\beq
\PS{m+2}{} =
\PS{m}{}\:[\rd p_{2}] =
\PS{m+1}{}\:[\rd p_{1}] =
\PS{m}{}\:[\rd p_{1}]\:[\rd p_{1}]
\eeq
(the precise meaning of the factors $[\rd p_{1}]$ and $[\rd p_{2}]$
will be given below).
%
%

\section{Singly-unresolved counterterms}
\label{sec:RR_A1}

The singly-unresolved counterterm $\bcA{1}{(0)} \M{m+2}{(0)}$ reads
\beq
\bcA{1}{(0)}\SME{m+2}{0}{\mom{}} =
\sum_{r} \left[\sum_{i\ne r} \frac{1}{2} \cC{ir}{(0,0)}(\mom{})
+ \left(\cS{r}{(0,0)}(\mom{}) - \sum_{i\ne r} \cCS{ir}{r}{(0,0)}(\mom{})\right) 
\right]\,.
\label{eq:A1}
\eeq
Here all three terms are functions of the original $m+2$ momenta that
enter the matrix element on the left hand side of \eqn{eq:A1}. To
shorten the notation we denote these momenta collectively as
$\mom{} \equiv \{p_1,\ldots,p_{m+2}\}$. Although the notation in
\eqn{eq:A1} is very similar to the operator notation introduced in
\Ref{Somogyi:2005xz}, it is important to understand that it is not
meant in the operator sense, for instance, the last term on
the right hand side does not refer to the collinear limit of anything.
Throughout this paper the subtraction terms are functions of the
original momenta for which the notation inherits the operator structure
of taking the various limits, but otherwise it has nothing to do with
taking limits.


\subsection{Collinear counterterm}
\label{ssec:RR_A1_Cir}

\subtitle{Counterterm}

The singly-collinear counterterm is 
\beq
\cC{ir}{(0,0)}(\mom{}) = 
8\pi\as\mu^{2\eps}\frac{1}{s_{ir}}
\bra{m+1}{(0)}{(\momt{(ir)}{+1})}
\hP_{f_i f_r}^{(0)}(\tzz{i}{r},\tzz{r}{i},\kTt{i,r};\eps)
\ket{m+1}{(0)}{(\momt{(ir)}{+1})}\,,
\label{eq:Cir00}
\eeq
where the kernels
$\hP_{f_i f_r}^{(0)}(\tzz{i}{r},\tzz{r}{i},\kTt{i,r};\eps)$ are defined
to coincide with the following specific forms of the Altarelli-Parisi
splitting functions (valid in the CDR scheme)
\beeq
\la r|\hP_{q_ig_r}^{(0)}(\tzz{i}{r},\tzz{r}{i};\eps)|s\ra \aand=
\delta_{rs}\CF\left[\frac{1+\tzz{i}{r}^2}{\tzz{r}{i}}-\eps \tzz{r}{i} \right]
\equiv \delta_{rs} P_{q_ig_r}^{(0)}(\tzz{i}{r},\tzz{r}{i};\eps)\,,
\label{eq:Pqg0}
\eeeq
\beeq
\la\mu|\hP_{\qb_i q_r}^{(0)}(\tzz{i}{r},\tzz{r}{i},\kTt{i,r}^\mu;\eps)|\nu\ra \aand=
\TR\left[
-g^{\mu\nu}+4\tzz{i}{r} \tzz{r}{i} \frac{\kTt{i,r}^{\mu}\kTt{i,r}^{\nu}}{\kTt{i,r}^2}
\right]\,,
\label{eq:Pqq0}
\\
\la\mu|\hP_{g_ig_r}^{(0)}(\tzz{i}{r},\tzz{r}{i},\kT{i,r}^\mu;\eps)|\nu\ra \aand=
2\CA\left[-g^{\mu\nu}\left(\frac{\tzz{i}{r}}{\tzz{r}{i}}+\frac{\tzz{r}{i}}{\tzz{i}{r}}\right)
-2(1-\eps)
\tzz{i}{r} \tzz{r}{i} \frac{\kTt{i,r}^{\mu}\kTt{i,r}^{\nu}}{\kTt{i,r}^2}\right]\,.
\nn\\
\label{eq:Pgg0}
\eeeq
In \eqn{eq:Cir00} the double superscipt on the
left hand side means that on the right hand side of the equation both the
matrix elements as well as the splitting kernels are at tree-level.

In \eqn{eq:Pqg0} we introduced our notation for the spin-averaged splitting 
function,
\beq
P^{(0)}_{f_i f_r}(\tzz{i}{r},\tzz{r}{i};\eps) 
\equiv 
\la\hP^{(0)}_{f_i f_r}(\tzz{i}{r},\tzz{r}{i},\kTt{i,r}^\mu;\eps)\ra\:.
\label{eq:Pav}
\eeq
The kernels are functions of the momentum fractions $\tzz{i}{r}$ and
$\tzz{r}{i}$ that we define as
\beq
\tzz{i}{r} = \frac{y_{iQ}}{y_{(ir)Q}}
\qquad\mbox{and}\qquad
\tzz{r}{i} = \frac{y_{rQ}}{y_{(ir)Q}}\,,
\label{eq:zt2}
\eeq
where $y_{(ir)Q} = y_{iQ} + y_{rQ}$ with $y_{iQ}=2p_i\cdot Q/Q^2$,
$y_{rQ}=2p_r\cdot Q/Q^2$ and $Q^\mu$ is the total four-momentum of the
incoming electron and positron.  With this definition $\tzz{i}{r} +
\tzz{r}{i} = 1$. Note that the momentum fractions are nothing but the
energy fractions of the daughter momenta of the splitting with respect
to the energy of the parent parton in the center-of-momentum frame. 
The transverse momentum $\kTt{i,r}$ is given by 
\beq
\kTt{i,r}^{\mu} = 
\zeta_{i,r} p_r^{\mu} - \zeta_{r,i} p_i^{\mu} + \zeta_{ir} \ti{p}_{ir}^{\mu}
\,,\qquad
\zeta_{i,r} = \tzz{i}{r}-\frac{y_{ir}}{\alpha_{ir}y_{(ir)Q}}
\,,\quad
\zeta_{r,i} =  \tzz{r}{i}-\frac{y_{ir}}{\alpha_{ir}y_{(ir)Q}}
\,.
\label{eq:kTtir}
\eeq
Here $y_{ir}=2p_i\cdot p_r/Q^2$ while  $\ti{p}_{ir}^{\mu}$ and
$\alpha_{ir}$ are defined below in \eqns{eq:PS_Cir}{eq:alphair}
respectively.  This choice for the transverse momentum is exactly
perpendicular to the parent momentum $\ti{p}_{ir}^{\mu}$ and ensures
that in the collinear limit $p_i^\mu || p_r^\mu $, the square of
$\kTt{i,r}^{\mu}$ behaves as
\beq
\kTt{i,r}^2 \simeq - s_{ir}\tzz{r}{i} \tzz{i}{r}
\,,
\label{eq:kTir2}
\eeq
as required (independently of $\zeta_{ir}$). In a NLO computation this
feature is sufficient to ensure the correct collinear behaviour of the
subtraction term. In a NNLO computation in addition to \eqn{eq:kTir2}
it is also important that $\kTt{i,r}^\mu$ itself vanishes in the
collinear limit,%
\footnote{If $\kTt{i,r}^\mu$ does not vanish in the collinear limit then
the iterated collinear-triple collinear counterterms of \sect{ssec:CktA2}
do not have the correct (strongly-ordered) collinear behaviour.} and it
is that convenient that $\kTt{i,r}^\mu$ is perpendicular to $Q^\mu$.
These conditions are fulfilled if we choose 
\beq
\zeta_{ir} =
\frac{y_{ir}}{\alpha_{ir} y_{\wti{ir}Q}}(\tzz{r}{i}-\tzz{i}{r})\,.
\label{eq:zetair}
\eeq
With this choice $\kTt{i,r}^\mu \to \kT{i}^\mu$ in the collinear limit
as can be shown by substituting the Sudakov parametrization of the
momenta into \eqn{eq:kTtir} (with properly chosen gauge vector).

\subtitle{Momentum mapping and phase space factorization}

The $m+1$ momenta,
$\momt{(ir)}{+1} \equiv \{\ti{p}_1,\ldots,\ti{p}_{ir},\ldots,\ti{p}_{m+2}\}$, 
entering the matrix elements on the right hand side of \eqn{eq:Cir00}
are defined as follows 
\beeq
\ti{p}_{ir}^{\mu} = \frac{1}{1-\alpha_{ir}}(p_i^{\mu} + p_r^{\mu} - \alpha_{ir} Q^{\mu})\,,
\qquad
\ti{p}_n^{\mu} = \frac{1}{1-\alpha_{ir}} p_n^{\mu}\,,
\qquad n\ne i,r\,,
\label{eq:PS_Cir}
\eeeq
where
\beq
\alpha_{ir} =
\frac12\left[y_{(ir)Q}-\sqrt{y_{(ir)Q}^2 - 4y_{ir}}\;\right]\,.
\label{eq:alphair}
\eeq
The total four-momentum is clearly conserved,
\beq
Q^\mu = p_i^\mu + p_r^\mu + \sum_n^m p_n^\mu
= \ti{p}_{ir}^{\mu} + \sum_n^m \ti{p}_n^{\mu}
\,.
\eeq
For further convenience let us denote the momentum mapping introduced above as
\beq
\mom{} \cmap{ir} \momt{(ir)}{+1}\,,
\label{eq:cmap}
\eeq
where $i$ and $r$ are any two labels of momenta that appear on the left
hand side, but not on the right hand side.  

The momentum mapping of \eqn{eq:PS_Cir} leads to exact phase space
factorization in the form 
\beq
\PS{m+2}(\mom{};Q)=\PS{m+1}(\momt{(ir)}{+1};Q)
\: [\rd p_{1;m+1}^{(ir)}(p_r,\ti{p}_{ir};Q)]\,,
\label{eq:PSfact_Cir}
\eeq
where, as indicated, the $m+1$ momenta in the first factor on the right
hand side of \eqn{eq:PSfact_Cir} are exactly those defined in \eqn{eq:PS_Cir}.
The explicit expression for $[\rd p_{1;m+1}^{(ir)}(p_r,\ti{p}_{ir};Q)]$ reads
\beq
[\rd p_{1;m+1}^{(ir)}(p_r,\ti{p}_{ir};Q)] =
\Jac{ir}{1;m+1}(p_r,\ti{p}_{ir};Q)
\,\frac{\rd^d p_r}{(2\pi)^{d-1}}\delta_{+}(p_r^2)\,,
\label{eq:dp_Cir}
\eeq
where the Jacobian is
\beq
\Jac{ir}{1;m+1}(p_r,\ti{p}_{ir};Q) =
y_{\wti{ir}Q}\,\frac{(1-\alpha_{ir})^{m(d-2)-1}\,\Theta(1-\alpha_{ir})}
{2 (1 - y_{\wti{ir}Q}) \alpha_{ir} + y_{r\wti{ir}}+y_{\wti{ir}Q}-y_{rQ}}
\,.
\label{eq:Jac_Cir}
\eeq
In this equation $\alpha_{ir}$ is the physical (falling between 0 and
1) solution of the constraint
\beq
\frac{p_i^2}{Q^2} = 
(1 - y_{\wti{ir}Q})\,\alpha_{ir}^2 
+ (y_{r\wti{ir}}+y_{\wti{ir}Q}-y_{rQ})\,\alpha_{ir}
- y_{r\wti{ir}} = 0\,.
\eeq
In order to implement the subtraction scheme, this solution is not
required, the momentum mapping given in \eqn{eq:PS_Cir} is sufficient.

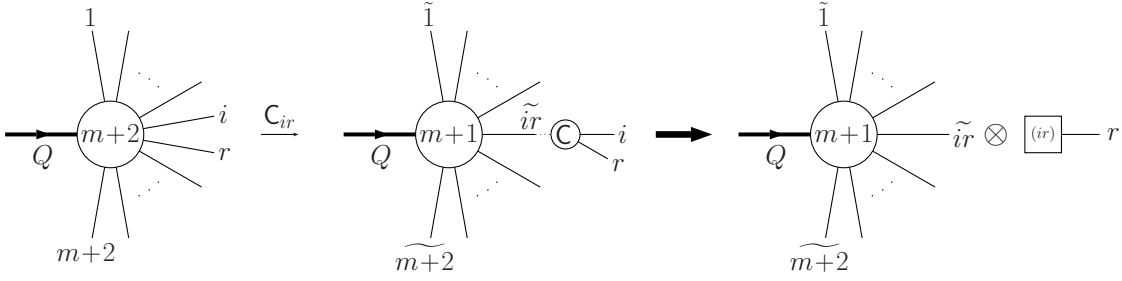
\begin{figure}
\begin{center}
\begin{pspicture}(0,0)(16,3.5)


\scalebox{0.5}{%
\psline[linewidth=3pt,arrowinset=0]{->}(0.2,4)(1.4,4)
\uput{0.3}[d](1.2,4){\LARGE $Q$}
\psline[linewidth=3pt](1.2,4)(2.2,4)

\SpecialCoor
\psline[origin={-3,-4}](0.8;100)(2.8;100)\uput{2.9}[100]{0}(3,4){\LARGE $1$}
\psline[origin={-3,-4}](0.8;80)(2.8;80)
\psdots*[origin={-3,-4},dotscale=0.3](1.8;65)(1.8;55)(1.8;45)
\psline[origin={-3,-4}](0.8;30)(2.8;30)
\psline[origin={-3,-4}](0.8;10)(2.8;10)\uput{2.9}[10]{0}(3,4){\LARGE $i$}
\psline[origin={-3,-4}](0.8;-10)(2.8;-10)\uput{2.9}[-10]{0}(3,4){\LARGE $r$}
\psline[origin={-3,-4}](0.8;-30)(2.8;-30)
\psdots*[origin={-3,-4},dotscale=0.3](1.8;-65)(1.8;-55)(1.8;-45)
\psline[origin={-3,-4}](0.8;-80)(2.8;-80)
\psline[origin={-3,-4}](0.8;-100)(2.8;-100)\uput{2.9}[-100]{0}(3,4){\LARGE $m\!+\!2$}

\pscircle[fillstyle=solid,fillcolor=white](3,4){0.9}\rput(3,4){\LARGE $m\!+\!2$}

\psline{->}(7,4)(8,4)\uput[u](7.5,4){{\LARGE $\mathsf{C}_{ir}$}}

\psline[linewidth=3pt,arrowinset=0]{->}(9.2,4)(10.4,4)\uput{0.3}[d](10.2,4){\LARGE $Q$}
\psline[linewidth=3pt](10.2,4)(11.2,4)

\SpecialCoor
\psline[origin={-12,-4}](0.8;100)(2.8;100)\uput{2.9}[100]{0}(12,4){\LARGE $\tilde{1}$}
\psline[origin={-12,-4}](0.8;80)(2.8;80)
\psdots*[origin={-12,-4},dotscale=0.3](1.8;65)(1.8;55)(1.8;45)
\psline[origin={-12,-4}](0.8;30)(2.8;30)
\psline[origin={-12,-4}](0.8;0)(2.3;0)\uput{1.9}[12]{0}(12,4){\LARGE $\widetilde{ir}$}
\psline[origin={-12,-4}](0.8;-30)(2.8;-30)
\psdots*[origin={-12,-4},dotscale=0.3](1.8;-65)(1.8;-55)(1.8;-45)
\psline[origin={-12,-4}](0.8;-80)(2.8;-80)
\psline[origin={-12,-4}](0.8;-100)(2.8;-100)\uput{2.9}[-100]{0}(12,4){\LARGE $\widetilde{m\!+\!2}$}

\pscircle[fillstyle=solid,fillcolor=white](12,4){0.9}\rput(12,4){\LARGE $m\!+\!1$}

\psline[origin={-12,-4},linestyle=dotted](2.3;0)(2.8;0)
\uput{2.95}[0]{0}(12,4){\LARGE $\mathsf{C}$}
\psline[origin={-15.1,-4}](0.3;0)(1.3;0)\uput{1.4}[0]{0}(15.1,4){\LARGE $i$}
\psline[origin={-15.1,-4}](0.3;-30)(1.3;-30)\uput{1.4}[-30]{0}(15.1,4){\LARGE $r$}

\pscircle[origin={-12,-4},fillstyle=solid,fillcolor=white](3.1;0){0.4}
\rput(15.1,4){\LARGE $\mathsf{C}\,$}


\psline[linewidth=5pt,arrowinset=0]{->}(17.5,4)(19,4)

\psline[linewidth=3pt,arrowinset=0]{->}(19.7,4)(20.9,4)\uput{0.3}[d](20.7,4){\LARGE $Q$}
\psline[linewidth=3pt](20.7,4)(21.7,4)

\SpecialCoor
\psline[origin={-22.5,-4}](0.8;100)(2.8;100)\uput{2.9}[100]{0}(22.5,4){\LARGE $\tilde{1}$}
\psline[origin={-22.5,-4}](0.8;80)(2.8;80)
\psdots*[origin={-22.5,-4},dotscale=0.3](1.8;65)(1.8;55)(1.8;45)
\psline[origin={-22.5,-4}](0.8;30)(2.8;30)
\psline[origin={-22.5,-4}](0.8;0)(2.8;0)\uput{2.9}[0]{0}(22.5,4){\LARGE $\widetilde{ir}$}
\psline[origin={-22.5,-4}](0.8;-30)(2.8;-30)
\psdots*[origin={-22.5,-4},dotscale=0.3](1.8;-65)(1.8;-55)(1.8;-45)
\psline[origin={-22.5,-4}](0.8;-80)(2.8;-80)
\psline[origin={-22.5,-4}](0.8;-100)(2.8;-100)\uput{2.9}[-100]{0}(22.5,4){\LARGE $\widetilde{m\!+\!2}$}

\pscircle[fillstyle=solid,fillcolor=white](22.5,4){0.9}\rput(22.5,4){\LARGE $m\!+\!1$}

\uput{3.7}[0]{0}(22.5,4){\Huge $\otimes$}
\psline[origin={-22.5,-4}](5.8;0)(6.8;0)
\psframe[origin={-22.5,-4},fillstyle=solid,fillcolor=white]%
(4.8,-0.5)(5.8,0.5)
\rput(27.8,4){$(ir)$}
\uput{7.0}[0]{0}(22.5,4){\LARGE $r$}
}
\end{pspicture}
\end{center}
\vskip-5mm
\caption{Graphical representation of the singly-collinear momentum
mapping and the implied factorization of the phase space.}
\label{fig:Cir}
\end{figure}

The collinear momentum mapping of \eqn{eq:PS_Cir} and the implied
phase-space factorization of \eqnss{eq:PSfact_Cir}{eq:Jac_Cir} are
represented graphically in \fig{fig:Cir}. The leftmost picture
represents the $(m+2)$-parton phase space $\PS{m+2}(\mom{};Q)$. In the
circle we denote the number of final-state partons. The picture in the
middle represents the result of the mapping of momenta in
\eqn{eq:PS_Cir}. The dots between the momentum $\ti{p}_{ir}$ and the
circle with the two momenta $p_i$ and $p_r$ means that the latter two are
replaced with $\ti{p}_{ir}$. This mapping implies the exact factorization
of the phase space, written in \eqn{eq:PSfact_Cir} and represented by
the picture on the right. The first factor is the phase space of
$(m+1)$-partons, $\PS{m+1}(\momt{(ir)}{+1};Q)$, and the second is 
$[\rd p_{1}^{(ir)}]$. In the latter the box represents the Jacobian on
\eqn{eq:Jac_Cir} and the line means the one-particle phase-space measure.
We shall use similar graphical representations of other momentum
mappings and implied factorizations of the phase space.

In the collinear limit, when $p_i^\mu || p_r^\mu$, the transverse
momentum behaves as in \eqn{eq:kTir2}, $\tzz{i}{r} \to z_i$ and
$\tzz{r}{i} \to z_r$. Furthermore, $\alpha_{ir}$ tends to zero so
$\ti{p}_{ir}^\mu \to p_i^\mu + p_r^\mu$ and $\ti{p}_n \to p_n$, i.e.~the
tildes disappear from \fig{fig:Cir}.  Consequently, the counterterm
reproduces the collinear behaviour of the squared matrix element in
this limit.


\subsection{Soft-type counterterms}
\label{ssec:RR_A1_Sr}

\subtitle{Counterterms}

The soft-type\footnote{The expression `soft-type' refers to the
momentum mapping used to define these terms.} terms are the singly-soft
and singly soft-collinear counterterms
\beeq
\cS{r_g}{(0,0)}(\mom{}) \aand= 
-8\pi\as\mu^{2\eps}\sum_{i}\sum_{k\ne i} \frac12 \calS_{ik}(r)
\SME{m+1,(i,k)}{0}{\momt{(r)}{+1}}\,,
\label{eq:Sr00}
\\
\cCS{ir_g}{r_g}{(0,0)}(\mom{}) \aand= 
8\pi\as\mu^{2\eps} \frac{1}{s_{ir}}\frac{2\tzz{i}{r}}{\tzz{r}{i}}\,\bT_i^2\,
\SME{m+1}{0}{\momt{(r)}{+1}}\,.
\label{eq:CirSr00}
\eeeq
If $r$ is a quark or antiquark, $\cS{r}{(0,0)}(\mom{})$ and $\cCS{ir}{r}{(0,0)}(\mom{})$
are both zero. The eikonal factor in \eqn{eq:Sr00} is
\beq
\calS_{ik}(r) = \frac{2 s_{ik}}{s_{ir} s_{rk}}\,,
\label{eq:Sikr}
\eeq
and the momentum fractions entering \eqn{eq:CirSr00} are given in
\eqn{eq:zt2}. The operator $\bT_i$ is the colour charge of parton $i$ 
\cite{Catani:1996vz}.

\subtitle{Momentum mapping and phase space factorization}

The $m+1$ momenta, $\momt{(r)}{+1} \equiv
\{\ti{p}_1,\ldots,\ti{p}_{r-1},\ti{p}_{r+1},\ldots,\ti{p}_{m+2}\}$
(the momentum with index $r$ is absent), entering the matrix elements
on the right hand sides of \eqns{eq:Sr00}{eq:CirSr00} are defined by
first rescaling all the hard momenta by a factor $1/\lambda_r$ and then
transforming all of the rescaled momenta as
\beq
\ti{p}_n^{\mu} = \Lambda^{\mu}_{\nu}[Q,(Q-p_r)/\lambda_r] (p_n^{\nu}/\lambda_r)\,,
\qquad n\ne r\,,
\label{eq:PS_Sr}
\eeq
where
\beq
\lambda_r = \sqrt{1-y_{rQ}}\,,
\label{eq:lambdar}
\eeq
and
\beq
\Lambda^{\mu}_{\nu}[K,\wti{K}] = g^{\mu}_{\nu}
- \frac{2(K+\wti{K})^{\mu}(K+\wti{K})_{\nu}}{(K+\wti{K})^{2}} 
+ \frac{2K^{\mu}\wti{K}_{\nu}}{K^2}\,.
\label{eq:LambdaKKt}
\eeq
The matrix $\Lambda^{\mu}_{\nu}[K,\wti{K}]$ generates a (proper) Lorentz
transformation, provided $K^2 = \wti{K}^2 \ne 0$. Since $p_r^\mu$ is
massless ($p_r^2 = 0$), the total four-momentum is again conserved,
\beq
Q^\mu = p_r^\mu + \sum_n^{m+1} p_n^\mu = \sum_n^{m+1} \ti{p}_n^{\mu}
\,.
\label{eq:conserve}
\eeq
We will find it convenient to introduce the notation
\beq
\mom{} \smap{r} \momt{(r)}{+1}
\label{eq:smap}
\eeq
to denote the above momentum mapping. Here $r$ is the label of any
momentum of the original momentum set $\mom{}$ that is absent from the
set on the right hand side.

The momentum mapping of \eqn{eq:PS_Sr} also leads to exact phase space
factorization 
\beq
\PS{m+2}{}(\mom{};Q)=\PS{m+1}{}(\momt{(r)}{+1};Q)
\: [\rd p_{1;m+1}^{(r)}(p_r;Q)]\,.
\label{eq:PSfact_Sr}
\eeq
The $m+1$ momenta in the first factor on the right hand side are those
of \eqn{eq:PS_Sr}.  The factorized one-parton phase space
$[\rd p_{1;m+1}^{(r)}(p_r;Q)]$ is
\beq 
[\rd p_{1;m+1}^{(r)}(p_r;Q)] =
\Jac{r}{1;m+1}(p_r;Q)
\,\frac{\rd^d p_r}{(2\pi)^{d-1}}\delta_{+}(p_r^2)\,,
\label{eq:dp_Sr}
\eeq
with Jacobian 
\beq
\Jac{r}{1;m+1}(p_r;Q) = \lambda_{r}^{m(d-2)-2}\Theta(\lambda_{r})\,.
\label{eq:Jac_Sr}
\eeq

\begin{figure}
\begin{center}
\begin{pspicture}(0,0)(16,4)

\scalebox{0.5}{%

\psline[linewidth=3pt,arrowinset=0]{->}(0.2,4)(1.4,4)\uput{0.3}[d](1.2,4){\LARGE $Q$}
\psline[linewidth=3pt](1.2,4)(2.2,4)

\SpecialCoor
\psline[origin={-3,-4}](0.8;100)(2.8;100)\uput{2.9}[100]{0}(3,4){\LARGE $1$}
\psline[origin={-3,-4}](0.8;80)(2.8;80)
\psdots*[origin={-3,-4},dotscale=0.3](1.8;65)(1.8;55)(1.8;45)
\psline[origin={-3,-4}](0.8;30)(2.8;30)
\psline[origin={-3,-4}](0.8;0)(2.8;0)\uput{2.9}[0]{0}(3,4){\LARGE $r$}
\psline[origin={-3,-4}](0.8;-30)(2.8;-30)
\psdots*[origin={-3,-4},dotscale=0.3](1.8;-65)(1.8;-55)(1.8;-45)
\psline[origin={-3,-4}](0.8;-80)(2.8;-80)
\psline[origin={-3,-4}](0.8;-100)(2.8;-100)\uput{2.9}[-100]{0}(3,4){\LARGE $m\!+\!2$}

\pscircle[fillstyle=solid,fillcolor=white](3,4){0.9}\rput(3,4){\LARGE $m\!+\!2$}

\psline{->}(7,4)(8,4)\uput[u](7.5,4){{\LARGE $\mathsf{S}_{r}$}}

\psline[linewidth=3pt,arrowinset=0]{->}(9.2,4)(10.4,4)\uput{0.3}[d](10.2,4){\LARGE $Q$}
\psline[linewidth=3pt](10.2,4)(11.2,4)

\SpecialCoor
\psline[origin={-12,-4}](0.8;100)(2.8;100)\uput{2.9}[100]{0}(12,4){\LARGE $\tilde{1}$}
\psline[origin={-12,-4}](0.8;80)(2.8;80)
\psdots*[origin={-12,-4},dotscale=0.3](1.8;65)(1.8;55)(1.8;45)
\psline[origin={-12,-4}](0.8;30)(2.8;30)
\psline[origin={-12,-4}](0.8;-30)(2.8;-30)
\psdots*[origin={-12,-4},dotscale=0.3](1.8;-65)(1.8;-55)(1.8;-45)
\psline[origin={-12,-4}](0.8;-80)(2.8;-80)
\psline[origin={-12,-4}](0.8;-100)(2.8;-100)\uput{2.9}[-100]{0}(12,4){\LARGE $\widetilde{m\!+\!2}$}
\psline[origin={-12,-4},linestyle=dotted](0.8;0)(2.8;0)

\pscircle[fillstyle=solid,fillcolor=white](12,4){0.9}\rput(12,4){\LARGE $m\!+\!1$}

\psline[origin={-12,-4}](3.4;0)(4.4;0)\uput{1.4}[0]{0}(15.1,4){\LARGE $r$}

\pscircle[origin={-12,-4},fillstyle=solid,fillcolor=white](3.1;0){0.4}
\rput(15.1,4){\LARGE $\mathsf{S}$}


\psline[linewidth=5pt,arrowinset=0]{->}(17.5,4)(19,4)

\psline[linewidth=3pt,arrowinset=0]{->}(19.7,4)(20.9,4)\uput{0.3}[d](20.7,4){\LARGE $Q$}
\psline[linewidth=3pt](20.7,4)(21.7,4)

\SpecialCoor
\psline[origin={-22.5,-4}](0.8;100)(2.8;100)\uput{2.9}[100]{0}(22.5,4){\LARGE $\tilde{1}$}
\psline[origin={-22.5,-4}](0.8;80)(2.8;80)
\psdots*[origin={-22.5,-4},dotscale=0.3](1.8;65)(1.8;55)(1.8;45)
\psline[origin={-22.5,-4}](0.8;30)(2.8;30)
\psline[origin={-22.5,-4}](0.8;-30)(2.8;-30)
\psdots*[origin={-22.5,-4},dotscale=0.3](1.8;-65)(1.8;-55)(1.8;-45)
\psline[origin={-22.5,-4}](0.8;-80)(2.8;-80)
\psline[origin={-22.5,-4}](0.8;-100)(2.8;-100)\uput{2.9}[-100]{0}(22.5,4){\LARGE $\widetilde{m\!+\!2}$}

\pscircle[fillstyle=solid,fillcolor=white](22.5,4){0.9}\rput(22.5,4){\LARGE $m\!+\!1$}

\uput{3.2}[0]{0}(22.5,4){\Huge $\otimes$}
\psline[origin={-22.5,-4}](5.8;0)(6.8;0)
\psframe[origin={-22.5,-4},fillstyle=solid,fillcolor=white]%
(4.8,-0.5)(5.8,0.5)
\rput(27.8,4){$(r)$}
\uput{7.0}[0]{0}(22.5,4){\LARGE $r$}
}
\end{pspicture}
\end{center}
\vskip-5mm
\caption{Graphical representation of the singly-soft momentum
mapping and the implied factorization of the phase space.}
\label{fig:Sr}
\end{figure}
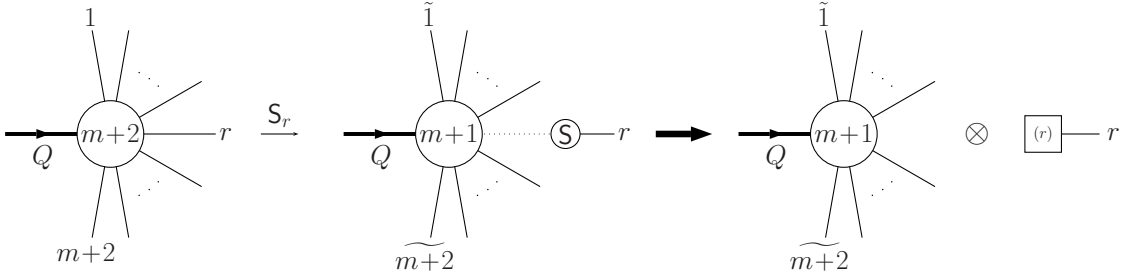

Similarly to the graphical representation of the collinear momentum
mapping and phase-space factorization, depicted in \fig{fig:Cir}, we
present the graphical representation of \eqnss{eq:PS_Sr}{eq:Jac_Sr}
in \fig{fig:Sr}. The middle picture in this figure represents the
soft-type mapping of \eqn{eq:PS_Sr}. The dots between the main circle and
the circle with momentum $p_r$ mean that the latter does not take away
momentum from $Q^\mu$ (see \eqn{eq:conserve}).

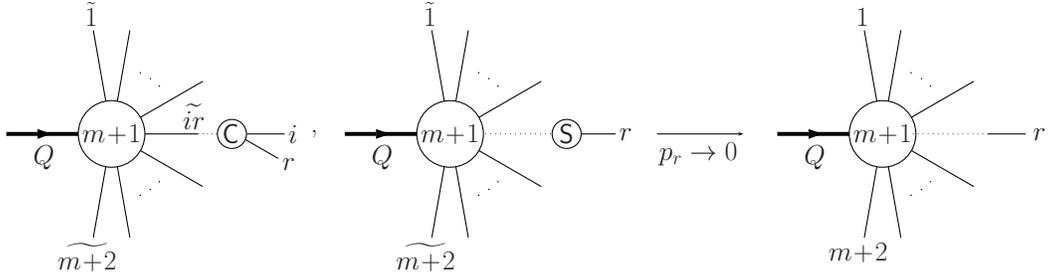
\begin{figure}[h]
\begin{center}
\begin{pspicture}(0,0)(14,4)

\scalebox{0.5}{%

\psline[linewidth=3pt,arrowinset=0]{->}(0.2,4)(1.4,4)
\uput{0.3}[d](1.2,4){\LARGE $Q$}
\psline[linewidth=3pt](1.2,4)(2.2,4)

\SpecialCoor
\psline[origin={-3,-4}](0.8;100)(2.8;100)\uput{2.9}[100]{0}(3,4){\LARGE $\tilde{1}$}
\psline[origin={-3,-4}](0.8;80)(2.8;80)
\psdots*[origin={-3,-4},dotscale=0.3](1.8;65)(1.8;55)(1.8;45)
\psline[origin={-3,-4}](0.8;30)(2.8;30)
\psline[origin={-3,-4}](0.8;0)(2.3;0)\uput{1.9}[12]{0}(3,4){\LARGE $\widetilde{ir}$}
\psline[origin={-3,-4}](0.8;-30)(2.8;-30)
\psdots*[origin={-3,-4},dotscale=0.3](1.8;-65)(1.8;-55)(1.8;-45)
\psline[origin={-3,-4}](0.8;-80)(2.8;-80)
\psline[origin={-3,-4}](0.8;-100)(2.8;-100)\uput{2.9}[-100]{0}(3,4){\LARGE $\widetilde{m\!+\!2}$}

\pscircle[fillstyle=solid,fillcolor=white](3,4){0.9}\rput(3,4){\LARGE $m\!+\!1$}

\psline[origin={-3,-4},linestyle=dotted](2.3;0)(2.8;0)
\pscircle[origin={-3,-4},fillstyle=solid,fillcolor=white](3.2;0){0.4}
\uput{2.95}[0]{0}(3,4){\LARGE $\mathsf{C}$}

\psline[origin={-6.3,-4}](0.3;0)(1.3;0)\uput{1.4}[0]{0}(6.3,4){\LARGE $i$}
\psline[origin={-6.3,-4}](0.3;-30)(1.3;-30)\uput{1.4}[-30]{0}(6.3,4){\LARGE $r$}
\uput{0}[0]{0}(8.3,4){\LARGE ,}

\psline[linewidth=3pt,arrowinset=0]{->}(9.2,4)(10.4,4)\uput{0.3}[d](10.2,4){\LARGE $Q$}
\psline[linewidth=3pt](10.2,4)(11.2,4)
\SpecialCoor
\psline[origin={-12,-4}](0.8;100)(2.8;100)\uput{2.9}[100]{0}(12,4){\LARGE $\tilde{1}$}
\psline[origin={-12,-4}](0.8;80)(2.8;80)
\psdots*[origin={-12,-4},dotscale=0.3](1.8;65)(1.8;55)(1.8;45)
\psline[origin={-12,-4}](0.8;30)(2.8;30)
\psline[origin={-12,-4}](0.8;-30)(2.8;-30)
\psdots*[origin={-12,-4},dotscale=0.3](1.8;-65)(1.8;-55)(1.8;-45)
\psline[origin={-12,-4}](0.8;-80)(2.8;-80)
\psline[origin={-12,-4}](0.8;-100)(2.8;-100)\uput{2.9}[-100]{0}(12,4){\LARGE $\widetilde{m\!+\!2}$}
\psline[origin={-12,-4},linestyle=dotted](0.8;0)(2.8;0)

\pscircle[fillstyle=solid,fillcolor=white](12,4){0.9}\rput(12,4){\LARGE $m\!+\!1$}

\psline[origin={-12,-4}](3.4;0)(4.4;0)\uput{1.4}[0]{0}(15.1,4){\LARGE $r$}

\pscircle[origin={-12,-4},fillstyle=solid,fillcolor=white](3.1;0){0.4}
\rput(15.1,4){\LARGE $\mathsf{S}$}


\psline{->}(17.5,4)(19.8,4)\uput[u](18.6,3){{\LARGE $p_{r}\to 0$}}

\psline[linewidth=3pt,arrowinset=0]{->}(20.7,4)(21.9,4)\uput{0.3}[d](21.7,4){\LARGE $Q$}
\psline[linewidth=3pt](21.7,4)(22.7,4)
\SpecialCoor
\psline[origin={-23.5,-4}](0.8;100)(2.8;100)\uput{2.9}[100]{0}(23.5,4){\LARGE 1}
\psline[origin={-23.5,-4}](0.8;80)(2.8;80)
\psdots*[origin={-23.5,-4},dotscale=0.3](1.8;65)(1.8;55)(1.8;45)
\psline[origin={-23.5,-4}](0.8;30)(2.8;30)
\psline[origin={-23.5,-4}](0.8;-30)(2.8;-30)
\psdots*[origin={-23.5,-4},dotscale=0.3](1.8;-65)(1.8;-55)(1.8;-45)
\psline[origin={-23.5,-4}](0.8;-80)(2.8;-80)
\psline[origin={-23.5,-4}](0.8;-100)(2.8;-100)\uput{2.9}[-100]{0}(23.5,4){\LARGE $m\!+\!2$}
\pscircle[fillstyle=solid,fillcolor=white](23.5,4){0.9}\rput(23.5,4){\LARGE $m\!+\!1$}
\psline[origin={-23.5,-4},linestyle=dotted](0.8;0)(2.8;0)
\psline[origin={-23.5,-4}](2.8;0)(3.8;0)
\uput{4.0}[0]{0}(23.5,4){\LARGE $r$}
}
\end{pspicture}
\end{center}
\caption{Graphical representation of the soft limit of collinear and
soft-type mappings.}
\label{fig:softlimit}
\end{figure}

The soft-type terms are defined on the same phase space, therefore, in
the collinear limit, when $p_i^\mu || p_r^\mu$, the soft-collinear
counterterm regularizes the kinematical singularity of the soft
counterterm by construction.  In the soft limit, when $p_r^\mu \to 0$,
in \eqns{eq:PS_Cir}{eq:alphair} $\alpha_{ir} \to 0$ and
$\ti{p}_{ir}^\mu \to p_i^\mu$, therefore, the momenta obtained in the
collinear mapping tend to the same momenta as those obtained in the
soft mapping, i.e.~all momenta with tilde tend to the corresponding
original momenta both in the case of collinear and soft momentum
mappings, with the soft momentum being dropped as shown in
\fig{fig:softlimit}. Thus the soft-collinear counterterm regularizes
the kinematical singularity of the collinear counterterm.  At the same
time the soft one regularizes the squared matrix element by construction.  

In closing this section, we note that the jet function $J_{m+1}$ that
multiplies $\dsiga{RR}{1}_{m+2}$ in \eqn{eq:sigmaNNLO_v3} is a function
of the $m+1$ momenta $\momt{(ir)}{+1}$ and $\momt{(r)}{+1}$ for the
collinear and soft-type countertems, respectively.

%
%

\section{Doubly-unresolved counterterms}
\label{sec:RR_A2}

The doubly-unresolved counterterm is
\beeq
\bcA{2}{(0)}\M{m+2}{(0)} \aand=
\sum_{r}\sum_{s}\Bigg\{
\sum_{i\ne r,s}\Bigg[\frac16\, \cC{irs}{(0,0)}(\mom{}) 
+ \sum_{j\ne i,r,s} \frac18\, \cC{ir;js}{(0,0)}(\mom{})
\nn\\ &&
+ \frac12\,\Bigg( \cSCS{ir;s}{(0,0)}(\mom{})
- \cC{irs}{}\cSCS{ir;s}{(0,0)}(\mom{}) 
- \sum_{j\ne i,r,s} \cC{ir;js}{} \cSCS{ir;s}{(0,0)}(\mom{}) \Bigg)
\nn\\ &&
- \cSCS{ir;s}{}\cS{rs}{(0,0)}(\mom{}) 
- \frac12\, \cC{irs}{}\cS{rs}{(0,0)}(\mom{})
+ \cC{irs}{}\cSCS{ir;s}{}\cS{rs}{(0,0)}(\mom{})
\nn\\ &&
+ \sum_{j\ne i,r,s} \frac12\, \cC{ir;js}{}\cS{rs}{(0,0)}(\mom{})\Bigg] 
+ \frac12\, \cS{rs}{(0,0)}(\mom{})
\Bigg\}
\,.
\label{eq:RR_A2}
\eeeq


\subsection{Triple collinear counterterm}
\label{ssec:RR_A2_Cirs}

\subtitle{Counterterm}

The triple collinear counterterm reads
\beq
\cC{irs}{(0,0)}(\mom{}) = 
(8\pi\as\mu^{2\eps})^2\frac{1}{s_{irs}^2}
\bra{m}{(0)}{(\momt{(irs)}{})}
\hP_{f_i f_r f_s}(\{\tzz{j}{kl},s_{jk},\kTt{j,kl}\};\eps)
\ket{m}{(0)}{(\momt{(irs)}{})}\,.
\label{eq:Cirs}
\eeq
The $\hP_{f_i f_r f_s}(\{\tzz{j}{kl},s_{jk},\kTt{j,kl}\};\eps)$ kernels
are defined to be the specific forms of the triple parton splitting
functions introduced in \Ref{Somogyi:2005xz}, with one important
modification: In the gluon splitting functions $\hP_{g_i q_r \qb_s}$
and $\hP_{g_i g_r g_s}$\footnote{In the case of the $g\to g g g$
splitting, \Ref{Somogyi:2005xz} uses the original definition of
\Ref{Catani:1999ss}, not spelled out explicitly.}, the
azimuth-dependent terms that depend on the transverse momenta have
always to be written in the form $\kT{j}^\mu\kT{k}^\nu/\kT{j}\cdot\kT{k}$
($k$ can be equal to $j$), otherwise the collinear behaviour of the
counterterm cannot be matched with that of the single collinear
counterterm in the singly-unresolved phase space region. The correct
azimuth-dependence can be achieved by making use of the following
identites:
\beeq
\kT{j}^\mu \kT{j}^\nu \aand= 
\Big(-z_j (1-z_j) s_{jkl} + z_j s_{kl}\Big)
\frac{\kT{j}^\mu \kT{j}^\nu}{\kT{j}^2}
\,,
\\[2mm]
2\,\kT{j}^\mu \kT{k}^\nu \aand= 
\Big(s_{jk} + 2\,z_j z_k s_{jkl} - z_i s_{j(ik)} - z_j s_{i(jk)}\Big)
\frac{\kT{j}^\mu \kT{k}^\nu}{\kT{j}\cdot \kT{k}}
\,,
\eeeq
where $\{k,l\} = \{i,r,s\}\setminus \{j\}$ and $j$ can be $i$, $r$ or
$s$.
Note that \Ref{Somogyi:2005xz} does not consider the splitting
function for the case of final-state fermions with identical flavours.
The reason is that it consists of two terms of the type of different
flavours plus a third term corresponding to an interference
contribution denoted by $\la \hP^{(\rm id)}_{\qb_1 q_2 q_3}\ra$ in
\Ref{Catani:1999ss}. The different flavour contributions were given in
\Ref{Somogyi:2005xz}, while the interference term can directly be taken
from \Ref{Catani:1999ss} with the simple substitution of indices $1 \to
r$, $2 \to s$ and $3 \to i$ in order to match our notation. This latter
term does not require any special care because it does not have a leading
singularity in any of the singly-, or other doubly-unresolved
regions of the phase space, apart from the triple collinear one.  

The momentum fractions in \eqn{eq:Cirs} are defined similarly as for
the collinear case given in \eqn{eq:zt2} 
\beq
\tzz{i}{rs} = \frac{y_{iQ}}{y_{(irs)Q}}\,,
\qquad
\tzz{r}{is} = \frac{y_{rQ}}{y_{(irs)Q}}\,,
\qquad
\tzz{s}{ir} = \frac{y_{sQ}}{y_{(irs)Q}}\,,
\label{eq:zt3}
\eeq
with $\tzz{i}{rs} + \tzz{r}{is} + \tzz{s}{ir} = 1$.  The transverse
momentum $\kTt{r,is}$ is
\beq
\kTt{r,is}^{\mu} = 
  \zeta_{r,is} p_i^{\mu}
- \zeta_{i,rs} p_r^{\mu}
+ \zeta_{r,is} p_s^{\mu}
- \zeta_{s,ir} p_r^{\mu}
+ \zeta_{ris}\,\ti{p}_{irs}^{\mu}\,,
\label{eq:kTtris}
\eeq
where\footnote{Note that the indices of $\zeta_{ris}$ are ordered!}
\beq
\zeta_{i,rs} = \tzz{i}{rs} - \frac{y_{(rs)i}}{\alpha_{irs}\,y_{(irs)Q}}
\,,\qquad
\zeta_{ris} =
\frac{y_{ir}-y_{rs}-2\tzz{r}{is}y_{irs}}{\alpha_{irs}\,y_{\wti{irs}Q}}
\,.
\eeq
The expression for $\kTt{s,ir}$ is obtained from \eqn{eq:kTtris} by
simply interchanging the indices $r$ and $s$, while
$\kTt{i,rs} = -\kTt{r,is}-\kTt{s,ir}$. Similarly, $\zeta_{r,is}$ and 
$\zeta_{s,ir}$ are obtained from $\zeta_{i,rs}$, while $\zeta_{irs}$
and $\zeta_{sir}$ from $\zeta_{ris}$ by interchanging the indices.
We define $\ti{p}_{irs}^{\mu}$ and $\alpha_{irs}$ in
\eqns{eq:PS_Cirs}{eq:alphairs}.  This choice for the transverse
momentum is also perpendicular to the parent momentum
$\ti{p}_{irs}^{\mu}$ as the one given in \eqn{eq:kTtir} and it ensures
also that in the triple collinear limit $p_i^\mu || p_r^\mu || p_s^\mu$,
relations of the type (see \Ref{Catani:1999ss})
\beq
s_{rs} \simeq - \tzz{r}{is}\,\tzz{s}{ir}
\left(\frac{\kTt{r,is}}{\tzz{r}{is}} - \frac{\kTt{s,ir}}{\tzz{s}{ir}}\right)^2
\,,
\label{eq:srs_TC}
\eeq
as well as $\kTt{r,is} \to \kTt{r}$ are fulfilled. In a NNLO
computation the longitudinal component in \eqn{eq:kTtris} does not give
any contribution due to gauge invariance, therefore, we may choose
$\zeta_{ris} = 0$.  

\subtitle{Momentum mapping and phase space factorization}

The matrix elements on the right hand side of \eqn{eq:Cirs} are
evaluated with the $m$ momenta
$\momt{(irs)}{} \equiv \{\ti{p}_1,\ldots,\ti{p}_{irs},\dots,\ti{p}_{m+2}\}$ 
defined as
\beeq
\ti{p}_{irs}^{\mu} = \frac{1}{1-\alpha_{irs}}(p_i^{\mu} + p_r^{\mu} + p_s^{\mu}
                                              - \alpha_{irs} Q^{\mu})\,,
\qquad
\ti{p}_n^{\mu} = \frac{1}{1-\alpha_{irs}} p_n^{\mu}\,,
\qquad n\ne i,r,s\,,
\label{eq:PS_Cirs}
\eeeq
with
\beq
\alpha_{irs} =
\frac12\left[y_{(irs)Q}-\sqrt{y_{(irs)Q}^2 - 4y_{irs}}\;\right]\,.
\label{eq:alphairs}
\eeq
Clearly \eqns{eq:PS_Cirs}{eq:alphairs} are generalizations of
\eqns{eq:PS_Cir}{eq:alphair}. The total four-momentum is again conserved,
\beq
Q^\mu = \ti{p}_{irs}^{\mu} + \sum_n \ti{p}_n^{\mu}
= p_i^\mu + p_r^\mu + p_s^{\mu} + \sum_n p_n^\mu\,.
\eeq

\begin{figure}
\begin{center}
\begin{pspicture}(0,0)(16,4)
\scalebox{0.5}{%

\psline[linewidth=3pt,arrowinset=0]{->}(0.2,4)(1.4,4)\uput{0.3}[d](1.2,4){\LARGE $Q$}
\psline[linewidth=3pt](1.2,4)(2.2,4)

\SpecialCoor
\psline[origin={-3,-4}](0.8;100)(2.8;100)\uput{2.9}[100]{0}(3,4){\LARGE $1$}
\psline[origin={-3,-4}](0.8;80)(2.8;80)
\psdots*[origin={-3,-4},dotscale=0.3](1.8;65)(1.8;55)(1.8;45)
\psline[origin={-3,-4}](0.8;30)(2.8;30)
\psline[origin={-3,-4}](0.8;15)(2.8;15)\uput{2.9}[15]{0}(3,4){\LARGE $i$}
\psline[origin={-3,-4}](0.8;0)(2.8;0)\uput{2.9}[0]{0}(3,4){\LARGE $r$}
\psline[origin={-3,-4}](0.8;-15)(2.8;-15)\uput{2.9}[-15]{0}(3,4){\LARGE $s$}
\psline[origin={-3,-4}](0.8;-30)(2.8;-30)
\psdots*[origin={-3,-4},dotscale=0.3](1.8;-65)(1.8;-55)(1.8;-45)
\psline[origin={-3,-4}](0.8;-80)(2.8;-80)
\psline[origin={-3,-4}](0.8;-100)(2.8;-100)\uput{2.9}[-100]{0}(3,4){\LARGE $m\!+\!2$}

\pscircle[fillstyle=solid,fillcolor=white](3,4){0.9}\rput(3,4){\LARGE $m\!+\!2$}

\psline{->}(7,4)(8,4)\uput[u](7.5,4){{\LARGE $\mathsf{C}_{irs}$}}

\psline[linewidth=3pt,arrowinset=0]{->}(9.2,4)(10.4,4)\uput{0.3}[d](10.2,4){\LARGE $Q$}
\psline[linewidth=3pt](10.2,4)(11.2,4)

\SpecialCoor
\psline[origin={-12,-4}](0.8;100)(2.8;100)\uput{2.9}[100]{0}(12,4){\LARGE $\tilde{1}$}
\psline[origin={-12,-4}](0.8;80)(2.8;80)
\psdots*[origin={-12,-4},dotscale=0.3](1.8;65)(1.8;55)(1.8;45)
\psline[origin={-12,-4}](0.8;30)(2.8;30)
\psline[origin={-12,-4}](0.8;0)(2.3;0)\uput{1.6}[12]{0}(12,4){\LARGE $\widetilde{irs}$}
\psline[origin={-12,-4}](0.8;-30)(2.8;-30)
\psdots*[origin={-12,-4},dotscale=0.3](1.8;-65)(1.8;-55)(1.8;-45)
\psline[origin={-12,-4}](0.8;-80)(2.8;-80)
\psline[origin={-12,-4}](0.8;-100)(2.8;-100)\uput{2.9}[-100]{0}(12,4){\LARGE $\widetilde{m\!+\!2}$}

\pscircle[fillstyle=solid,fillcolor=white](12,4){0.9}\rput(12,4){\LARGE $m$}

\psline[origin={-12,-4},linestyle=dotted](2.3;0)(2.8;0)
\psline[origin={-15.1,-4}](0.3;0)(1.3;0)\uput{1.4}[0]{0}(15.1,4){\LARGE $i$}
\psline[origin={-15.1,-4}](0.3;-30)(1.3;-30)\uput{1.4}[-30]{0}(15.1,4){\LARGE $r$}
\psline[origin={-15.1,-4}](0.3;-60)(1.3;-60)\uput{1.4}[-60]{0}(15.1,4){\LARGE $s$}

\pscircle[origin={-12,-4},fillstyle=solid,fillcolor=white](3.1;0){0.4}
\rput(15.1,4){\LARGE $\mathsf{C}\,$}


\psline[linewidth=5pt,arrowinset=0]{->}(17.5,4)(19,4)

\psline[linewidth=3pt,arrowinset=0]{->}(19.7,4)(20.9,4)\uput{0.3}[d](20.7,4){\LARGE $Q$}
\psline[linewidth=3pt](20.7,4)(21.7,4)

\SpecialCoor
\psline[origin={-22.5,-4}](0.8;100)(2.8;100)\uput{2.9}[100]{0}(22.5,4){\LARGE $\tilde{1}$}
\psline[origin={-22.5,-4}](0.8;80)(2.8;80)
\psdots*[origin={-22.5,-4},dotscale=0.3](1.8;65)(1.8;55)(1.8;45)
\psline[origin={-22.5,-4}](0.8;30)(2.8;30)
\psline[origin={-22.5,-4}](0.8;0)(2.8;0)\uput{2.9}[0]{0}(22.5,4){\LARGE $\widetilde{irs}$}
\psline[origin={-22.5,-4}](0.8;-30)(2.8;-30)
\psdots*[origin={-22.5,-4},dotscale=0.3](1.8;-65)(1.8;-55)(1.8;-45)
\psline[origin={-22.5,-4}](0.8;-80)(2.8;-80)
\psline[origin={-22.5,-4}](0.8;-100)(2.8;-100)\uput{2.9}[-100]{0}(22.5,4){\LARGE $\widetilde{m\!+\!2}$}

\pscircle[fillstyle=solid,fillcolor=white](22.5,4){0.9}\rput(22.5,4){\LARGE $m$}

\uput{3.85}[0]{0}(22.5,4){\Huge $\otimes$}
\psline[origin={-27.3,-4}](1;10)(2;10)\uput{2.2}[10]{0}(27.3,4){\LARGE $r$}
\psline[origin={-27.3,-4}](1;-10)(2;-10)\uput{2.2}[-10]{0}(27.3,4){\LARGE $s$}
\psframe[origin={-22.5,-4},fillstyle=solid,fillcolor=white]%
(4.8,-0.5)(5.8,0.5)
\rput(27.8,4){$(irs)$}
}
\end{pspicture}
\end{center}
\caption{Graphical representation of the triply-collinear momentum
mapping and the implied factorization of the phase space.}
\label{fig:Cirs}
\end{figure}
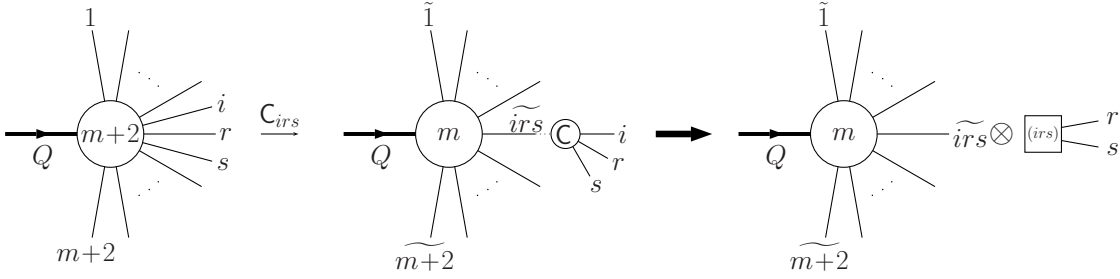

The momentum mapping of \eqn{eq:PS_Cirs} leads to exact phase-space
factorization in a form very similar to \eqn{eq:PSfact_Cir}
\beq
\PS{m+2}(\mom{};Q)=\PS{m}(\momt{(irs)}{};Q)
\: [\rd p_{2;m}^{(irs)}(p_r,p_s,\ti{p}_{irs};Q)]\,,
\label{eq:PSfact_Cirs}
\eeq
where the $m$ momenta in the first factor on the right hand side of
\eqn{eq:PSfact_Cirs} are those given in \eqn{eq:PS_Cirs}.  The explicit
expression for $[\rd p_{2;m}^{(irs)}(p_r,p_s,\ti{p}_{irs};Q)]$ is
\beq
[\rd p_{2;m}^{(irs)}(p_r,p_s,\ti{p}_{irs};Q)] =
\Jac{irs}{2;m}(p_r,p_s,\ti{p}_{irs};Q)
\,\frac{\rd^d p_r}{(2\pi)^{d-1}}\delta_{+}(p_r^2)
\,\frac{\rd^d p_s}{(2\pi)^{d-1}}\delta_{+}(p_s^2)\,,
\label{eq:dp_Cirs}
\eeq
where the Jacobian is
\beq
\Jac{irs}{2;m}(p_r,p_s,\ti{p}_{irs};Q) =
y_{\wti{irs}Q}\,
\frac{(1-\alpha_{irs})^{(m-1)(d-2)-1}\,\Theta(1-\alpha_{irs})}
{2 (1 - y_{\wti{irs}Q}) \alpha_{irs} +
y_{(rs)\wti{irs}}+y_{\wti{irs}Q}-y_{(rs)Q}}
\,.
\label{eq:Jac_Cirs}
\eeq
In this equation $\alpha_{irs}$ is the physical (falling between 0 and
1) solution of the constraint
\beq
\frac{p_i^2}{Q^2} = 
(1 - y_{\wti{irs}Q})\,\alpha_{irs}^2 
+ (y_{(rs)\wti{irs}}+y_{\wti{irs}Q}-y_{(rs)Q})\,\alpha_{irs}
+ y_{rs} - y_{(rs)\wti{irs}} = 0\,.
\eeq
We present the graphical representation of \eqnss{eq:PS_Cirs}{eq:Jac_Cirs}
in \fig{fig:Cirs}. The meaning of the various graphical elements is
analogous to those in \fig{fig:Cir}.

In the triply-collinear limit, when $p_i^\mu || p_r^\mu || p_s^\mu$,
the transverse momenta behave as in \eqn{eq:srs_TC}, $\tzz{i}{rs} \to
z_i$, $\tzz{r}{is} \to z_r$ and $\tzz{s}{ir} \to z_s$. Furthermore,
$\alpha_{irs}$ tends to zero so
$\ti{p}_{irs}^\mu \to p_i^\mu + p_r^\mu + p_s^\mu$
and $\ti{p}_n \to p_n$ (the tildes disappear from \fig{fig:Cirs}).
Consequently, the counterterm reproduces the
collinear behaviour of the squared matrix element in this limit.


\newpage
\subsection{Double collinear counterterm}
\label{ssec:RR_A2_Cirjs}

\subtitle{Counterterm}

The double collinear subtraction term reads
\beeq
&&
\cC{ir;js}{(0,0)}(\mom{}) = \label{eq:Cirjs}
(8\pi\as\mu^{2\eps})^2\frac{1}{s_{ir}s_{js}}
\\&&\times
\bra{m}{(0)}{(\momt{(ir;js)}{})}
\hP_{f_i f_r}^{(0)}(\tzz{i}{r},\tzz{r}{i},\kTt{ir;js};\eps)
\hP_{f_j f_s}^{(0)}(\tzz{j}{s},\tzz{s}{j},\kTt{js;ir};\eps)
\ket{m}{(0)}{(\momt{(ir;js)}{})}\,,
\nn
\eeeq
where $\hP_{f_k f_l}^{(0)}(z_k,z_l,\kT{};\eps)$ are given in
\eqnss{eq:Pqg0}{eq:Pgg0}.  The momentum fractions are defined in
\eqn{eq:zt2}, and the transverse momentum $\kTt{ir;js}$ is
\beq
\kTt{ir;js}^{\mu} = \zeta_{i,r} p_r^{\mu} - \zeta_{r,i} p_i^{\mu}
+ \zeta_{ir} \ti{p}_{ir}^{\mu}\,,
\label{eq:kTtirjs}
\eeq
i.e., it is formally identical to the $\kTt{i,r}^{\mu}$ defined in
\eqn{eq:kTtir}, although the definition of $\ti{p}_{ir}^{\mu}$ is
different in the two cases (cf.~\eqns{eq:PS_Cir}{eq:PS_Cirjs}).  In the
NNLO computation we may choose $\zeta_{ir}=0$ in \eqn{eq:kTtirjs}. Of
course, $\kTt{js;ir}$ given by \eqn{eq:kTtirjs} after the interchange of
indices $(i\leftrightarrow j,r\leftrightarrow s)$. \eqn{eq:alphair}
defines $\alpha_{ir}$ and $\alpha_{js}$, the latter with the same change
of indices as before.  

\subtitle{Momentum mapping and phase space factorization}

The $m$ momenta $\momt{(irjs)}{}\equiv
\{\ti{p},\ldots,\ti{p}_{ir},\ldots,\ti{p}_{js},\ldots,\ti{p}_{m+2}\}$
entering the matrix element on the right hand side of \eqn{eq:Cirjs}
are again given by a simple generalization of \eqn{eq:PS_Cir}
\beeq
&&
\ti{p}_{ir}^{\mu} = \frac{1}{1-\alpha_{ir}-\alpha_{js}}
                      (p_i^{\mu} + p_r^{\mu} - \alpha_{ir} Q^{\mu})\,,
\qquad
\ti{p}_{js}^{\mu} = \frac{1}{1-\alpha_{ir}-\alpha_{js}}
                      (p_j^{\mu} + p_s^{\mu} - \alpha_{js} Q^{\mu})\,,
\nn\\&&
\qquad\qquad\qquad\qquad\qquad
\ti{p}_n^{\mu} = \frac{1}{1-\alpha_{ir}-\alpha_{js}} p_n^{\mu}\,,
\qquad n\ne i,r,j,s\,.
\label{eq:PS_Cirjs}
\eeeq
The total four-momentum is clearly conserved.

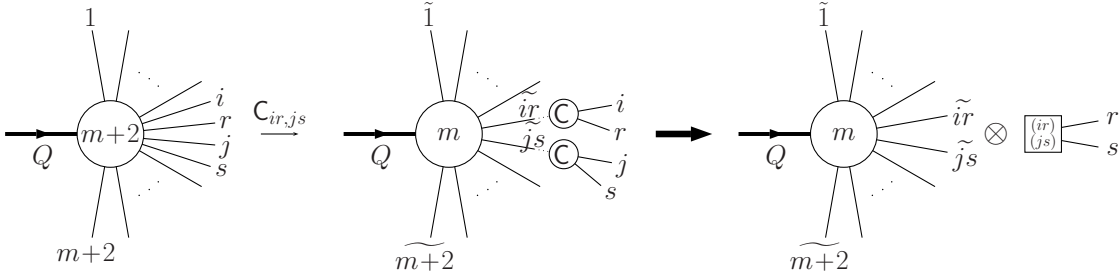
\begin{figure}
\begin{center}
\begin{pspicture}(0,0)(16,4)
\scalebox{0.5}{%

\psline[linewidth=3pt,arrowinset=0]{->}(0.2,4)(1.4,4)\uput{0.3}[d](1.2,4){\LARGE $Q$}
\psline[linewidth=3pt](1.2,4)(2.2,4)

\SpecialCoor
\psline[origin={-3,-4}](0.8;100)(2.8;100)\uput{2.9}[100]{0}(3,4){\LARGE $1$}
\psline[origin={-3,-4}](0.8;80)(2.8;80)
\psdots*[origin={-3,-4},dotscale=0.3](1.8;65)(1.8;55)(1.8;45)
\psline[origin={-3,-4}](0.8;30)(2.8;30)
\psline[origin={-3,-4}](0.8;18)(2.8;18)\uput{2.9}[18]{0}(3,4){\LARGE $i$}
\psline[origin={-3,-4}](0.8;6)(2.8;6)\uput{2.9}[6]{0}(3,4){\LARGE $r$}
\psline[origin={-3,-4}](0.8;-8)(2.8;-6)\uput{2.9}[-6]{0}(3,4){\LARGE $j$}
\psline[origin={-3,-4}](0.8;-18)(2.8;-18)\uput{2.9}[-18]{0}(3,4){\LARGE $s$}
\psline[origin={-3,-4}](0.8;-30)(2.8;-30)
\psdots*[origin={-3,-4},dotscale=0.3](1.8;-65)(1.8;-55)(1.8;-45)
\psline[origin={-3,-4}](0.8;-80)(2.8;-80)
\psline[origin={-3,-4}](0.8;-100)(2.8;-100)\uput{2.9}[-100]{0}(3,4){\LARGE $m\!+\!2$}

\pscircle[fillstyle=solid,fillcolor=white](3,4){0.9}\rput(3,4){\LARGE $m\!+\!2$}

\psline{->}(7,4)(8,4)\uput[u](7.5,4){{\LARGE $\mathsf{C}_{ir,js}$}}

\psline[linewidth=3pt,arrowinset=0]{->}(9.2,4)(10.4,4)\uput{0.3}[d](10.2,4){\LARGE $Q$}
\psline[linewidth=3pt](10.2,4)(11.2,4)

\SpecialCoor
\psline[origin={-12,-4}](0.8;100)(2.8;100)\uput{2.9}[100]{0}(12,4){\LARGE $\tilde{1}$}
\psline[origin={-12,-4}](0.8;80)(2.8;80)
\psdots*[origin={-12,-4},dotscale=0.3](1.8;65)(1.8;55)(1.8;45)
\psline[origin={-12,-4}](0.8;30)(2.8;30)
\psline[origin={-12,-4}](0.8;10)(2.3;10)\uput{1.9}[20]{0}(12,4){\LARGE $\widetilde{ir}$}
\psline[origin={-12,-4}](0.8;-10)(2.3;-10)\uput{1.9}[-2]{0}(12,4){\LARGE $\widetilde{js}$}
\psline[origin={-12,-4}](0.8;-30)(2.8;-30)
\psdots*[origin={-12,-4},dotscale=0.3](1.8;-65)(1.8;-55)(1.8;-45)
\psline[origin={-12,-4}](0.8;-80)(2.8;-80)
\psline[origin={-12,-4}](0.8;-100)(2.8;-100)\uput{2.9}[-100]{0}(12,4){\LARGE $\widetilde{m\!+\!2}$}

\pscircle[fillstyle=solid,fillcolor=white](12,4){0.9}\rput(12,4){\LARGE $m$}

\psline[origin={-12,-4},linestyle=dotted](2.3;10)(2.8;10)
\psline[origin={-12,-4},linestyle=dotted](2.3;-10)(2.8;-10)

\psline[origin={-12,-4}](3.4;10)(4.4;10)\uput{1.4}[10]{0}(15.05,4.54){\LARGE $i$}
\psline[origin={-15.05,-4.54}](0.29;-20)(1.3;-20)\uput{1.4}[-20]{0}(15.05,4.54){\LARGE $r$}

\pscircle[origin={-12,-4},fillstyle=solid,fillcolor=white](3.1;10){0.4}
\rput(15.05,4.54){\LARGE $\mathsf{C}\,$}
\psline[origin={-12,-4}](3.4;-10)(4.4;-10)\uput{1.4}[-10]{0}(15.05,3.46){\LARGE $j$}
\psline[origin={-15.05,-3.46}](0.29;-40)(1.3;-40)\uput{1.4}[-40]{0}(15.05,3.46){\LARGE $s$}

\pscircle[origin={-12,-4},fillstyle=solid,fillcolor=white](3.1;-10){0.4}
\rput(15.05,3.46){\LARGE $\mathsf{C}\,$}


\psline[linewidth=5pt,arrowinset=0]{->}(17.5,4)(19,4)

\psline[linewidth=3pt,arrowinset=0]{->}(19.7,4)(20.9,4)\uput{0.3}[d](20.7,4){\LARGE $Q$}
\psline[linewidth=3pt](20.7,4)(21.7,4)

\SpecialCoor
\psline[origin={-22.5,-4}](0.8;100)(2.8;100)\uput{2.9}[100]{0}(22.5,4){\LARGE $\tilde{1}$}
\psline[origin={-22.5,-4}](0.8;80)(2.8;80)
\psdots*[origin={-22.5,-4},dotscale=0.3](1.8;65)(1.8;55)(1.8;45)
\psline[origin={-22.5,-4}](0.8;30)(2.8;30)
\psline[origin={-22.5,-4}](0.8;10)(2.8;10)\uput{2.9}[10]{0}(22.5,4){\LARGE $\widetilde{ir}$}
\psline[origin={-22.5,-4}](0.8;-10)(2.8;-10)\uput{2.9}[-10]{0}(22.5,4){\LARGE $\widetilde{js}$}
\psline[origin={-22.5,-4}](0.8;-30)(2.8;-30)
\psdots*[origin={-22.5,-4},dotscale=0.3](1.8;-65)(1.8;-55)(1.8;-45)
\psline[origin={-22.5,-4}](0.8;-80)(2.8;-80)
\psline[origin={-22.5,-4}](0.8;-100)(2.8;-100)\uput{2.9}[-100]{0}(22.5,4){\LARGE $\widetilde{m\!+\!2}$}

\pscircle[fillstyle=solid,fillcolor=white](22.5,4){0.9}\rput(22.5,4){\LARGE $m$}

\uput{3.7}[0]{0}(22.5,4){\Huge $\otimes$}
\psline[origin={-27.3,-4}](1;10)(2;10)\uput{2.2}[10]{0}(27.3,4){\LARGE $r$}
\psline[origin={-27.3,-4}](1;-10)(2;-10)\uput{2.2}[-10]{0}(27.3,4){\LARGE $s$}
\psframe[origin={-22.5,-4},fillstyle=solid,fillcolor=white]%
(4.8,-0.5)(5.8,0.5)
\rput(27.8,4.225){$(ir)$}
\rput(27.8,3.775){$(js)$}
}
\end{pspicture}
\end{center}
\caption{Graphical representation of the doubly-collinear momentum
mapping and the implied factorization of the phase space.}
\label{fig:Cirjs}
\end{figure}

The momentum mapping of \eqn{eq:PS_Cirjs} leads to the following
exact factorization of the phase space:
\beq
\PS{m+2}(\mom{};Q)=\PS{m}(\momt{(ir;js)}{};Q) 
\: [\rd p_{2;m}^{(ir;js)}(p_r,p_s,\ti{p}_{ir},\ti{p}_{js};Q)]\,.
\label{eq:PSfact_Cirjs}
\eeq
where the $m$ momenta in the first factor on the right hand side of
\eqn{eq:PSfact_Cirjs} are given by \eqn{eq:PS_Cirjs} and
$[\rd p_{2;m}^{(ir;js)}(p_r,p_s,\ti{p}_{ir},\ti{p}_{js};Q)]$ reads
\beeq
[\rd p_{2;m}^{(ir;js)}(p_r,p_s,\ti{p}_{ir},\ti{p}_{js};Q)] \aand= 
\Jac{ir;js}{2;m}(p_r,p_s,\ti{p}_{ir},\ti{p}_{js};Q)
\nn \\[3mm] && \times
  \frac{\rd^d p_r}{(2\pi)^{d-1}}\delta_{+}(p_r^2)
\,\frac{\rd^d p_s}{(2\pi)^{d-1}}\delta_{+}(p_s^2)
\,.
\label{eq:dp_Cirjs}
\eeeq
In this equation the Jacobian factor can be written as, for instance,
\beeq
&&
\Jac{ir;js}{2;m}(p_r,p_s,\ti{p}_{ir},\ti{p}_{js};Q) =
y_{\wti{ir}Q}\,y_{\wti{js}Q}
\,(1-\alpha_{ir}-\alpha_{js})^{(m-1)(d-2)}
\,\Theta(1-\alpha_{ir}-\alpha_{js})
\nn \\ && \qquad \times
\,\Bigg(
4 (1 - y_{\wti{ir}Q} - y_{\wti{js}Q}) \alpha_{ir} \alpha_{js}
\nn \\ &&\qquad \quad 
+\Big[y_{\wti{js}Q} (2 - y_{\wti{ir}Q} + y_{r\wti{ir}} - y_{rQ})
+ y_{s\wti{js}} (2 + y_{\wti{ir}Q})
- 2 y_{sQ} (1 - y_{\wti{ir}Q})\Big] \alpha_{ir}
\nn \\[1mm] &&\qquad \quad
+\Big[y_{\wti{ir}Q} (2 - y_{\wti{js}Q} + y_{s\wti{js}} - y_{sQ})
+ y_{r\wti{ir}} (2 + y_{\wti{js}Q})
- 2 y_{rQ} (1 - y_{\wti{js}Q})\Big] \alpha_{js}
\nn \\ &&\qquad \quad
+ (y_{\wti{ir}Q} - y_{rQ}) (y_{\wti{js}Q} - y_{sQ})
- (y_{\wti{ir}Q} + y_{rQ}) y_{s\wti{js}}
- (y_{\wti{js}Q} + y_{sQ}) y_{r\wti{ir}}
\Bigg)^{-1}
\,,
\qquad~
\label{eq:Jac_Cirjs}
\eeeq
where $\alpha_{ir}$ and $\alpha_{js}$ are the physical (both are positive
and their sum falling between 0 and 1) solutions of the coupled
constraints
\beeq
\frac{p_i^2}{Q^2} \aand = 
(1 - y_{\wti{ir}Q})\,\alpha_{ir}^2 
- y_{\wti{ir}Q}\,\alpha_{ir}\alpha_{js}
+ (y_{r\wti{ir}}+y_{\wti{ir}Q}-y_{rQ})\,\alpha_{ir}
+ y_{r\wti{ir}} \alpha_{js}
- y_{r\wti{ir}} = 0
\,,
\nn \\
\frac{p_j^2}{Q^2} \aand = 
(1 - y_{\wti{js}Q})\,\alpha_{js}^2 
- y_{\wti{js}Q}\,\alpha_{ir}\alpha_{js}
+ (y_{s\wti{js}}+y_{\wti{js}Q}-y_{sQ})\,\alpha_{js}
+ y_{s\wti{js}} \alpha_{is}
- y_{s\wti{js}} = 0
\,.
\qquad~
\eeeq
The analytical solution of these equations is rather complicated. In
a numerical calculation one can always find the solution numerically.
However, for implementing the subtraction scheme, we need only the
momentum mapping given in \eqn{eq:PS_Cirjs}. In order to integrate the
subtraction term over the factorized phase space 
$[\rd p_{2;m}^{(ir;js)}(p_r,p_s,\ti{p}_{ir},\ti{p}_{js};Q)]$,
we can choose $\alpha_{ir}$ and $\alpha_{js}$ as integration variables
and thus the solution of the coupled quadratic equations can be avoided.
We present the graphical representation of \eqnss{eq:PS_Cirjs}{eq:Jac_Cirjs}
in \fig{fig:Cirjs}.

In the doubly-collinear limit, when $p_i^\mu || p_r^\mu$ and 
$p_j^\mu || p_s^\mu$, the transverse momenta behave as in \eqn{eq:kTir2},
$\tzz{k}{l} \to z_k$ ($k,\;l = i,\;j,\;r,\;s$).  Both $\alpha_{ir}$ and
$\alpha_{js}$ tend to zero so 
$\ti{p}_{ir}^\mu \to p_i^\mu + p_r^\mu$,
$\ti{p}_{js}^\mu \to p_j^\mu + p_s^\mu$, $\ti{p}_n^\mu \to p_n^\mu$.
Consequently, the counterterm properly regularizes the squared matrix
element in this limit.


\subsection{Double soft-collinear-type counterterms}
\label{ssec:RR_A2_CSirs}

\subtitle{Counterterms}

We refer to the three terms on the second line of \eqn{eq:RR_A2} as
double soft-collinear-type terms because they are all defined using the
same momentum mapping, that of the soft-collinear subtraction term.
Explicitly these terms read
\beeq
\cSCS{ir;s}{(0,0)}(\mom{}) \aand=
-(8\pi\as\mu^{2\eps})^2
\sum_{j}\sum_{k\ne j}\frac12 \calS_{jk}(s)
\nn\\&&\times
\frac{1}{s_{ir}}
\bra{m}{(0)}{(\momt{(\ha{s},ir)}{})}
\bT_j \bT_k \hP_{f_i
f_r}^{(0)}(\tzz{i}{r},\tzz{r}{i},\kappatir;\eps)
\ket{m}{(0)}{(\momt{(\ha{s},ir)}{})}\,,
\qquad~
\label{eq:CSirs}
\\
\cC{irs}{}\cSCS{ir;s}{(0,0)}(\mom{}) \aand=
(8\pi\as\mu^{2\eps})^2
\frac{2}{s_{(ir)s}}\frac{1-\tzz{s}{ir}}{\tzz{s}{ir}}
\,\bT^2_{ir}
\nn\\&&\times
\frac{1}{s_{ir}}
\bra{m}{(0)}{(\momt{(\ha{s},ir)}{})}
\hP_{f_i f_r}^{(0)}(\tzz{i}{r},\tzz{r}{i},\kappatir;\eps)
\ket{m}{(0)}{(\momt{(\ha{s},ir)}{})}\,,
\label{eq:CirsCSirs}
\eeeq
\beeq
\cC{ir;js}{}\cSCS{ir;s}{(0,0)}(\mom{}) \aand=
(8\pi\as\mu^{2\eps})^2
\frac{2}{s_{js}}\frac{\tzz{j}{s}}{\tzz{s}{j}} \,\bT^2_{j}
\nn\\&&\times
\frac{1}{s_{ir}}
\bra{m}{(0)}{(\momt{(\ha{s},ir)}{})}
\hP_{f_i f_r}^{(0)}(\tzz{i}{r},\tzz{r}{i},\kappatir;\eps)
\ket{m}{(0)}{(\momt{(\ha{s},ir)}{})}\,.
\label{eq:CirjsCSirs}
\eeeq
Here we remind the reader of what we wrote below \eqn{eq:A1} about the
notation. For instance, on the left hand side of \eqn{eq:CirsCSirs}
$\cC{irs}{}\cSCS{ir;s}{(0,0)}$ denotes a function that is defined by the
function of the original momenta given on the right hand side.
The eikonal factor entering the soft-collinear counterterm
$\cSCS{ir;s}{(0,0)}(\mom{})$ is given in \eqn{eq:Sikr}. Nevertheless,
we record explicitly that whenever $j$ or $l$ in \eqn{eq:CSirs} is
equal to $(ir)$, \eqn{eq:Sikr} evaluates to
\beq
\calS_{(ir)l}(s) = \frac{2s_{(ir)l}}{s_{(ir)s} s_{ls}} 
= \frac{2(s_{il} + s_{rl})}{(s_{is} + s_{rs}) s_{ls}}\,. 
\label{eq:Sirls}
\eeq
The momentum fractions appearing in \eqnss{eq:CSirs}{eq:CirjsCSirs} are
the same as those for the singly-collinear counterterm (see
\eqn{eq:zt2}), while transverse components $\kappatir$ are the image of
the transverse momentum $\kTt{i,r}$ in \eqn{eq:kTtir}
under the soft map of \eqn{eq:smap}
\beq
\kTt{i,r} \smap{\ha{s}} \kappatir\,.
\label{eq:kappatir}
\eeq
Explicitly we have
\beq
\kappatir^{\mu} = 
\Lambda^{\mu}_{\nu}[Q,(Q-p_{\ha{s}})/\lambda_{\ha{s}}]
(\kTt{i,r}^{\nu}/\lambda_{\ha{s}})\,,
\label{eq:kappatirmu}
\eeq
where $\lambda_{\ha{s}}$ and the matrix $\Lambda^{\mu}_{\nu}$ are
defined in \eqns{eq:lambdar}{eq:LambdaKKt}. The hatted momentum
$\ha{p}_s$ is given in \eqn{eq:PS_CSirs_ir} with $n=s$.

\subtitle{Momentum mapping and phase space factorization}

The $m$ momenta $\momt{(\ha{s},ir)}{} \equiv
\{\ti{p}_1,\ldots,\ti{p}_{ir},\ldots,\ti{p}_{m+2}\}$ ($p_s$ is absent)
entering the matrix elements on the right hand sides of
\eqnss{eq:CSirs}{eq:CirjsCSirs} are defined by the composition of a
single collinear and a single soft mapping as follows:
\beq
\mom{} \cmap{ir} \momh{(ir)}{+1} \smap{\ha{s}}
\momt{(\ha{s},ir)}{}\,.
\label{eq:PS_CSirs}
\eeq
The first mapping is a collinear-type one leading to the hatted momenta
\beq
\ha{p}_{ir}^{\mu} = \frac{1}{1-\alpha_{ir}}
(p_i^{\mu} + p_r^{\mu} - \alpha_{ir} Q^{\mu})\,,
\qquad
\ha{p}_n^{\mu} = \frac{1}{1-\alpha_{ir}} p_n^{\mu}\,,
\qquad n\ne i,r\,.
\label{eq:PS_CSirs_ir}
\eeq
where $\alpha_{ir}$ is given in \eqn{eq:alphair}, followed by a soft
mapping 
\beq
\ti{p}_n^{\mu} =
\Lambda^{\mu}_{\nu}[Q,(Q-\ha{p}_s)/\lambda_{\ha{s}}]
(\ha{p}_n^{\nu}/\lambda_{\ha{s}})\,,
\qquad n\ne \ha{s}\,,
\label{eq:PS_CSirs_h}
\eeq
where $\lambda_{\ha{s}}$ is given in \eqn{eq:lambdar} (with $r \to \ha{s}$).
The other of the soft and collinear mappings is not crucial. The order
chosen here leads to simpler integrals of the singular factors over the
unresolved phase space.

\begin{figure}
\begin{center}
\begin{pspicture}(0,0)(14,4)
\scalebox{0.5}{%

\psline[linewidth=3pt,arrowinset=0]{->}(0.2,4)(1.4,4)\uput{0.3}[d](1.2,4){\LARGE $Q$}
\psline[linewidth=3pt](1.2,4)(2.2,4)

\SpecialCoor
\psline[origin={-3,-4}](0.8;100)(2.8;100)\uput{2.9}[100]{0}(3,4){\LARGE $1$}
\psline[origin={-3,-4}](0.8;80)(2.8;80)
\psdots*[origin={-3,-4},dotscale=0.3](1.8;65)(1.8;55)(1.8;45)
\psline[origin={-3,-4}](0.8;30)(2.8;30)
\psline[origin={-3,-4}](0.8;15)(2.8;15)\uput{2.9}[15]{0}(3,4){\LARGE $i$}
\psline[origin={-3,-4}](0.8;0)(2.8;0)\uput{2.9}[0]{0}(3,4){\LARGE $r$}
\psline[origin={-3,-4}](0.8;-15)(2.8;-15)\uput{2.9}[-15]{0}(3,4){\LARGE $s$}
\psline[origin={-3,-4}](0.8;-30)(2.8;-30)
\psdots*[origin={-3,-4},dotscale=0.3](1.8;-65)(1.8;-55)(1.8;-45)
\psline[origin={-3,-4}](0.8;-80)(2.8;-80)
\psline[origin={-3,-4}](0.8;-100)(2.8;-100)\uput{2.9}[-100]{0}(3,4){\LARGE $m\!+\!2$}

\pscircle[fillstyle=solid,fillcolor=white](3,4){0.9}\rput(3,4){\LARGE $m\!+\!2$}

\psline{->}(7,4)(8,4)\uput[u](7.5,4){{\LARGE $\mathsf{C}_{ir}$}}

\psline[linewidth=3pt,arrowinset=0]{->}(9.2,4)(10.4,4)\uput{0.3}[d](10.2,4){\LARGE $Q$}
\psline[linewidth=3pt](10.2,4)(11.2,4)

\SpecialCoor
\psline[origin={-12,-4}](0.8;100)(2.8;100)\uput{2.9}[100]{0}(12,4){\LARGE $\hat{1}$}
\psline[origin={-12,-4}](0.8;80)(2.8;80)
\psdots*[origin={-12,-4},dotscale=0.3](1.8;65)(1.8;55)(1.8;45)
\psline[origin={-12,-4}](0.8;30)(2.8;30)
\psline[origin={-12,-4}](0.8;10)(2.3;10)\uput{1.9}[20]{0}(12,4){\LARGE $\widehat{ir}$}
\psline[origin={-12,-4}](0.8;-10)(2.8;-10)\uput{3.0}[-10]{0}(12,4){\LARGE $\widehat{s}$}
\psline[origin={-12,-4}](0.8;-30)(2.8;-30)
\psdots*[origin={-12,-4},dotscale=0.3](1.8;-65)(1.8;-55)(1.8;-45)
\psline[origin={-12,-4}](0.8;-80)(2.8;-80)
\psline[origin={-12,-4}](0.8;-100)(2.8;-100)\uput{2.9}[-100]{0}(12,4){\LARGE $\widehat{m\!+\!2}$}

\pscircle[fillstyle=solid,fillcolor=white](12,4){0.9}\rput(12,4){\LARGE $m\!+\!1$}

\psline[origin={-12,-4},linestyle=dotted](2.3;10)(2.8;10)
\psline[origin={-12,-4},linestyle=dotted](2.3;-10)(2.8;-10)

\psline[origin={-12,-4}](3.4;10)(4.4;10)\uput{1.4}[10]{0}(15.05,4.54){\LARGE $i$}
\psline[origin={-15.05,-4.54}](0.29;-20)(1.3;-20)\uput{1.4}[-20]{0}(15.05,4.54){\LARGE $r$}

\pscircle[origin={-12,-4},fillstyle=solid,fillcolor=white](3.1;10){0.4}
\rput(15.05,4.54){\LARGE $\mathsf{C}\,$}

\psline{->}(17.5,4)(18.5,4)\uput[u](18,4){{\LARGE $\mathsf{S}_{\hat{s}}$}}

\psline[linewidth=3pt,arrowinset=0]{->}(19.7,4)(20.9,4)\uput{0.3}[d](20.7,4){\LARGE $Q$}
\psline[linewidth=3pt](20.7,4)(21.7,4)

\SpecialCoor
\psline[origin={-22.5,-4}](0.8;100)(2.8;100)\uput{2.9}[100]{0}(22.5,4){\LARGE $\tilde{1}$}
\psline[origin={-22.5,-4}](0.8;80)(2.8;80)
\psdots*[origin={-22.5,-4},dotscale=0.3](1.8;65)(1.8;55)(1.8;45)
\psline[origin={-22.5,-4}](0.8;30)(2.8;30)
\psline[origin={-22.5,-4}](0.8;10)(2.3;10)\uput{1.9}[20]{0}(22.5,4){\LARGE $\widetilde{ir}$}
\psline[origin={-22.5,-4}](0.8;-30)(2.8;-30)
\psdots*[origin={-22.5,-4},dotscale=0.3](1.8;-65)(1.8;-55)(1.8;-45)
\psline[origin={-22.5,-4}](0.8;-80)(2.8;-80)
\psline[origin={-22.5,-4}](0.8;-100)(2.8;-100)\uput{2.9}[-100]{0}(22.5,4){\LARGE $\widetilde{m\!+\!2}$}

\psline[origin={-22.5,-4},linestyle=dotted](2.3;10)(2.8;10)
\psline[origin={-22.5,-4},linestyle=dotted](0.8;-10)(2.8;-10)

\pscircle[fillstyle=solid,fillcolor=white](22.5,4){0.9}\rput(22.5,4){\LARGE $m$}

\psline[origin={-22.5,-4}](3.4;10)(4.4;10)\uput{1.4}[10]{0}(25.55,4.54){\LARGE $i$}
\psline[origin={-25.55,-4.54}](0.29;-20)(1.3;-20)\uput{1.4}[-20]{0}(25.55,4.54){\LARGE $r$}

\pscircle[origin={-22.5,-4},fillstyle=solid,fillcolor=white](3.1;10){0.4}
\rput(25.55,4.54){\LARGE $\mathsf{C}\,$}

\psline[origin={-22.5,-4}](3.4;-10)(4.4;-10)\uput{1.4}[-10]{0}(25.55,3.46){\LARGE $\hat{s}$}

\pscircle[origin={-22.5,-4},fillstyle=solid,fillcolor=white](3.1;-10){0.4}
\rput(25.55,3.46){\LARGE $\mathsf{S}$}


\psline[linewidth=5pt,arrowinset=0]{->}(7,-4)(8.5,-4)

\psline[origin={10.5,8},linewidth=3pt,arrowinset=0]{->}(19.7,4)(20.9,4)\uput{0.3}[d](20.7,4){\LARGE $Q$}
\psline[origin={10.5,8},linewidth=3pt](20.7,4)(21.7,4)

\SpecialCoor
\psline[origin={-12,4}](0.8;100)(2.8;100)\uput{2.9}[100]{0}(12,-4){\LARGE $\tilde{1}$}
\psline[origin={-12,4}](0.8;80)(2.8;80)
\psdots*[origin={-12,4},dotscale=0.3](1.8;65)(1.8;55)(1.8;45)
\psline[origin={-12,4}](0.8;30)(2.8;30)
\psline[origin={-12,4}](0.8;0)(2.8;0)\uput{2.9}[0]{0}(12,-4){\LARGE $\widetilde{ir}$}
\psline[origin={-12,4}](0.8;-30)(2.8;-30)
\psdots*[origin={-12,4},dotscale=0.3](1.8;-65)(1.8;-55)(1.8;-45)
\psline[origin={-12,4}](0.8;-80)(2.8;-80)
\psline[origin={-12,4}](0.8;-100)(2.8;-100)\uput{2.9}[-100]{0}(12,-4){\LARGE $\widetilde{m\!+\!2}$}

\pscircle[fillstyle=solid,fillcolor=white](12,-4){0.9}\rput(12,-4){\LARGE $m$}

\uput{3.7}[0]{0}(12,-4){\Huge $\otimes$}
\psline[origin={-12,4}](5.8;0)(6.8;0)
\psframe[origin={-12,4},fillstyle=solid,fillcolor=white]%
(4.8,-0.5)(5.8,0.5)
\rput(17.3,-4){$(ir)$}
\uput{7}[0]{0}(12,-4){\LARGE $r$}

\uput{3.7}[0]{0}(16,-4){\Huge $\otimes$}
\psline[origin={-16,4}](5.8;0)(6.8;0)
\psframe[origin={-16,4},fillstyle=solid,fillcolor=white]%
(4.8,-0.5)(5.8,0.5)
\rput(21.3,-4){$(\hat{s})$}
\uput{7}[0]{0}(16,-4){\LARGE $\hat{s}$}
}
\end{pspicture}
\end{center}
~\vskip 30mm
\caption{Graphical representation of the soft-collinear momentum
mapping and the implied factorization of the phase space.}
\label{fig:CSirs}
\end{figure}
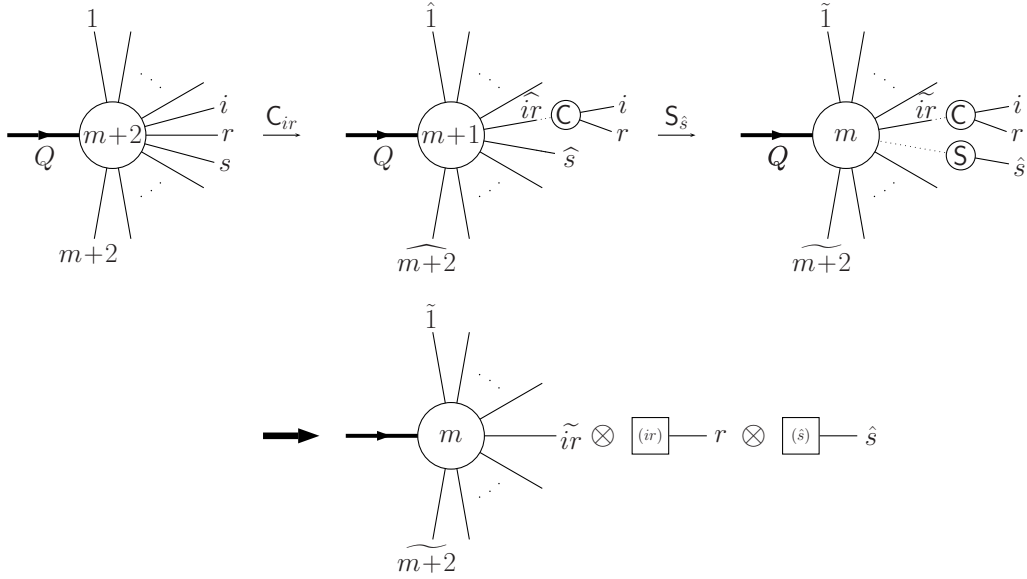

This momentum mapping leads to exact phase-space factorization
\beq
\PS{m+2}{}(\mom{};Q) = \PS{m}{}(\momt{(\ha{s},ir)}{};Q)
\,[\rd p_{1;m+1}^{(ir)}(p_{r},\ha{p}_{ir};Q)]
\,[\rd p_{1;m}^{(\ha{s})}(\ha{p}_s;Q)]\,.
\label{eq:PSfact_CSirs}
\eeq
The collinear and soft one-particle factorized phase space measures 
$[\rd p_{1;m+1}^{(ir)}(p_{r},\ha{p}_{ir};Q)]$ and
$[\rd p_{1;m}^{(\ha{s})}(\ha{p}_s;Q)]$ are given respectively in
\eqns{eq:dp_Cir}{eq:dp_Sr}.  
We show a graphical representation for \eqnss{eq:PS_CSirs}{eq:PSfact_CSirs}
in \fig{fig:CSirs}.

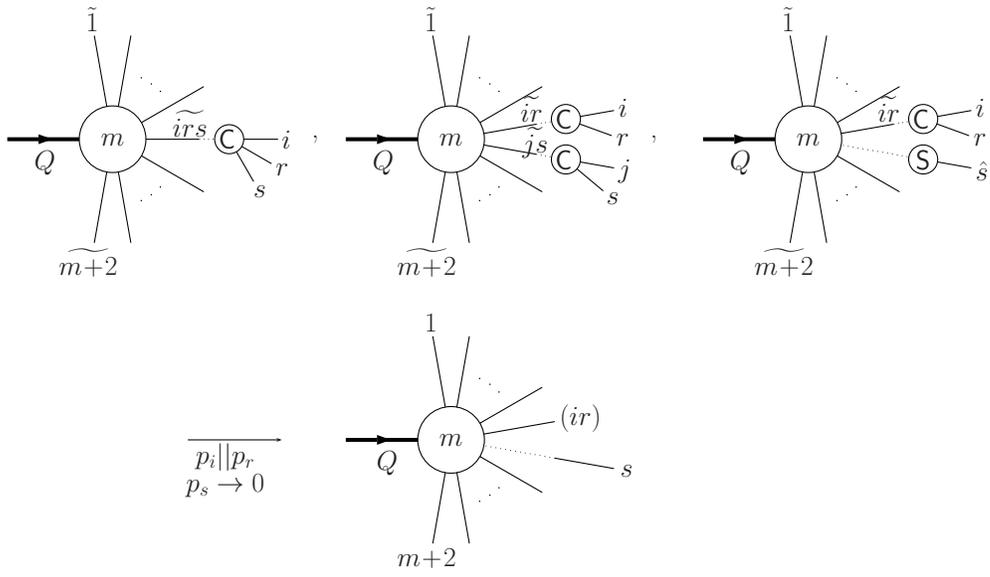
\begin{figure}
\begin{center}
\begin{pspicture}(0,0)(14,4)
\scalebox{0.5}{%

\psline[linewidth=3pt,arrowinset=0]{->}(0.2,4)(1.4,4)\uput{0.3}[d](1.2,4){\LARGE $Q$}
\psline[linewidth=3pt](1.2,4)(2.2,4)

\SpecialCoor
\psline[origin={-3,-4}](0.8;100)(2.8;100)\uput{2.9}[100]{0}(3,4){\LARGE $\tilde{1}$}
\psline[origin={-3,-4}](0.8;80)(2.8;80)
\psdots*[origin={-3,-4},dotscale=0.3](1.8;65)(1.8;55)(1.8;45)
\psline[origin={-3,-4}](0.8;30)(2.8;30)
\psline[origin={-3,-4}](0.8;0)(2.3;0)\uput{1.6}[12]{0}(3,4){\LARGE $\widetilde{irs}$}
\psline[origin={-3,-4}](0.8;-30)(2.8;-30)
\psdots*[origin={-3,-4},dotscale=0.3](1.8;-65)(1.8;-55)(1.8;-45)
\psline[origin={-3,-4}](0.8;-80)(2.8;-80)
\psline[origin={-3,-4}](0.8;-100)(2.8;-100)\uput{2.9}[-100]{0}(3,4){\LARGE $\widetilde{m\!+\!2}$}

\pscircle[fillstyle=solid,fillcolor=white](3,4){0.9}\rput(3,4){\LARGE $m$}

\psline[origin={-3,-4},linestyle=dotted](2.3;0)(2.8;0)
\psline[origin={-6.1,-4}](0.3;0)(1.3;0)\uput{1.4}[0]{0}(6.1,4){\LARGE $i$}
\psline[origin={-6.1,-4}](0.3;-30)(1.3;-30)\uput{1.4}[-30]{0}(6.1,4){\LARGE $r$}
\psline[origin={-6.1,-4}](0.3;-60)(1.3;-60)\uput{1.4}[-60]{0}(6.1,4){\LARGE $s$}

\pscircle[origin={-3,-4},fillstyle=solid,fillcolor=white](3.1;0){0.4}
\rput(6.1,4){\LARGE $\mathsf{C}\,$}

\uput{0}[0]{0}(8.3,4){\LARGE ,}

\psline[linewidth=3pt,arrowinset=0]{->}(9.2,4)(10.4,4)\uput{0.3}[d](10.2,4){\LARGE $Q$}
\psline[linewidth=3pt](10.2,4)(11.2,4)

\SpecialCoor
\psline[origin={-12,-4}](0.8;100)(2.8;100)\uput{2.9}[100]{0}(12,4){\LARGE $\tilde{1}$}
\psline[origin={-12,-4}](0.8;80)(2.8;80)
\psdots*[origin={-12,-4},dotscale=0.3](1.8;65)(1.8;55)(1.8;45)
\psline[origin={-12,-4}](0.8;30)(2.8;30)
\psline[origin={-12,-4}](0.8;10)(2.3;10)\uput{1.9}[20]{0}(12,4){\LARGE $\widetilde{ir}$}
\psline[origin={-12,-4}](0.8;-10)(2.3;-10)\uput{1.9}[-2]{0}(12,4){\LARGE $\widetilde{js}$}
\psline[origin={-12,-4}](0.8;-30)(2.8;-30)
\psdots*[origin={-12,-4},dotscale=0.3](1.8;-65)(1.8;-55)(1.8;-45)
\psline[origin={-12,-4}](0.8;-80)(2.8;-80)
\psline[origin={-12,-4}](0.8;-100)(2.8;-100)\uput{2.9}[-100]{0}(12,4){\LARGE $\widetilde{m\!+\!2}$}

\pscircle[fillstyle=solid,fillcolor=white](12,4){0.9}\rput(12,4){\LARGE $m$}

\psline[origin={-12,-4},linestyle=dotted](2.3;10)(2.8;10)
\psline[origin={-12,-4},linestyle=dotted](2.3;-10)(2.8;-10)

\psline[origin={-12,-4}](3.4;10)(4.4;10)\uput{1.4}[10]{0}(15.05,4.54){\LARGE $i$}
\psline[origin={-15.05,-4.54}](0.29;-20)(1.3;-20)\uput{1.4}[-20]{0}(15.05,4.54){\LARGE $r$}

\pscircle[origin={-12,-4},fillstyle=solid,fillcolor=white](3.1;10){0.4}
\rput(15.05,4.54){\LARGE $\mathsf{C}\,$}
\psline[origin={-12,-4}](3.4;-10)(4.4;-10)\uput{1.4}[-10]{0}(15.05,3.46){\LARGE $j$}
\psline[origin={-15.05,-3.46}](0.29;-40)(1.3;-40)\uput{1.4}[-40]{0}(15.05,3.46){\LARGE $s$}

\pscircle[origin={-12,-4},fillstyle=solid,fillcolor=white](3.1;-10){0.4}
\rput(15.05,3.46){\LARGE $\mathsf{C}\,$}

\uput{0}[0]{0}(17.3,4){\LARGE ,}

\psline[linewidth=3pt,arrowinset=0]{->}(18.7,4)(19.9,4)\uput{0.3}[d](19.7,4){\LARGE $Q$}
\psline[linewidth=3pt](19.7,4)(20.7,4)

\SpecialCoor
\psline[origin={-21.5,-4}](0.8;100)(2.8;100)\uput{2.9}[100]{0}(21.5,4){\LARGE $\tilde{1}$}
\psline[origin={-21.5,-4}](0.8;80)(2.8;80)
\psdots*[origin={-21.5,-4},dotscale=0.3](1.8;65)(1.8;55)(1.8;45)
\psline[origin={-21.5,-4}](0.8;30)(2.8;30)
\psline[origin={-21.5,-4}](0.8;10)(2.3;10)\uput{1.9}[20]{0}(21.5,4){\LARGE $\widetilde{ir}$}
\psline[origin={-21.5,-4}](0.8;-30)(2.8;-30)
\psdots*[origin={-21.5,-4},dotscale=0.3](1.8;-65)(1.8;-55)(1.8;-45)
\psline[origin={-21.5,-4}](0.8;-80)(2.8;-80)
\psline[origin={-21.5,-4}](0.8;-100)(2.8;-100)\uput{2.9}[-100]{0}(21.5,4){\LARGE $\widetilde{m\!+\!2}$}

\psline[origin={-21.5,-4},linestyle=dotted](2.3;10)(2.8;10)
\psline[origin={-21.5,-4},linestyle=dotted](0.8;-10)(2.8;-10)

\pscircle[fillstyle=solid,fillcolor=white](21.5,4){0.9}\rput(21.5,4){\LARGE $m$}

\psline[origin={-21.5,-4}](3.4;10)(4.4;10)\uput{1.4}[10]{0}(24.55,4.54){\LARGE $i$}
\psline[origin={-24.55,-4.54}](0.29;-20)(1.3;-20)\uput{1.4}[-20]{0}(24.55,4.54){\LARGE $r$}

\pscircle[origin={-21.5,-4},fillstyle=solid,fillcolor=white](3.1;10){0.4}
\rput(24.55,4.54){\LARGE $\mathsf{C}\,$}

\psline[origin={-21.5,-4}](3.4;-10)(4.4;-10)\uput{1.4}[-10]{0}(24.55,3.46){\LARGE $\hat{s}$}

\pscircle[origin={-21.5,-4},fillstyle=solid,fillcolor=white](3.1;-10){0.4}
\rput(24.55,3.46){\LARGE $\mathsf{S}$}


\psline{->}(5,-4)(7.5,-4)\uput[u](6,-5){{\LARGE $p_{i}||p_{r}$}}
\uput[u](6,-5.7){{\LARGE $p_{s}\to 0$}}

\psline[origin={10.5,8},linewidth=3pt,arrowinset=0]{->}(19.7,4)(20.9,4)\uput{0.3}[d](10.3,-4){\LARGE $Q$}
\psline[origin={10.5,8},linewidth=3pt](20.7,4)(21.7,4)

\SpecialCoor
\psline[origin={-12,4}](0.8;100)(2.8;100)\uput{2.9}[100]{0}(12,-4){\LARGE 1}
\psline[origin={-12,4}](0.8;80)(2.8;80)
\psdots*[origin={-12,4},dotscale=0.3](1.8;65)(1.8;55)(1.8;45)
\psline[origin={-12,4}](0.8;30)(2.8;30)
\psline[origin={-12,4}](0.8;10)(2.8;10)\uput{3}[20]{0}(12,-4.6){\LARGE $(ir)$}
\psline[origin={-12,4}](0.8;-30)(2.8;-30)
\psline[origin={-12,4},linestyle=dotted](0.8;-10)(2.8;-10)
\psline[origin={-12,4}](2.8;-10)(4.4;-10)\uput{0}[-10]{0}(16.5,-4.8){\LARGE $s$}
\psdots*[origin={-12,4},dotscale=0.3](1.8;-65)(1.8;-55)(1.8;-45)
\psline[origin={-12,4}](0.8;-80)(2.8;-80)
\psline[origin={-12,4}](0.8;-100)(2.8;-100)\uput{2.9}[-100]{0}(12,-4)
{\LARGE $m\!+\!2$}

\pscircle[fillstyle=solid,fillcolor=white](12,-4){0.9}\rput(12,-4){\LARGE $m$}
}
\end{pspicture}
\end{center}
~\vskip 30mm
\caption{Graphical representation of the soft-collinear limit of the
triply-, doubly- and soft-collinear momentum mappings.}
\label{fig:CSlimit}
\end{figure}

The counterterms in \eqnss{eq:CSirs}{eq:CirjsCSirs} are defined on the
same phase spaces. Therefore, in the triply-collinear limit,
when $p_i^\mu || p_r^\mu || p_s^\mu$, the term in \eqn{eq:CirsCSirs}
regularizes that in \eqn{eq:CSirs}, while \eqn{eq:CirjsCSirs} is
integrable. Conversely, in the doubly-collinear limit, when $p_i^\mu ||
p_r^\mu$ and $p_j^\mu || p_s^\mu$, the term in \eqn{eq:CirjsCSirs}
regularizes that in \eqn{eq:CSirs}, while \eqn{eq:CirsCSirs} is
integrable. In the soft-collinear limit, when $p_i^\mu || p_r^\mu$ and
$p_s^\mu \to 0$ simultaneously, both the triply-collinear mapping,
defined in \eqns{eq:PS_Cirs}{eq:alphairs}, and the doubly-collinear
mapping, defined in \eqn{eq:PS_Cirjs}, approach the corresponding limit
of the iterated mappings of \eqnss{eq:PS_CSirs}{eq:PS_CSirs_h}.  This
limit of the momentum
mappings is depicted graphically in \fig{fig:CSlimit}.  We then conclude
that the subtraction term in \eqn{eq:CirsCSirs} regularizes the
kinematical singularities in \eqn{eq:Cirs}, while \eqn{eq:CirjsCSirs}
regularizes the singularities in the doubly-collinear subtraction in
\eqn{eq:Cirjs}. In this limit, the subtraction term in \eqn{eq:CSirs}
is a local counterterm to the squared matrix element by construction.


\subsection{Double soft-type counterterms}
\label{ssec:RR_A2_Srs}

All terms on the last two lines of \eqn{eq:RR_A2} are constructed using
the same momentum mapping as the doubly soft counterterm
$\cS{rs}{(0,0)}(\mom{})$ itself. We refer to these as doubly soft-type terms.

\subtitle{Counterterms}

We begin by exhibiting the soft-type terms explicitly for the case of
double soft gluon emission and then present counterterms to it in the
various doubly-unresolved regions. We have
\beeq
\cS{r_gs_g}{(0,0)}(\mom{}) \aand= (8\pi\as\mu^{2\eps})^2
\Bigg[\frac18
\sum_{i,j,k,l}\calS_{ik}(r)\calS_{jl}(s)\SME{m;(i,k)(j,l)}{0}{\momt{(rs)}{}}
\nn\\&&\qquad\qquad\qquad
-\frac14 \CA \sum_{i,k}\calS_{ik}(r,s)\SME{m;(i,k)}{0}{\momt{(rs)}{}}
\Bigg]\,,
\label{eq:Srsgg}
\\[3mm]
\cC{ir_gs_g}{}\cS{r_gs_g}{(0,0)}(\mom{}) \aand= (8\pi\as\mu^{2\eps})^2
\Bigg\{\bT_i^2\,\frac{4\tzz{i}{rs}^2}{s_{ir}s_{is}\tzz{r}{is}\tzz{s}{ir}}
+
\CA \Bigg[
\frac{(1 - \eps)}{s_{i(rs)} s_{rs}}
\frac{(s_{ir} \tzz{s}{ir} - s_{is} \tzz{r}{is})^2}{s_{i(rs)} s_{rs} (\tzz{r}{is} + \tzz{s}{ir})^2}
\nn\\[2mm]&&
- \frac{\tzz{i}{rs}}{s_{i(rs)} s_{rs}}
  \left(\frac{4}{\tzz{r}{is}+\tzz{s}{ir}} - \frac1{\tzz{r}{is}} \right)
- \frac{1}{s_{i(rs)}  s_{ir}} \frac{2\tzz{i}{rs}^2}{\tzz{r}{is} (\tzz{r}{is} + \tzz{s}{ir})}
\nn\\[2mm]&&
- \frac{\tzz{i}{rs}^2}{s_{i(rs)}  s_{is}} \frac{1}{\tzz{r}{is} (\tzz{r}{is} + \tzz{s}{ir})}
+ \frac{\tzz{i}{rs}}{s_{ir} s_{rs}}
  \left(\frac{1}{\tzz{s}{ir}} + \frac{1}{\tzz{r}{is} + \tzz{s}{ir}}\right)
+ (r \leftrightarrow s) \Bigg]\Bigg\}
\nn\\[2mm]&&
\times\,\bT_i^2 \, \SME{m}{0}{\momt{(rs)}{}}\,,
\label{eq:CirsSrsgg}
\eeeq
\beeq
\cSCS{ir_g;s_g}{}\cS{r_gs_g}{(0,0)}(\mom{}) \aand= -(8\pi\as\mu^{2\eps})^2 
\frac{1}{s_{ir}}\frac{2\tzz{i}{r}}{\tzz{r}{i}}\bT_{i}^2
\sum_{j}\sum_{l\ne j}\frac12 \calS_{jl}(s)\SME{m;(j,l)}{0}{\momt{(rs)}{}}\,,
\label{eq:CSirsSrsgg}
\\[3mm]
\cC{ir_g;js_g}{}\cS{r_gs_g}{(0,0)}(\mom{}) \aand= (8\pi\as\mu^{2\eps})^2 
\frac{1}{s_{ir}}\frac{2\tzz{i}{r}}{\tzz{r}{i}}\bT_{i}^2
\frac{1}{s_{js}}\frac{2\tzz{j}{s}}{\tzz{s}{j}}\bT_{j}^2
\SME{m}{0}{\momt{(rs)}{}}\,,
\label{eq:CirjsSrsgg}
\\[3mm]
\cC{ir_gs_g}{}\cSCS{ir_g;s_g}{}\cS{r_gs_g}{(0,0)} \aand= (8\pi\as\mu^{2\eps})^2 
\frac{4 \tzz{i}{rs}(\tzz{i}{rs} + \tzz{r}{is})}
     {s_{ir}s_{(ir)s}\tzz{r}{is}\tzz{s}{ir}}
\bT_i^2 \bT_i^2 \SME{m}{0}{\momt{(rs)}{}}\,.
\label{eq:CirsCSirsSrsgg}
\eeeq
The momentum fractions were defined in \eqns{eq:zt2}{eq:zt3}. 
The two-gluon soft function appearing in \eqn{eq:Srsgg} is
\beq
\calS_{ik}(r, s) = \calS_{ik}^{(\rm s.o.)}(r, s) 
+ 4 \frac{s_{ir} s_{ks} + s_{is} s_{kr}}{s_{i(rs)} s_{k(rs)}}
\left[\frac{1-\eps}{s_{rs}^2} - \frac18 \calS_{ik}^{(\rm s.o.)}(r, s)\right]
- \frac{4}{s_{rs}} \calS_{ik}(rs)\ ,
\label{eq:softggnab}
\eeq
where
\beq
\calS_{ik}^{(\rm s.o.)}(r, s) = 
\calS_{ik}(s)\left(\calS_{is}(r) + \calS_{ks}(r) - \calS_{ik}(r)\right)
\label{eq:softggnabso}
\eeq
is the form of this function in the strongly-ordered approximation and 
$\calS_{ik}(rs)$ is given by \eqn{eq:Sikr},
\beq
\calS_{ik}(rs) = \frac{2 s_{ik}}{s_{i(rs)} s_{k(rs)}}\,.
\label{eq:Sikrs}
\eeq
The discussion below \eqn{eq:CirjsCSirs} about the eikonal factor
appering in the double soft-collinear subtraction term (\eqn{eq:CSirs})
and especially \eqn{eq:Sirls} apply also to the soft-collinear limit
of the doubly-soft subtraction, \eqn{eq:CSirsSrsgg}.  

The only nonzero terms for the case of emission of a soft
quark-antiquark pair are $\cS{r_{\qb}s_q}{(0,0)}(\mom{})$ and
$\cC{ir_{\qb}s_q}{}\cS{r_{\qb}s_q}{(0,0)}(\mom{})$, the remaining terms
all vanish. Explicit expressions for these nonzero terms are
\beeq
\cS{r_{\qb}s_q}{(0,0)}(\mom{}) \aand= (8\pi\as\mu^{2\eps})^2
\,\frac{1}{s_{rs}^2}\,\TR
\nn\\[2mm]&&\times
\sum_i \sum_{k\ne i}
\Bigg(
\frac{s_{ir}s_{ks}+s_{kr}s_{is}-s_{ik}s_{rs}}{s_{i(rs)}s_{k(rs)}}
-2\frac{s_{ir}s_{is}}{s_{i(rs)}^2}
\Bigg)\SME{m;(i,k)}{0}{\momt{(rs)}{}}\,,\quad~
\label{eq:Srsqq}
\\[3mm]
\cC{ir_{\qb}s_q}{}\cS{r_{\qb}s_q}{(0,0)}(\mom{}) \aand=
(8\pi\as\mu^{2\eps})^2
\,\bT_{i}^2\,\TR 
\nn\\[2mm]&&\times
\frac{2}{s_{i(rs)}s_{rs}}
\Bigg(
\frac{\tzz{i}{rs}}{\tzz{r}{is} + \tzz{s}{ir}}
-\frac{(s_{ir}\tzz{s}{ir} - s_{is}\tzz{r}{is})^2}
      {s_{i(rs)}s_{rs}(\tzz{r}{is} + \tzz{s}{ir})^2}
\Bigg)\SME{m}{0}{\momt{(rs)}{}}\,.
\label{eq:CirsSrsqq}
\eeeq

\subtitle{Momentum mapping and phase space factorization}

The $m$ momenta $\momt{(rs)}{} \equiv \{\ti{p}_1,\ldots,\ti{p}_{m+2}\}$
($p_r$ and $p_s$ are omitted from the set) entering the matrix elements
on the right hand sides of \eqnss{eq:Srsgg}{eq:CirsCSirsSrsgg} and
\eqnss{eq:Srsqq}{eq:CirsSrsqq} are given by a generalization of the
singly-soft momentum mapping of \eqn{eq:PS_Sr} to the case of two soft
momenta,
\beq
\ti{p}_n^{\mu} =
\Lambda^{\mu}_{\nu}[Q,(Q-p_r-p_s)/\lambda_{rs}] (p_n^{\nu}/\lambda_{rs})\,,
\qquad n\ne r,s\,,
\label{eq:PS_Srs}
\eeq
where
\beq
\lambda_{rs} = \sqrt{1-\left(y_{(rs)Q} - y_{rs}\right)}
\label{eq:lambdars}
\eeq
and $\Lambda^{\mu}_{\nu}[Q,(Q-p_r-p_s)/\lambda_{rs}]$ is the matrix given in 
\eqn{eq:LambdaKKt}. This momentum mapping again conserves total four-momentum.

The $m+2$ parton phase space factorizes exactly under the mapping of
\eqn{eq:PS_Srs}.  We find
\beq
\PS{m+2}{}(\mom{};Q) =
\PS{m}{}(\momt{(rs)}{};Q)[\rd p_{2;m}^{(rs)}(p_r,p_s;Q)]\,,
\label{eq:PSfact_Srs}
\eeq
where the $m$ momenta in the first factor on the right hand side are those
defined in \eqn{eq:PS_Srs}. The explicit form of the factorized two-parton
phase space is
\beq
[\rd p_{2;m}^{(rs)}(p_r,p_s;Q)] =
\Jac{rs}{2;m}(p_r,p_s;Q)
\,\frac{\rd^d p_r}{(2\pi)^{d-1}}\delta_{+}(p_r^2)
\,\frac{\rd^d p_s}{(2\pi)^{d-1}}\delta_{+}(p_s^2)\,,
\label{eq:dp_Srs}
\eeq
with Jacobian 
\beq
\Jac{rs}{2;m}(p_r,p_s;Q) = \lambda_{rs}^{m(d-2)-2}\Theta(\lambda_{rs})\,.
\label{eq:Jac_Srs}
\eeq

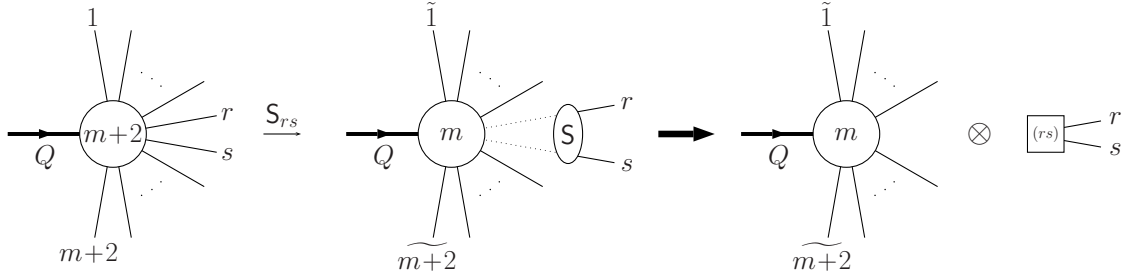
\begin{figure}
\begin{center}
\begin{pspicture}(0,0)(15,4)
\scalebox{0.5}{%

\psline[linewidth=3pt,arrowinset=0]{->}(0.2,4)(1.4,4)\uput{0.3}[d](1.2,4){\LARGE $Q$}
\psline[linewidth=3pt](1.2,4)(2.2,4)

\SpecialCoor
\psline[origin={-3,-4}](0.8;100)(2.8;100)\uput{2.9}[100]{0}(3,4){\LARGE $1$}
\psline[origin={-3,-4}](0.8;80)(2.8;80)
\psdots*[origin={-3,-4},dotscale=0.3](1.8;65)(1.8;55)(1.8;45)
\psline[origin={-3,-4}](0.8;30)(2.8;30)
\psline[origin={-3,-4}](0.8;10)(2.8;10)\uput{2.9}[10]{0}(3,4){\LARGE $r$}
\psline[origin={-3,-4}](0.8;-10)(2.8;-10)\uput{2.9}[-10]{0}(3,4){\LARGE $s$}
\psline[origin={-3,-4}](0.8;-30)(2.8;-30)
\psdots*[origin={-3,-4},dotscale=0.3](1.8;-65)(1.8;-55)(1.8;-45)
\psline[origin={-3,-4}](0.8;-80)(2.8;-80)
\psline[origin={-3,-4}](0.8;-100)(2.8;-100)\uput{2.9}[-100]{0}(3,4){\LARGE $m\!+\!2$}

\pscircle[fillstyle=solid,fillcolor=white](3,4){0.9}\rput(3,4){\LARGE $m\!+\!2$}

\psline{->}(7,4)(8,4)\uput[u](7.5,4){{\LARGE $\mathsf{S}_{rs}$}}

\psline[linewidth=3pt,arrowinset=0]{->}(9.2,4)(10.4,4)\uput{0.3}[d](10.2,4){\LARGE $Q$}
\psline[linewidth=3pt](10.2,4)(11.2,4)

\SpecialCoor
\psline[origin={-12,-4}](0.8;100)(2.8;100)\uput{2.9}[100]{0}(12,4){\LARGE $\tilde{1}$}
\psline[origin={-12,-4}](0.8;80)(2.8;80)
\psdots*[origin={-12,-4},dotscale=0.3](1.8;65)(1.8;55)(1.8;45)
\psline[origin={-12,-4}](0.8;30)(2.8;30)
\psline[origin={-12,-4}](0.8;-30)(2.8;-30)
\psdots*[origin={-12,-4},dotscale=0.3](1.8;-65)(1.8;-55)(1.8;-45)
\psline[origin={-12,-4}](0.8;-80)(2.8;-80)
\psline[origin={-12,-4}](0.8;-100)(2.8;-100)\uput{2.9}[-100]{0}(12,4){\LARGE $\widetilde{m\!+\!2}$}

\psline[origin={-12,-4},linestyle=dotted](0.8;10)(2.8;10)
\psline[origin={-12,-4},linestyle=dotted](0.8;-10)(2.8;-10)

\pscircle[fillstyle=solid,fillcolor=white](12,4){0.9}\rput(12,4){\LARGE $m$}

\psline[origin={-12,-4}](3.35;10)(4.4;10)\uput{4.55}[10]{0}(12,4){\LARGE $r$}
\psline[origin={-12,-4}](3.35;-10)(4.4;-10)\uput{4.55}[-10]{0}(12,4){\LARGE $s$}

\psellipse[origin={-12,-4},fillstyle=solid,fillcolor=white](3.1;0)(0.4,0.8)
\rput(15.1,4){\LARGE $\mathsf{S}$}


\psline[linewidth=5pt,arrowinset=0]{->}(17.5,4)(19,4)

\psline[linewidth=3pt,arrowinset=0]{->}(19.7,4)(20.9,4)\uput{0.3}[d](20.7,4){\LARGE $Q$}
\psline[linewidth=3pt](20.7,4)(21.7,4)

\SpecialCoor
\psline[origin={-22.5,-4}](0.8;100)(2.8;100)\uput{2.9}[100]{0}(22.5,4){\LARGE $\tilde{1}$}
\psline[origin={-22.5,-4}](0.8;80)(2.8;80)
\psdots*[origin={-22.5,-4},dotscale=0.3](1.8;65)(1.8;55)(1.8;45)
\psline[origin={-22.5,-4}](0.8;30)(2.8;30)
\psline[origin={-22.5,-4}](0.8;-30)(2.8;-30)
\psdots*[origin={-22.5,-4},dotscale=0.3](1.8;-65)(1.8;-55)(1.8;-45)
\psline[origin={-22.5,-4}](0.8;-80)(2.8;-80)
\psline[origin={-22.5,-4}](0.8;-100)(2.8;-100)\uput{2.9}[-100]{0}(22.5,4){\LARGE $\widetilde{m\!+\!2}$}

\pscircle[fillstyle=solid,fillcolor=white](22.5,4){0.9}\rput(22.5,4){\LARGE $m$}

\uput{3.2}[0]{0}(22.5,4){\Huge $\otimes$}
\psline[origin={-27.3,-4}](1;10)(2;10)\uput{2.2}[10]{0}(27.3,4){\LARGE $r$}
\psline[origin={-27.3,-4}](1;-10)(2;-10)\uput{2.2}[-10]{0}(27.3,4){\LARGE $s$}
\psframe[origin={-22.5,-4},fillstyle=solid,fillcolor=white]%
(4.8,-0.5)(5.8,0.5)
\rput(27.8,4){$(rs)$}
}
\end{pspicture}
\end{center}
\caption{Graphical representation of the doubly-soft momentum
mapping and the implied factorization of the phase space.}
\label{fig:Srs}
\end{figure}

We show a graphical representation for \eqnss{eq:PS_Srs}{eq:Jac_Srs}
in \fig{fig:Srs}.

The counterterms in \eqnss{eq:Srsgg}{eq:CirsCSirsSrsgg} are all defined
on the same phase spaces. Therefore, all the cancellations among these
terms in the triply-collinear limit, when $p_i^\mu || p_r^\mu || p_s^\mu$, 
in the doubly-collinear limit, when $p_i^\mu || p_r^\mu$ and
$p_j^\mu || p_s^\mu$, and in the soft-collinear limit, when
$p_i^\mu || p_r^\mu$ and $p_s^\mu \to 0$, take place in just the same way as
for the QCD factorization formulae described in \Ref{Somogyi:2005xz}.
Consequently, the combination of the terms in
\eqnss{eq:Srsgg}{eq:CirsCSirsSrsgg} as present in \eqn{eq:RR_A2} is
integrable in four dimensions in all of these limits. The same is true
for the difference of the two terms in \eqns{eq:Srsqq}{eq:CirsSrsqq}.

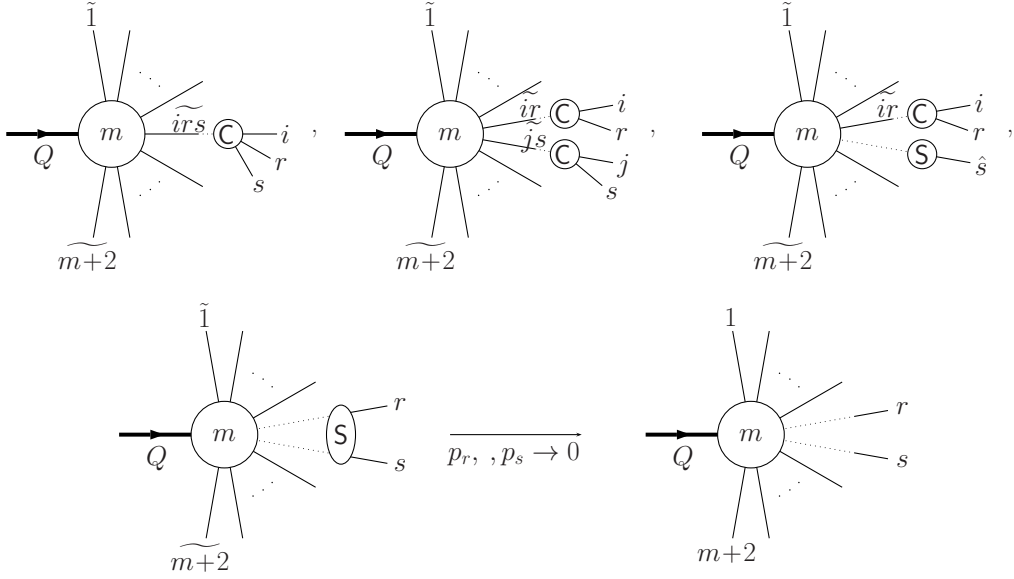
\begin{figure}
\begin{center}
\begin{pspicture}(0,0)(14,4)
\scalebox{0.5}{%

\psline[linewidth=3pt,arrowinset=0]{->}(0.2,4)(1.4,4)\uput{0.3}[d](1.2,4){\LARGE $Q$}
\psline[linewidth=3pt](1.2,4)(2.2,4)

\SpecialCoor
\psline[origin={-3,-4}](0.8;100)(2.8;100)\uput{2.9}[100]{0}(3,4){\LARGE $\tilde{1}$}
\psline[origin={-3,-4}](0.8;80)(2.8;80)
\psdots*[origin={-3,-4},dotscale=0.3](1.8;65)(1.8;55)(1.8;45)
\psline[origin={-3,-4}](0.8;30)(2.8;30)
\psline[origin={-3,-4}](0.8;0)(2.3;0)\uput{1.6}[12]{0}(3,4){\LARGE $\widetilde{irs}$}
\psline[origin={-3,-4}](0.8;-30)(2.8;-30)
\psdots*[origin={-3,-4},dotscale=0.3](1.8;-65)(1.8;-55)(1.8;-45)
\psline[origin={-3,-4}](0.8;-80)(2.8;-80)
\psline[origin={-3,-4}](0.8;-100)(2.8;-100)\uput{2.9}[-100]{0}(3,4){\LARGE $\widetilde{m\!+\!2}$}

\pscircle[fillstyle=solid,fillcolor=white](3,4){0.9}\rput(3,4){\LARGE $m$}

\psline[origin={-3,-4},linestyle=dotted](2.3;0)(2.8;0)
\psline[origin={-6.1,-4}](0.3;0)(1.3;0)\uput{1.4}[0]{0}(6.1,4){\LARGE $i$}
\psline[origin={-6.1,-4}](0.3;-30)(1.3;-30)\uput{1.4}[-30]{0}(6.1,4){\LARGE $r$}
\psline[origin={-6.1,-4}](0.3;-60)(1.3;-60)\uput{1.4}[-60]{0}(6.1,4){\LARGE $s$}

\pscircle[origin={-3,-4},fillstyle=solid,fillcolor=white](3.1;0){0.4}
\rput(6.1,4){\LARGE $\mathsf{C}\,$}

\uput{0}[0]{0}(8.3,4){\LARGE ,}

\psline[linewidth=3pt,arrowinset=0]{->}(9.2,4)(10.4,4)\uput{0.3}[d](10.2,4){\LARGE $Q$}
\psline[linewidth=3pt](10.2,4)(11.2,4)

\SpecialCoor
\psline[origin={-12,-4}](0.8;100)(2.8;100)\uput{2.9}[100]{0}(12,4){\LARGE $\tilde{1}$}
\psline[origin={-12,-4}](0.8;80)(2.8;80)
\psdots*[origin={-12,-4},dotscale=0.3](1.8;65)(1.8;55)(1.8;45)
\psline[origin={-12,-4}](0.8;30)(2.8;30)
\psline[origin={-12,-4}](0.8;10)(2.3;10)\uput{1.9}[20]{0}(12,4){\LARGE $\widetilde{ir}$}
\psline[origin={-12,-4}](0.8;-10)(2.3;-10)\uput{1.9}[-2]{0}(12,4){\LARGE $\widetilde{js}$}
\psline[origin={-12,-4}](0.8;-30)(2.8;-30)
\psdots*[origin={-12,-4},dotscale=0.3](1.8;-65)(1.8;-55)(1.8;-45)
\psline[origin={-12,-4}](0.8;-80)(2.8;-80)
\psline[origin={-12,-4}](0.8;-100)(2.8;-100)\uput{2.9}[-100]{0}(12,4){\LARGE $\widetilde{m\!+\!2}$}

\pscircle[fillstyle=solid,fillcolor=white](12,4){0.9}\rput(12,4){\LARGE $m$}

\psline[origin={-12,-4},linestyle=dotted](2.3;10)(2.8;10)
\psline[origin={-12,-4},linestyle=dotted](2.3;-10)(2.8;-10)

\psline[origin={-12,-4}](3.4;10)(4.4;10)\uput{1.4}[10]{0}(15.05,4.54){\LARGE $i$}
\psline[origin={-15.05,-4.54}](0.29;-20)(1.3;-20)\uput{1.4}[-20]{0}(15.05,4.54){\LARGE $r$}

\pscircle[origin={-12,-4},fillstyle=solid,fillcolor=white](3.1;10){0.4}
\rput(15.05,4.54){\LARGE $\mathsf{C}\,$}
\psline[origin={-12,-4}](3.4;-10)(4.4;-10)\uput{1.4}[-10]{0}(15.05,3.46){\LARGE $j$}
\psline[origin={-15.05,-3.46}](0.29;-40)(1.3;-40)\uput{1.4}[-40]{0}(15.05,3.46){\LARGE $s$}

\pscircle[origin={-12,-4},fillstyle=solid,fillcolor=white](3.1;-10){0.4}
\rput(15.05,3.46){\LARGE $\mathsf{C}\,$}

\uput{0}[0]{0}(17.3,4){\LARGE ,}

\psline[linewidth=3pt,arrowinset=0]{->}(18.7,4)(19.9,4)\uput{0.3}[d](19.7,4){\LARGE $Q$}
\psline[linewidth=3pt](19.7,4)(20.7,4)

\SpecialCoor
\psline[origin={-21.5,-4}](0.8;100)(2.8;100)\uput{2.9}[100]{0}(21.5,4){\LARGE $\tilde{1}$}
\psline[origin={-21.5,-4}](0.8;80)(2.8;80)
\psdots*[origin={-21.5,-4},dotscale=0.3](1.8;65)(1.8;55)(1.8;45)
\psline[origin={-21.5,-4}](0.8;30)(2.8;30)
\psline[origin={-21.5,-4}](0.8;10)(2.3;10)\uput{1.9}[20]{0}(21.5,4){\LARGE $\widetilde{ir}$}
\psline[origin={-21.5,-4}](0.8;-30)(2.8;-30)
\psdots*[origin={-21.5,-4},dotscale=0.3](1.8;-65)(1.8;-55)(1.8;-45)
\psline[origin={-21.5,-4}](0.8;-80)(2.8;-80)
\psline[origin={-21.5,-4}](0.8;-100)(2.8;-100)\uput{2.9}[-100]{0}(21.5,4){\LARGE $\widetilde{m\!+\!2}$}

\psline[origin={-21.5,-4},linestyle=dotted](2.3;10)(2.8;10)
\psline[origin={-21.5,-4},linestyle=dotted](0.8;-10)(2.8;-10)

\pscircle[fillstyle=solid,fillcolor=white](21.5,4){0.9}\rput(21.5,4){\LARGE $m$}

\psline[origin={-21.5,-4}](3.4;10)(4.4;10)\uput{1.4}[10]{0}(24.55,4.54){\LARGE $i$}
\psline[origin={-24.55,-4.54}](0.29;-20)(1.3;-20)\uput{1.4}[-20]{0}(24.55,4.54){\LARGE $r$}

\pscircle[origin={-21.5,-4},fillstyle=solid,fillcolor=white](3.1;10){0.4}
\rput(24.55,4.54){\LARGE $\mathsf{C}\,$}

\psline[origin={-21.5,-4}](3.4;-10)(4.4;-10)\uput{1.4}[-10]{0}(24.55,3.46){\LARGE $\hat{s}$}

\pscircle[origin={-21.5,-4},fillstyle=solid,fillcolor=white](3.1;-10){0.4}
\rput(24.55,3.46){\LARGE $\mathsf{S}$}

\uput{0}[0]{0}(26.8,4){\LARGE ,}

\psline[linewidth=3pt,arrowinset=0]{->}(3.2,-4)(4.4,-4)\uput{0.3}[d](4.2,-4){\LARGE $Q$}
\psline[linewidth=3pt](4.2,-4)(5.2,-4)

\SpecialCoor
\psline[origin={-6,4}](0.8;100)(2.8;100)\uput{2.9}[100]{0}(6,-4){\LARGE $\tilde{1}$}
\psline[origin={-6,4}](0.8;80)(2.8;80)
\psdots*[origin={-6,4},dotscale=0.3](1.8;65)(1.8;55)(1.8;45)
\psline[origin={-6,4}](0.8;30)(2.8;30)
\psline[origin={-6,4}](0.8;-30)(2.8;-30)
\psdots*[origin={-6,4},dotscale=0.3](1.8;-65)(1.8;-55)(1.8;-45)
\psline[origin={-6,4}](0.8;-80)(2.8;-80)
\psline[origin={-6,4}](0.8;-100)(2.8;-100)\uput{2.9}[-100]{0}(6,-4){\LARGE $\widetilde{m\!+\!2}$}

\psline[origin={-6,4},linestyle=dotted](0.8;10)(2.8;10)
\psline[origin={-6,4},linestyle=dotted](0.8;-10)(2.8;-10)

\pscircle[fillstyle=solid,fillcolor=white](6,-4){0.9}\rput(6,-4){\LARGE $m$}

\psline[origin={-6,4}](3.35;10)(4.4;10)\uput{4.55}[10]{0}(6,-4){\LARGE $r$}
\psline[origin={-6,4}](3.35;-10)(4.4;-10)\uput{4.55}[-10]{0}(6,-4){\LARGE $s$}

\psellipse[origin={-6,4},fillstyle=solid,fillcolor=white](3.1;0)(0.4,0.8)
\rput(9.1,-4){\LARGE $\mathsf{S}$}


\psline{->}(12,-4)(15.5,-4) \uput[u](13.7,-5){{\LARGE $p_r,\:,p_s\to 0$}}

\psline[linewidth=3pt,arrowinset=0]{->}(17.2,-4)(18.4,-4)\uput{0.3}[d](18.2,-4){\LARGE $Q$}
\psline[linewidth=3pt](18.2,-4)(19.2,-4)

\SpecialCoor
\psline[origin={-20,4}](0.8;100)(2.8;100)\uput{2.9}[100]{0}(20,-4){\LARGE 1}
\psline[origin={-20,4}](0.8;80)(2.8;80)
\psdots*[origin={-20,4},dotscale=0.3](1.8;65)(1.8;55)(1.8;45)
\psline[origin={-20,4}](0.8;30)(2.8;30)
\psline[origin={-20,4}](0.8;-30)(2.8;-30)
\psdots*[origin={-20,4},dotscale=0.3](1.8;-65)(1.8;-55)(1.8;-45)
\psline[origin={-20,4}](0.8;-80)(2.8;-80)
\psline[origin={-20,4}](0.8;-100)(2.8;-100)\uput{2.9}[-100]{0}(20,-4){\LARGE $m\!+\!2$}

\psline[origin={-20,4},linestyle=dotted](0.8;10)(2.8;10)
\psline[origin={-20,4},linestyle=dotted](0.8;-10)(2.8;-10)

\pscircle[fillstyle=solid,fillcolor=white](20,-4){0.9}\rput(20,-4){\LARGE $m$}

\psline[origin={-20,4}](2.8;10)(3.7;10)\uput{3.9}[10]{0}(20,-4){\LARGE $r$}
\psline[origin={-20,4}](2.8;-10)(3.7;-10)\uput{3.9}[-10]{0}(20,-4){\LARGE $s$}
}
\end{pspicture}
\end{center}
~\vskip 30mm
\caption{Graphical representation of the doubly-soft limit of the
triply-, doubly-, soft-collinear and doubly-soft momentum mappings.}
\label{fig:Srslimit}
\end{figure}

In the doubly-soft regions of the phase space, when $p_r^\mu$ and
$p_s^\mu \to 0$, the momentum mappings in Eqns.~(\ref{eq:PS_Cirs}),
(\ref{eq:PS_Cirjs}) and in (\ref{eq:PS_CSirs}) approach the
corresponding limit of the double soft-type mapping in \eqn{eq:PS_Srs},
shown graphically in \fig{fig:Srslimit}.  Therefore, the cancellation
of kinematical singularities among the various subtraction terms in
\eqn{eq:RR_A2} takes place in just the same way as for the QCD
factorization formulae described in \Ref{Somogyi:2005xz}, with the
exception of the double soft subtraction terms in
\eqns{eq:Srsgg}{eq:Srsqq}.  In this limit, the latter provide local
counterterms to the squared matrix element by construction.

%
%

\section{Iterated singly-unresolved counterterms}
\label{sec:RR_A12}

In \Ref{Somogyi:2005xz}, we introduced the term $\bA{12} \M{m+2}{(0)}$
that reproduced simultaneously both the singly-unresolved limits of the
doubly-unresolved limits and the doubly-unresolved limits of the
singly-unresolved limits of the squared matrix element. The structure of
these terms was such that $\bA{12} \M{m+2}{(0)}$ was defined to be
$\bA{1}\bA{2} \M{m+2}{(0)}$. When extending the terms
$\bA{1} \M{m+2}{(0)}$ and $\bA{2} \M{m+2}{(0)}$ over the whole phase
space to obtain the subtraction terms $\bcA{1}{(0)} \M{m+2}{(0)}$ and
$\bcA{2}{(0)} \M{m+2}{(0)}$, we defined the momentum mappings such that
this structure could be preserved, i.e., the iterated singly-unresolved
counterterm $\bcA{12}{(0)} \M{m+2}{(0)}$ is just the singly-unresolved
subtraction for the doubly-unresolved counterterm $\bcA{2}{(0)} \M{m+2}{(0)}$.
Thus formally we can write
\beeq
\bcA{12}{(0)}\SME{m+2}{0}{\mom{}} \aand=
\sum_{t}\Bigg[
\sum_{k\ne t} \frac12 \,\cC{kt}{}\bcA{2}{(0)}\SME{m+2}{0}{\mom{}}
\nn \\ &&
+\left(\cS{t}{}\bcA{2}{(0)}\SME{m+2}{0}{\mom{}} -
\sum_{k\ne t} \cC{kt}{}\cS{t}{}\bcA{2}{(0)}\SME{m+2}{0}{\mom{}}\right)\Bigg]\,,
\label{eq:RR_A12}
\eeeq
where the three terms in \eqn{eq:RR_A12} each evaluate further into the
expressions,%
\footnote{At the level of the factorization formulae in
\Ref{Somogyi:2005xz} we neglected subleading terms in the
triply-collinear limit of the soft-collinear formula (see Eqns.~(4.33)
and (4.34) in \cite{Somogyi:2005xz}), which we do not apply here.
Therefore, the soft limit of the doubly-real subtraction is slightly
different from that in \Ref{Somogyi:2005xz}.}
\beeq
\cC{kt}{}\bcA{2}{(0)} \aand= 
\sum_{r\ne k,t} \Bigg[\cC{kt}{}\cC{ktr}{(0,0)} +\cC{kt}{}\cSCS{kt;r}{(0,0)} 
- \cC{kt}{}\cC{ktr}{}\cSCS{kt;r}{(0,0)}
- \cC{kt}{}\cC{rkt}{}\cS{kt}{(0,0)}
\nn \\ &&\qquad\qquad
+ \sum_{i\ne r,k,t}\Bigg(\frac12\,\cC{kt}{}\cC{ir;kt}{(0,0)} 
- \cC{kt}{}\cC{ir;kt}{}\cSCS{kt;r}{(0,0)}\Bigg)\Bigg]
+ \cC{kt}{}\cS{kt}{(0,0)}\,,
\label{eq:CktA2}
\eeeq
\beeq
\cS{t}{}\bcA{2}{(0)} \aand= 
\sum_{r\ne t}\Bigg\{\sum_{i\ne r,t} 
\Bigg[\frac12\Bigg(\cS{t}{}\cC{irt}{(0,0)} +\cS{t}{}\cSCS{ir;t}{(0,0)} 
- \cS{t}{}\cC{irt}{}\cSCS{ir;t}{(0,0)}\Bigg)
\nn\\ &&\qquad\qquad\qquad
- \cS{t}{}\cC{irt}{}\cS{rt}{(0,0)} - \cS{t}{}\cSCS{ir;t}{}\cS{rt}{(0,0)} 
+ \cS{t}{}\cC{irt}{}\cSCS{ir;t}{}\cS{rt}{(0,0)}\Bigg] 
+ \cS{t}{}\cS{rt}{(0,0)}\Bigg\}
\label{eq:StA2}
\eeeq
and 
\beeq
\cC{kt}{}\cS{t}{}\bcA{2}{(0)} \aand= 
  \sum_{r\ne k,t} \Bigg[\cC{kt}{}\cS{t}{}\cC{krt}{(0,0)} 
+ \sum_{i\ne r,k,t}\Bigg(\frac12\cC{kt}{}\cS{t}{}\cSCS{ir;t}{(0,0)} 
- \cC{kt}{}\cS{t}{}\cSCS{ir;t}{}\cS{rt}{(0,0)}\Bigg)
\nn \\ &&\qquad\qquad
- \cC{kt}{}\cS{t}{}\cC{krt}{}\cS{rt}{(0,0)}
- \cC{kt}{}\cS{t}{}\cC{rkt}{}\cS{kt}{(0,0)}
+ \cC{kt}{}\cS{t}{}\cS{rt}{(0,0)}\Bigg]
+ \cC{kt}{}\cS{t}{}\cS{kt}{(0,0)}\,.
\label{eq:CktStA2}
\eeeq
In this section we spell out explicitly all subtraction terms that
appear in \eqnss{eq:CktA2}{eq:CktStA2}. These subtraction terms are
defined on factorized phase spaces that are obtained using two
singly-unresolved collinear and/or soft-type mappings iteratively. 


\subsection{Iterated collinear counterterms}
\label{ssec:CktA2}

\subsubsection{Iterated collinear--triple collinear counterterm}

\subtitle{Counterterm}

The counterterm corresponding to the collinear limit of the triple
collinear subtraction is
\beeq
&&
\cC{kt}{}\cC{ktr}{(0,0)}(\mom{}) =
(8\pi\as\mu^{2\eps})^2\frac{1}{s_{kt}}\frac{1}{s_{\wha{kt}\ha{r}}}
\label{eq:CktCktr}
\\ \nn &&\qquad\times
\bra{m}{(0)}{(\momt{(\wha{kt}\ha{r},kt)}{})}
\hP_{f_k f_t f_r}^{{\rm s.o.\,(0)}}
(\{\tzz{j}{l},\tzz{\ha{j}}{\ha{l}},\kTt{j,l},\kTt{\ha{j},\ha{l}},s_{jl};\eps\})
\ket{m}{(0)}{(\momt{(\wha{kt}\ha{r},kt)}{})}\,,
\eeeq
where $j,\:l = k,\:t$ and $\ha{j},\:\ha{l} = \wha{kt},\:\ha{r}$,
defined in \eqnss{eq:PS_CktCktr}{eq:PS_CktCktr_t}. Note that in our
convention the ordering of the labels on the splitting-kernels is
usually meaningless, but in the strongly ordered kernels 
$\hP_{f_k f_t f_r}^{{\rm s.o.\,(0)}}$ the ordering matters.
As a result, the same triple-parton splitting function may
have different strongly-ordered limits, which can be distinguished 
by the momentum labels in the kernel, once the ordering of the limits is
fixed by the momentum mapping, given in \eqn{eq:PS_CktCktr}. Thus, in the 
following kernels, we do not further mark the abelian and non-abelian
limits as we did at the level of the factorization formulae in
\Ref{Somogyi:2005xz}. For quark splitting we have
\beeq
&&
\la s|\hP_{q_r \qb'_k q'_t}^{{\rm s.o.\,(0)}}
(\tzz{k}{t},\tzz{t}{k},\kTt{k,t},\tzz{\wha{kt}}{\ha{r}}
,\tzz{\ha{r}}{\wha{kt}},\kTt{\wha{kt},\ha{r}};\eps)|s'\ra =
\nn \\[2mm] &&\qquad
=\TR\Bigg[
P_{q_{\ha{r}} g_{\wha{kt}}}^{(0)}
(\tzz{\ha{r}}{\wha{kt}},\tzz{\wha{kt}}{\ha{r}};\eps)
-2\CF \tzz{k}{t}\tzz{t}{k}\Bigg(\tzz{\wha{kt}}{\ha{r}}
-\frac{s_{\ha{r}\kTt{k,t}}^2}{\kTt{k,t}^2 s_{\wha{kt}\ha{r}}}\Bigg)
\Bigg]\delta_{ss'}\,,
\label{eq:PqrqbkqtSO}
\\[2mm]
&&
\la s|\hP_{q_k g_t g_r}^{{\rm s.o.\,(0)}}(\tzz{k}{t},\tzz{t}{k},
\tzz{\wha{kt}}{\ha{r}},\tzz{\ha{r}}{\wha{kt}};\eps)|s'\ra =
\nn\\[2mm] &&\qquad=
P_{q_k g_t}^{(0)}(\tzz{k}{t},\tzz{t}{k};\eps)
P_{q_{\wha{kt}} g_{\ha{r}}}^{(0)}(\tzz{\wha{kt}}{\ha{r}},\tzz{\ha{r}}{\wha{kt}};\eps)
\delta_{ss'}\,,
\label{eq:PqkgtgrSO}
\\[2mm]
&&
\la s|P_{q_r g_k g_t}^{{\rm s.o.\,(0)}}
(\tzz{k}{t},\tzz{t}{k},\kTt{k,t},\tzz{\wha{kt}}{\ha{r}}
,\tzz{\ha{r}}{\wha{kt}},\kTt{\wha{kt},\ha{r}};\eps)|s'\ra =
2\CA\Bigg[
P_{q_{\ha{r}} g_{\wha{kt}}}^{(0)}
(\tzz{\ha{r}}{\wha{kt}},\tzz{\wha{kt}}{\ha{r}};\eps)
\nn \\[2mm]&&\qquad\times\,
\Bigg(\frac{\tzz{k}{t}}{\tzz{t}{k}} + \frac{\tzz{t}{k}}{\tzz{k}{t}}\Bigg)
+\CF(1-\eps) \tzz{k}{t}\tzz{t}{k}\Bigg(\tzz{\wha{kt}}{\ha{r}}
-\frac{s_{\ha{r}\kTt{k,t}}^2}{\kTt{k,t}^2 s_{\wha{kt}\ha{r}}}\Bigg)\Bigg]
\delta_{ss'}\,,
\label{eq:qrgkgtSO}
\eeeq
where $s_{\ha{r}\kTt{k,t}} = 2p_{\ha{r}}\cdot \kTt{k,t}$. For gluon
splitting we define 
\beeq
&&
\la\mu |\hP_{g_k q_t \qb_r}^{{\rm s.o.\,(0)}}
(\tzz{k}{t},\tzz{t}{k},\tzz{\wha{kt}}{\ha{r}}
,\tzz{\ha{r}}{\wha{kt}},\kTt{\wha{kt},\ha{r}};\eps)|\nu\ra =
\label{eq:PgkqtqbrSO}
\nn \\[3mm]&&\qquad
=P_{q_t g_k}^{(0)}(\tzz{t}{k},\tzz{k}{t};\eps)
\la\mu |
\hP_{\qb_{\ha{r}}q_{(\wha{tk})}}^{(0)}
(\tzz{\ha{r}}{(\wha{tk})},\tzz{(\wha{tk})}{\ha{r}},\kTt{\ha{r}(\wha{tk})};\eps)
|\nu\ra\,,
\\[3mm]&&
\la\mu |\hP_{g_r q_k \qb_t}^{{\rm s.o.\,(0)}}
(\tzz{k}{t},\tzz{t}{k},\kTt{k,t},\tzz{\wha{kt}}{\ha{r}}
,\tzz{\ha{r}}{\wha{kt}},\kTt{\wha{kt},\ha{r}};\eps)|\nu\ra =
\nn \\[3mm] &&\qquad=
2\CA\TR\Bigg[
-g^{\mu\nu}\Bigg(\frac{\tzz{\ha{r}}{\wha{kt}}}{\tzz{\wha{kt}}{\ha{r}}} 
+ \frac{\tzz{\wha{kt}}{\ha{r}}}{\tzz{\ha{r}}{\wha{kt}}}
 +\tzz{k}{t}\tzz{t}{k}
  \frac{s_{\ha{r}\kTt{k,t}}^2}{\kTt{k,t}^2 s_{\wha{kt}\ha{r}}}\Bigg)
+ 4\tzz{k}{t}\tzz{t}{k}\frac{\tzz{\wha{kt}}{\ha{r}}}{\tzz{\ha{r}}{\wha{kt}}}
  \frac{\kTtm{k,t}{\mu}\kTtm{k,t}{\nu}}{\kTt{k,t}^2}\Bigg]
\nn \\[3mm] &&\qquad
-\,4\CA(1-\eps)\tzz{\ha{r}}{\wha{kt}}\tzz{\wha{kt}}{\ha{r}}
P_{q_k \qb_t}^{(0)}(\tzz{k}{t},\tzz{t}{k},\kTt{k,t};\eps)
\frac{\kTtm{\ha{r},\wha{kt}}{\mu}\kTtm{\ha{r},\wha{kt}}{\nu}}
     {\kTt{\ha{r},\wha{kt}}^2}
\label{eq:PgrqkqbtSO}
\eeeq
and
\beeq
&&
\la\mu |\hP_{g_k g_t g_r}^{{\rm s.o.\,(0)}}
(\tzz{k}{t},\tzz{t}{k},\kTt{k,t},\tzz{\wha{kt}}{\ha{r}}
,\tzz{\ha{r}}{\wha{kt}},\kTt{\wha{kt},\ha{r}};\eps)|\nu\ra =
4\CA^2\Bigg[
- g^{\mu\nu}\Bigg(\frac{\tzz{\ha{r}}{\wha{kt}}}{\tzz{\wha{kt}}{\ha{r}}}
+ \frac{\tzz{\wha{kt}}{\ha{r}}}{\tzz{\ha{r}}{\wha{kt}}}\Bigg)
\nn \\[2mm] &&\qquad
\times\,
\Bigg(\frac{\tzz{k}{t}}{\tzz{t}{k}} + \frac{\tzz{t}{k}}{\tzz{k}{t}}\Bigg)
+ g^{\mu\nu} \tzz{k}{t}\tzz{t}{k}\frac{1-\eps}{2}
  \frac{s_{\ha{r}\kTt{k,t}}^2}{\kTt{k,t}^2 s_{\wha{kt}\ha{r}}}
- 2(1-\eps)\tzz{k}{t}\tzz{t}{k}
  \frac{\tzz{\wha{kt}}{\ha{r}}}{\tzz{\ha{r}}{\wha{kt}}}
\frac{\kTtm{k,t}{\mu}\kTtm{k,t}{\nu}}{\kTt{k,t}^2}\Bigg]
\nn \\[2mm] &&\qquad
-\,4\CA(1-\eps)\tzz{\ha{r}}{\wha{kt}}\tzz{\wha{kt}}{\ha{r}}
   P_{g_k g_t}^{(0)}(\tzz{k}{t},\tzz{t}{k},\kTt{k,t};\eps)
   \frac{\kTtm{\ha{r},\wha{kt}}{\mu}\kTtm{\ha{r},\wha{kt}}{\nu}}
        {\kTt{\ha{r},\wha{kt}}^2}\,.
\label{eq:PgkgtgrSO}
\eeeq
The momentum fractions $\tzz{j}{l}$ and the transverse momenta
$\kTt{j,l}$ are defined in \eqns{eq:zt2}{eq:kTtir}, respectively, with
$\zeta_{ir}$ in \eqn{eq:zetair}.  The hatted momenta that also appear
in the definition $\tzz{\ha{j}}{\ha{l}}$ and $\kTt{\ha{j},\ha{l}}$ are
defined in \eqn{eq:PS_CktCktr_h}. 

\subtitle{Momentum mapping and phase space factorization}

The $m$ momenta $\momt{(\wha{kt}\ha{r},kt)}{} \equiv 
\{\ti{p}_1,\ldots,\ti{p}_{\wha{kt}\ha{r}},\ldots,\ti{p}_{m+2}\}$ 
appearing in the matrix elements on the right hand side of
\eqn{eq:CktCktr} are defined in two successive collinear mappings
through an intermediate set of $m+1$ momenta 
$\momh{(kt)}{+1} \equiv \{\ha{p}_1,\ldots,\ha{p}_{kt},\ldots,\ha{p}_{m+2}\}$,
\beq
\mom{} \cmap{kt} \momh{(kt)}{+1}
\cmap{\wha{kt}\ha{r}} \momt{(\wha{kt}\ha{r},kt)}{}\,,
\label{eq:PS_CktCktr}
\eeq
or explicitly
\beq
\ha{p}_{kt}^{\mu} = \frac{1}{1-\alpha_{kt}}
(p_k^{\mu} + p_t^{\mu} - \alpha_{kt} Q^{\mu})\,,
\qquad
\ha{p}_n^{\mu} = \frac{1}{1-\alpha_{kt}} p_n^{\mu}\,,
\qquad n\ne k,t
\label{eq:PS_CktCktr_h}
\eeq
and
\beq
\ti{p}_{\wha{kt}\ha{r}}^{\mu} = \frac{1}{1-\alpha_{\wha{kt}\ha{r}}}
(\ha{p}_{kt}^{\mu} + \ha{p}_r^{\mu} - \alpha_{\wha{kt}\ha{r}} Q^{\mu})\,,
\qquad
\ti{p}_n^{\mu} = \frac{1}{1-\alpha_{\wha{kt}\ha{r}}} \ha{p}_n^{\mu}\,,
\qquad n\ne \wha{kt},\ha{r}\,,
\label{eq:PS_CktCktr_t}
\eeq
where $\alpha_{kt}$ and $\alpha_{\wha{kt}\ha{r}}$ are given by
\eqn{eq:alphair}, the latter being defined on the hatted momenta. 

\begin{figure}
\begin{center}
\begin{pspicture}(0,0)(15,4)
\scalebox{0.5}{%

\psline[linewidth=3pt,arrowinset=0]{->}(0.2,4)(1.4,4)\uput{0.3}[d](1.2,4){\LARGE $Q$}
\psline[linewidth=3pt](1.2,4)(2.2,4)

\SpecialCoor
\psline[origin={-3,-4}](0.8;100)(2.8;100)\uput{2.9}[100]{0}(3,4){\LARGE $1$}
\psline[origin={-3,-4}](0.8;80)(2.8;80)
\psdots*[origin={-3,-4},dotscale=0.3](1.8;65)(1.8;55)(1.8;45)
\psline[origin={-3,-4}](0.8;30)(2.8;30)
\psline[origin={-3,-4}](0.8;15)(2.8;15)\uput{2.9}[15]{0}(3,4){\LARGE $k$}
\psline[origin={-3,-4}](0.8;0)(2.8;0)\uput{2.9}[0]{0}(3,4){\LARGE $t$}
\psline[origin={-3,-4}](0.8;-15)(2.8;-15)\uput{2.9}[-15]{0}(3,4){\LARGE $r$}
\psline[origin={-3,-4}](0.8;-30)(2.8;-30)
\psdots*[origin={-3,-4},dotscale=0.3](1.8;-65)(1.8;-55)(1.8;-45)
\psline[origin={-3,-4}](0.8;-80)(2.8;-80)
\psline[origin={-3,-4}](0.8;-100)(2.8;-100)\uput{2.9}[-100]{0}(3,4){\LARGE $m\!+\!2$}

\pscircle[fillstyle=solid,fillcolor=white](3,4){0.9}\rput(3,4){\LARGE $m\!+\!2$}

\psline{->}(7,4)(8,4)\uput[u](7.5,4){{\LARGE $\mathsf{C}_{kt}$}}

\psline[linewidth=3pt,arrowinset=0]{->}(9.2,4)(10.4,4)\uput{0.3}[d](10.2,4){\LARGE $Q$}
\psline[linewidth=3pt](10.2,4)(11.2,4)

\SpecialCoor
\psline[origin={-12,-4}](0.8;100)(2.8;100)\uput{2.9}[100]{0}(12,4){\LARGE $\hat{1}$}
\psline[origin={-12,-4}](0.8;80)(2.8;80)
\psdots*[origin={-12,-4},dotscale=0.3](1.8;65)(1.8;55)(1.8;45)
\psline[origin={-12,-4}](0.8;30)(2.8;30)
\psline[origin={-12,-4}](0.8;10)(2.3;10)\uput{1.9}[20]{0}(12,4){\LARGE $\widehat{kt}$}
\psline[origin={-12,-4}](0.8;-10)(2.8;-10)\uput{3.0}[-10]{0}(12,4){\LARGE $\widehat{r}$}
\psline[origin={-12,-4}](0.8;-30)(2.8;-30)
\psdots*[origin={-12,-4},dotscale=0.3](1.8;-65)(1.8;-55)(1.8;-45)
\psline[origin={-12,-4}](0.8;-80)(2.8;-80)
\psline[origin={-12,-4}](0.8;-100)(2.8;-100)\uput{2.9}[-100]{0}(12,4){\LARGE $\widehat{m\!+\!2}$}

\pscircle[fillstyle=solid,fillcolor=white](12,4){0.9}\rput(12,4){\LARGE $m\!+\!1$}

\psline[origin={-12,-4},linestyle=dotted](2.3;10)(2.8;10)
\psline[origin={-12,-4},linestyle=dotted](2.3;-10)(2.8;-10)

\psline[origin={-12,-4}](3.4;10)(4.4;10)\uput{1.4}[10]{0}(15.05,4.54){\LARGE $k$}
\psline[origin={-15.05,-4.54}](0.29;-20)(1.3;-20)\uput{1.4}[-20]{0}(15.05,4.54){\LARGE $t$}

\pscircle[origin={-12,-4},fillstyle=solid,fillcolor=white](3.1;10){0.4}
\rput(15.05,4.54){\LARGE $\mathsf{C}\,$}

\psline{->}(17.5,4)(18.5,4)\uput[u](18,4){{\LARGE $\mathsf{C}_{\widehat{kt}\hat{r}}$}}

\psline[linewidth=3pt,arrowinset=0]{->}(19.7,4)(20.9,4)\uput{0.3}[d](20.7,4){\LARGE $Q$}
\psline[linewidth=3pt](20.7,4)(21.7,4)

\SpecialCoor
\psline[origin={-22.5,-4}](0.8;100)(2.8;100)\uput{2.9}[100]{0}(22.5,4){\LARGE $\tilde{1}$}
\psline[origin={-22.5,-4}](0.8;80)(2.8;80)
\psdots*[origin={-22.5,-4},dotscale=0.3](1.8;65)(1.8;55)(1.8;45)
\psline[origin={-22.5,-4}](0.8;30)(2.8;30)
\psline[origin={-22.5,-4}](0.8;0)(2.3;0)\uput{1.6}[14]{0}(22.5,4){\LARGE $\widetilde{ktr}$}
\psline[origin={-22.5,-4}](0.8;-30)(2.8;-30)
\psdots*[origin={-22.5,-4},dotscale=0.3](1.8;-65)(1.8;-55)(1.8;-45)
\psline[origin={-22.5,-4}](0.8;-80)(2.8;-80)
\psline[origin={-22.5,-4}](0.8;-100)(2.8;-100)\uput{2.9}[-100]{0}(22.5,4){\LARGE $\widetilde{m\!+\!2}$}

\psline[origin={-22.5,-4},linestyle=dotted](2.3;0)(2.8;0)

\pscircle[fillstyle=solid,fillcolor=white](22.5,4){0.9}\rput(22.5,4){\LARGE $m$}

\psline[origin={-22.5,-4}](3.4;0)(3.9;0)\uput{0.6}[29]{0}(25.6,4){\LARGE $\hat{kt}$}
\psline[origin={-22.5,-4},linestyle=dotted](3.9;0)(4.4;0)
\psline[origin={-25.6,-4}](0.29;-30)(1.3;-30)\uput{1.4}[-30]{0}(25.6,4){\LARGE $\hat{r}$}

\pscircle[origin={-22.5,-4},fillstyle=solid,fillcolor=white](3.1;0){0.4}
\rput(25.6,4){\LARGE $\mathsf{C}\,$}
\psline[origin={-27.4,-4}](0.4;0)(1.4;0)\uput{1.6}[0]{0}(27.4,4){\LARGE $k$}
\psline[origin={-27.4,-4}](0.4;-30)(1.4;-30)\uput{1.6}[-30]{0}(27.4,4){\LARGE $t$}

\pscircle[origin={-27.4,-4},fillstyle=solid,fillcolor=white](0;0){0.4}
\rput(27.4,4){\LARGE $\mathsf{C}\,$}


\psline[linewidth=5pt,arrowinset=0]{->}(7,-4)(8.5,-4)

\psline[origin={10.5,8},linewidth=3pt,arrowinset=0]{->}(19.7,4)(20.9,4)\uput{0.3}[d](10.2,-4){\LARGE $Q$}
\psline[origin={10.5,8},linewidth=3pt](20.7,4)(21.7,4)

\SpecialCoor
\psline[origin={-12,4}](0.8;100)(2.8;100)\uput{2.9}[100]{0}(12,-4){\LARGE $\tilde{1}$}
\psline[origin={-12,4}](0.8;80)(2.8;80)
\psdots*[origin={-12,4},dotscale=0.3](1.8;65)(1.8;55)(1.8;45)
\psline[origin={-12,4}](0.8;30)(2.8;30)
\psline[origin={-12,4}](0.8;0)(2.8;0)\uput{2.9}[0]{0}(12,-4){\LARGE $\widetilde{ktr}$}
\psline[origin={-12,4}](0.8;-30)(2.8;-30)
\psdots*[origin={-12,4},dotscale=0.3](1.8;-65)(1.8;-55)(1.8;-45)
\psline[origin={-12,4}](0.8;-80)(2.8;-80)
\psline[origin={-12,4}](0.8;-100)(2.8;-100)\uput{2.9}[-100]{0}(12,-4){\LARGE $\widetilde{m\!+\!2}$}

\pscircle[fillstyle=solid,fillcolor=white](12,-4){0.9}\rput(12,-4){\LARGE $m$}

\uput{3.7}[0]{0}(12.5,-4){\Huge $\otimes$}
\psline[origin={-12.5,4}](5.8;0)(6.8;0)
\psframe[origin={-12.5,4},fillstyle=solid,fillcolor=white]%
(4.8,-0.5)(5.8,0.5)
\rput(17.8,-4){$(kt)$}
\uput{7}[0]{0}(12.5,-4){\LARGE $t$}

\uput{3.7}[0]{0}(16.5,-4){\Huge $\otimes$}
\psline[origin={-16.5,4}](5.8;0)(6.8;0)
\psframe[origin={-16.5,4},fillstyle=solid,fillcolor=white]%
(4.8,-0.5)(5.8,0.5)
\rput(21.8,-4){$(\widehat{kt}\hat{r})$}
\uput{7}[0]{0}(16.5,-4){\LARGE $\hat{r}$}
}
\end{pspicture}
\end{center}
~\vskip 26mm
\caption{Graphical representation of the iterated collinear momentum
mapping and the implied factorization of the phase space.}
\label{fig:CktCktr}.
\end{figure}
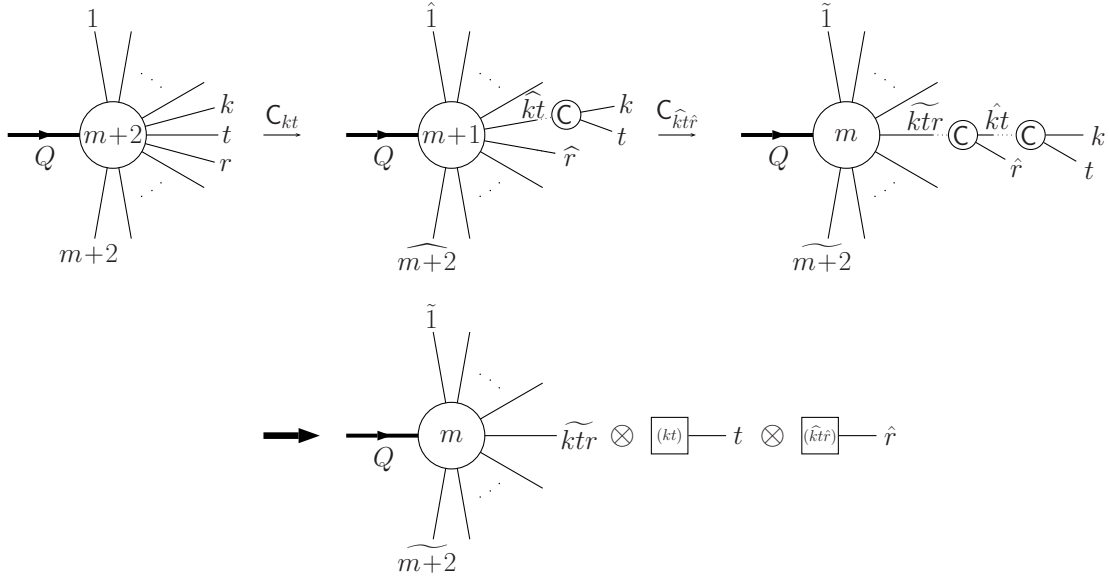

This iterated momentum mapping leads to exact phase space factorization
in the iterated form
\beq
\PS{m+2}{}(\mom{};Q) = \PS{m}{}(\momt{(\wha{kt}\ha{r},kt)}{};Q)
[\rd p_{1;m}^{(\wha{kt}\ha{r})}(\ha{p}_{r},\ti{p}_{\wha{kt}\ha{r}};Q)]
[\rd p_{1;m+1}^{(kt)}(p_k,\ha{p}_{kt};Q)]\,.
\label{eq:PSfact_CktCktr}
\eeq
The one-parton factorized phase spaces
$[\rd p_{1;m+1}^{(kt)}(p_k,\ha{p}_{kt};Q)]$ and
$[\rd p_{1;m}^{(\wha{kt}\ha{r})}(\ha{p}_{r},\ti{p}_{\wha{kt}\ha{r}};Q)]$ are
given explicitly by \eqn{eq:dp_Cir}.  
The graphical representation of this iterated momentum mapping and the
corresponding factorization of the phase space is shown in \fig{fig:CktCktr}.

\subsubsection{Iterated collinear--double collinear counterterm}

\subtitle{Counterterm}

Corresponding to the collinear limit of the double collinear
subtraction is the counterterm
\beeq
&&
\cC{kt}{}\cC{ir;kt}{(0,0)}(\mom{}) =
(8\pi\as\mu^{2\eps})^2\frac{1}{s_{kt}}\frac{1}{s_{\ha{i}\ha{r}}}
\label{eq:CktCirkt}
\\ \nn&&\qquad\times\,
\bra{m}{(0)}{(\momt{(\ha{i}\ha{r},kt)}{})}
\hP_{f_k f_t}^{(0)}(\tzz{k}{t},\tzz{t}{k},\kTt{k,t};\eps)
\hP_{f_{\ha{i}} f_{\ha{r}}}^{(0)}(\tzz{\ha{i}}{\ha{r}},\tzz{\ha{r}}{\ha{i}},\kTt{\ha{i},\ha{r}}
;\eps)\ket{m}{(0)}{(\momt{(\ha{i}\ha{r},kt)}{})}\,.
\eeeq
The variables of the Altarelli-Parisi kernels, the momentum fractions
and transverse momenta, are given by \eqns{eq:zt2}{eq:kTtir} while the
kernels themselves are recorded in \eqnss{eq:Pqg0}{eq:Pgg0}.  

\subtitle{Momentum mapping and phase space factorization}

\begin{figure}
\begin{center}
\begin{pspicture}(0,0)(15,4)
\scalebox{0.5}{%

\psline[linewidth=3pt,arrowinset=0]{->}(0.2,4)(1.4,4)\uput{0.3}[d](1.2,4){\LARGE $Q$}
\psline[linewidth=3pt](1.2,4)(2.2,4)

\SpecialCoor
\psline[origin={-3,-4}](0.8;100)(2.8;100)\uput{2.9}[100]{0}(3,4){\LARGE $1$}
\psline[origin={-3,-4}](0.8;80)(2.8;80)
\psdots*[origin={-3,-4},dotscale=0.3](1.8;65)(1.8;55)(1.8;45)
\psline[origin={-3,-4}](0.8;30)(2.8;30)
\psline[origin={-3,-4}](0.8;18)(2.8;18)\uput{2.9}[18]{0}(3,4){\LARGE $k$}
\psline[origin={-3,-4}](0.8;6)(2.8;6)\uput{2.9}[6]{0}(3,4){\LARGE $t$}
\psline[origin={-3,-4}](0.8;-6)(2.8;-6)\uput{2.9}[-6]{0}(3,4){\LARGE $i$}
\psline[origin={-3,-4}](0.8;-18)(2.8;-18)\uput{2.9}[-18]{0}(3,4){\LARGE $r$}
\psline[origin={-3,-4}](0.8;-30)(2.8;-30)
\psdots*[origin={-3,-4},dotscale=0.3](1.8;-65)(1.8;-55)(1.8;-45)
\psline[origin={-3,-4}](0.8;-80)(2.8;-80)
\psline[origin={-3,-4}](0.8;-100)(2.8;-100)\uput{2.9}[-100]{0}(3,4){\LARGE $m\!+\!2$}

\pscircle[fillstyle=solid,fillcolor=white](3,4){0.9}\rput(3,4){\LARGE $m\!+\!2$}

\psline{->}(7,4)(8,4)\uput[u](7.5,4){{\LARGE $\mathsf{C}_{kt}$}}

\psline[linewidth=3pt,arrowinset=0]{->}(9.2,4)(10.4,4)\uput{0.3}[d](10.2,4){\LARGE $Q$}
\psline[linewidth=3pt](10.2,4)(11.2,4)

\SpecialCoor
\psline[origin={-12,-4}](0.8;100)(2.8;100)\uput{2.9}[100]{0}(12,4){\LARGE $\hat{1}$}
\psline[origin={-12,-4}](0.8;80)(2.8;80)
\psdots*[origin={-12,-4},dotscale=0.3](1.8;65)(1.8;55)(1.8;45)
\psline[origin={-12,-4}](0.8;30)(2.8;30)
\psline[origin={-12,-4}](0.8;10)(2.3;10)\uput{1.9}[20]{0}(12,4){\LARGE $\widehat{kt}$}
\psline[origin={-12,-4}](0.8;-3.33)(2.8;-3.33)\uput{3.0}[-3.33]{0}(12,4){\LARGE $\widehat{i}$}
\psline[origin={-12,-4}](0.8;-16.67)(2.8;-16.67)\uput{3.0}[-16.67]{0}(12,4){\LARGE $\widehat{r}$}
\psline[origin={-12,-4}](0.8;-30)(2.8;-30)
\psdots*[origin={-12,-4},dotscale=0.3](1.8;-65)(1.8;-55)(1.8;-45)
\psline[origin={-12,-4}](0.8;-80)(2.8;-80)
\psline[origin={-12,-4}](0.8;-100)(2.8;-100)\uput{2.9}[-100]{0}(12,4){\LARGE $\widehat{m\!+\!2}$}

\pscircle[fillstyle=solid,fillcolor=white](12,4){0.9}\rput(12,4){\LARGE $m\!+\!1$}

\psline[origin={-12,-4},linestyle=dotted](2.3;10)(2.8;10)

\psline[origin={-12,-4}](3.4;10)(4.4;10)\uput{1.4}[10]{0}(15.05,4.54){\LARGE $k$}
\psline[origin={-15.05,-4.54}](0.29;-20)(1.3;-20)\uput{1.4}[-20]{0}(15.05,4.54){\LARGE $t$}

\pscircle[origin={-12,-4},fillstyle=solid,fillcolor=white](3.1;10){0.4}
\rput(15.05,4.54){\LARGE $\mathsf{C}\,$}

\psline{->}(17.5,4)(18.5,4)\uput[u](18,4){{\LARGE $\mathsf{C}_{\widehat{i}\hat{r}}$}}

\psline[linewidth=3pt,arrowinset=0]{->}(19.7,4)(20.9,4)\uput{0.3}[d](20.7,4){\LARGE $Q$}
\psline[linewidth=3pt](20.7,4)(21.7,4)

\SpecialCoor
\psline[origin={-22.5,-4}](0.8;100)(2.8;100)\uput{2.9}[100]{0}(22.5,4){\LARGE $\tilde{1}$}
\psline[origin={-22.5,-4}](0.8;80)(2.8;80)
\psdots*[origin={-22.5,-4},dotscale=0.3](1.8;65)(1.8;55)(1.8;45)
\psline[origin={-22.5,-4}](0.8;30)(2.8;30)
\psline[origin={-22.5,-4}](0.8;10)(2.3;10)\uput{1.9}[20]{0}(22.5,4){\LARGE $\widetilde{kt}$}
\psline[origin={-22.5,-4}](0.8;-10)(2.3;-10)\uput{1.9}[-2]{0}(22.5,4){\LARGE $\widetilde{ir}$}
\psline[origin={-22.5,-4}](0.8;-30)(2.8;-30)
\psdots*[origin={-22.5,-4},dotscale=0.3](1.8;-65)(1.8;-55)(1.8;-45)
\psline[origin={-22.5,-4}](0.8;-80)(2.8;-80)
\psline[origin={-22.5,-4}](0.8;-100)(2.8;-100)\uput{2.9}[-100]{0}(22.5,4){\LARGE $\widetilde{m\!+\!2}$}

\psline[origin={-22.5,-4},linestyle=dotted](2.3;10)(2.8;10)
\psline[origin={-22.5,-4},linestyle=dotted](2.3;-10)(2.8;-10)

\pscircle[fillstyle=solid,fillcolor=white](22.5,4){0.9}\rput(22.5,4){\LARGE $m$}

\psline[origin={-22.5,-4}](3.4;10)(4.4;10)\uput{1.4}[10]{0}(25.55,4.54){\LARGE $k$}
\psline[origin={-25.55,-4.54}](0.29;-20)(1.3;-20)\uput{1.4}[-20]{0}(25.55,4.54){\LARGE $t$}

\pscircle[origin={-22.5,-4},fillstyle=solid,fillcolor=white](3.1;10){0.4}
\rput(25.55,4.54){\LARGE $\mathsf{C}\,$}

\psline[origin={-22.5,-4}](3.4;-10)(4.4;-10)\uput{1.4}[-10]{0}(25.55,3.46){\LARGE $\hat{i}$}
\psline[origin={-25.55,-3.46}](0.29;-40)(1.3;-40)\uput{1.4}[-40]{0}(25.55,3.46){\LARGE $\hat{r}$}

\pscircle[origin={-22.5,-4},fillstyle=solid,fillcolor=white](3.1;-10){0.4}
\rput(25.55,3.46){\LARGE $\mathsf{C}\,$}


\psline[linewidth=5pt,arrowinset=0]{->}(7,-4)(8.5,-4)

\psline[origin={10.5,8},linewidth=3pt,arrowinset=0]{->}(19.7,4)(20.9,4)\uput{0.3}[d](10.2,-4){\LARGE $Q$}
\psline[origin={10.5,8},linewidth=3pt](20.7,4)(21.7,4)

\SpecialCoor
\psline[origin={-12,4}](0.8;100)(2.8;100)\uput{2.9}[100]{0}(12,-4){\LARGE $\tilde{1}$}
\psline[origin={-12,4}](0.8;80)(2.8;80)
\psdots*[origin={-12,4},dotscale=0.3](1.8;65)(1.8;55)(1.8;45)
\psline[origin={-12,4}](0.8;30)(2.8;30)
\psline[origin={-12,4}](0.8;10)(2.8;10)\uput{2.9}[10]{0}(12,-4){\LARGE $\widetilde{kt}$}
\psline[origin={-12,4}](0.8;-10)(2.8;-10)\uput{2.9}[-10]{0}(12,-4){\LARGE $\widetilde{ir}$}
\psline[origin={-12,4}](0.8;-30)(2.8;-30)
\psdots*[origin={-12,4},dotscale=0.3](1.8;-65)(1.8;-55)(1.8;-45)
\psline[origin={-12,4}](0.8;-80)(2.8;-80)
\psline[origin={-12,4}](0.8;-100)(2.8;-100)\uput{2.9}[-100]{0}(12,-4){\LARGE $\widetilde{m\!+\!2}$}

\pscircle[fillstyle=solid,fillcolor=white](12,-4){0.9}\rput(12,-4){\LARGE $m$}

\uput{3.7}[0]{0}(12,-4){\Huge $\otimes$}
\psline[origin={-12,4}](5.8;0)(6.8;0)
\psframe[origin={-12,4},fillstyle=solid,fillcolor=white]%
(4.8,-0.5)(5.8,0.5)
\rput(17.3,-4){$(kt)$}
\uput{7}[0]{0}(12,-4){\LARGE $t$}

\uput{3.7}[0]{0}(16,-4){\Huge $\otimes$}
\psline[origin={-16,4}](5.8;0)(6.8;0)
\psframe[origin={-16,4},fillstyle=solid,fillcolor=white]%
(4.8,-0.5)(5.8,0.5)
\rput(21.3,-4){$(\hat{i}\hat{r})$}
\uput{7.0}[0]{0}(16,-4){\LARGE $\hat{r}$}
}
\end{pspicture}
\end{center}
~\vskip 26mm
\caption{Graphical representation of the iterated collinear momentum
mapping and the implied factorization of the phase space.}
\label{fig:CktCir}
\end{figure}
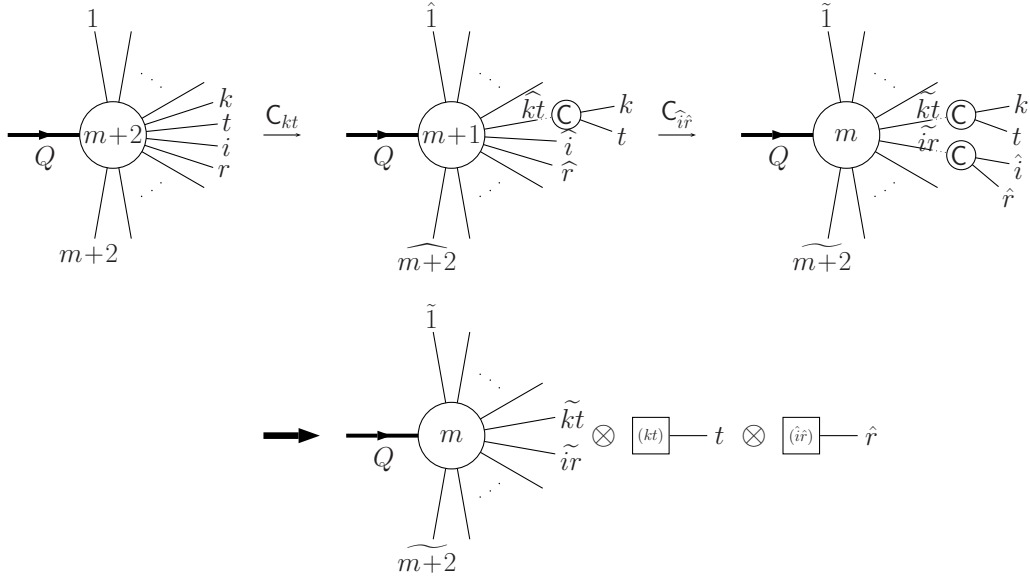

The $m$ momenta $\momt{(\ha{i}\ha{r},kt)}{} \equiv 
\{\ti{p}_1,\ldots,\ti{p}_{\ha{i}\ha{r}},\ldots,\ti{p}_{kt},\ldots,
\ti{p}_{m+2}\}$ in the matrix element on the right hand side of
\eqn{eq:CktCirkt} are again given by an iteration of the collinear
momentum mapping of \eqn{eq:PS_Cir},
\beq
\mom{} \cmap{kt} \momh{(kt)}{+1} \cmap{\ha{i}\ha{r}}
\momt{(\ha{i}\ha{r},kt)}{}\,.
\label{eq:PS_CktCirkt}
\eeq
The new momenta are defined as
\beq
\ti{p}_{\ha{i}\ha{r}}^{\mu} = \frac{1}{1-\alpha_{\ha{i}\ha{r}}}
(\ha{p}_{i}^{\mu} + \ha{p}_{r}^{\mu} - \alpha_{\ha{i}\ha{r}} Q^{\mu})\,,
\qquad
\ti{p}_n^{\mu} = \frac{1}{1-\alpha_{\ha{i}\ha{r}}} \ha{p}_n^{\mu}\,,
\qquad n\ne \ha{i},\ha{r}\,,
\label{eq:PS_CktCirkt_t}
\eeq
where the hatted momenta that also appear in the definition
$\tzz{\ha{i}}{\ha{r}}$, $\tzz{\ha{r}}{\ha{i}}$ and
$\kTt{\ha{i},\ha{r}}$ in \eqn{eq:CktCirkt} are defined in
\eqn{eq:PS_CktCktr_h}. Again $\alpha_{\ha{i}\ha{r}}$ is given by
\eqn{eq:alphair}. This above iterated momentum mapping leads to exact
phase space factorization in the following form:
\beq
\PS{m+2}{}(\mom{};Q) = \PS{m}{}(\momt{(\ha{i}\ha{r},kt)}{};Q)
[\rd p_{1;m}^{(\ha{i}\ha{r})}(\ha{p}_{r},\ti{p}_{\ha{i}\ha{r}};Q)]
[\rd p_{1;m+1}^{(kt)}(p_k,\ha{p}_{kt};Q)]\,,
\label{eq:PSfact_CktCirkt}
\eeq
where the factorized phase space measures
$[\rd p_{1;m+1}^{(kt)}(p_k,\ha{p}_{kt};Q)]$ and
$[\rd p_{1;m}^{(\ha{i}\ha{r})}(\ha{p}_{r},\ti{p}_{\ha{i}\ha{r}};Q)]$ are
given in \eqn{eq:dp_Cir}.  
The graphical representation of this iterated momentum mapping and the
corresponding factorization of the phase space is shown in \fig{fig:CktCir}.

\subsubsection{Iterated collinear--soft-collinear-type counterterms}

\subtitle{Counterterms}

The following three counterterms all use the same momentum mapping,
that of the collinear--soft-collinear term
$\cC{kt}{}\cSCS{kt;r}{(0,0)}(\mom{})$ and we present these together. We
define 
\beeq
\cC{kt}{}\cSCS{kt;r}{(0,0)}(\mom{}) \aand=
-(8\pi\as\mu^{2\eps})^2
\sum_{j}\sum_{l\ne j}\frac12 \calS_{\hat{j}\hat{l}}(\hat{r})
\label{eq:CktCSktr}
\\ \nn &&\times
\frac{1}{s_{kt}}
\bra{m}{(0)}{(\momt{(\ha{r},kt)}{})}
\bT_j \bT_l \hP_{f_k f_t}^{(0)}
(\tzz{k}{t},\tzz{t}{k},\kappatkt;\eps)
\ket{m}{(0)}{(\momt{(\ha{r},kt)}{})}\,,
\\
\cC{kt}{}\cC{ir;kt}{}\cSCS{kt;r}{(0,0)}(\mom{}) \aand=
(8\pi\as\mu^{2\eps})^2 
\frac{2}{s_{\ha{i}\ha{r}}}
\frac{\tzz{\ha{i}}{\ha{r}}}{\tzz{\ha{r}}{\ha{i}}}\bT_i^2
\label{eq:CktCirktCSktr}
\\  \nn &&\times
\frac{1}{s_{kt}}
\bra{m}{(0)}{(\momt{(\ha{r},kt)}{})}
\hP_{f_k f_t}^{(0)}(\tzz{k}{t},\tzz{t}{k},\kappatkt;\eps)
\ket{m}{(0)}{(\momt{(\ha{r},kt)}{})}\,,
\\
\cC{kt}{}\cC{ktr}{}\cSCS{kt;r}{(0,0)}(\mom{}) \aand=
(8\pi\as\mu^{2\eps})^2
\frac{2}{s_{\wha{kt}\ha{r}}}
\frac{\tzz{\wha{kt}}{\ha{r}}}{\tzz{\ha{r}}{\wha{kt}}}\bT_{kt}^2
\label{eq:CktCktrCSktr}
\\  \nn &&\times
\frac{1}{s_{kt}}
\bra{m}{(0)}{(\momt{(\ha{r},kt)}{})}
\hP_{f_k f_t}^{(0)}(\tzz{k}{t},\tzz{t}{k},\kappatkt;\eps)
\ket{m}{(0)}{(\momt{(\ha{r},kt)}{})}\,.
\eeeq
As usual \eqn{eq:zt2} defines the momentum fractions and $\kappatkt$ is
defined in \eqn{eq:kappatir}.

\subtitle{Momentum mapping and phase space factorization}

The set of $m$ momenta $\momt{(\ha{r},kt)}{} \equiv
\{\ti{p}_1,\ldots,\ti{p}_{kt},\ldots,\ti{p}_{m+2}\}$
($p_r$ is absent) that enter the matrix elements on the right hand
sides of \eqnss{eq:CktCSktr}{eq:CktCktrCSktr} are constructed by
applying the collinear momentum mapping of \eqn{eq:cmap} followed by
the soft mapping of \eqn{eq:smap} to the original set of $m+2$ momenta
$\mom{}$,
\beq
\mom{} \cmap{kt} \momh{(kt)}{+1} \smap{\ha{r}} \momt{(\ha{r},kt)}{}\,,
\label{eq:PS_CktCSktr}
\eeq
which we have already discussed under \eqn{eq:PS_CSirs} and shown in
\fig{fig:CSirs}.

\subsubsection{Iterated collinear--double-soft-type counterterms}

\subtitle{Counterterms}

The subtraction terms presented below use the same momentum mapping,
thus they are discussed together. We set
\beeq
\cC{kt}{}\cS{kt}{(0,0)}(\mom{}) \aand=
(8\pi\as\mu^{2\eps})^2 
\sum_j \sum_{l} \frac12 \calS_{\ha{j}\ha{l}}^{\mu\nu}(\wha{kt})
\nn \\ &&\times
\label{eq:CktSkt}
\frac{1}{s_{kt}}
\la\mu|\hP_{f_k f_t}^{(0)}(\tzz{k}{t},\tzz{t}{k},\kTt{k,t};\eps)|\nu\ra
\SME{m;(j,l)}{0}{\momt{(\wha{kt},kt)}{}}\,,
\\
\cC{k_gt_g}{}\cC{rk_gt_g}{}\cS{k_gt_g}{(0,0)}(\mom{}) \aand=
(8\pi\as\mu^{2\eps})^2 
\frac{2}{s_{kt}s_{\wha{kt}\ha{r}}}\bT_r^2\CA
\nn \\ &&\times
\Bigg[
\frac{2\tzz{\ha{r}}{\wha{kt}}}{\tzz{\wha{kt}}{\ha{r}}}
\Bigg(\frac{\tzz{k}{t}}{\tzz{t}{k}} + \frac{\tzz{t}{k}}{\tzz{k}{t}}\Bigg)
-(1-\eps)\tzz{k}{t}\tzz{t}{k}
\frac{s_{\ha{r}\kTt{k,t}}^2}{\kTt{k,t}^2 s_{\wha{kt}\ha{r}}}\Bigg]
\nn \\[2mm] &&\times
\SME{m}{0}{\momt{(\wha{kt},kt)}{}}\,,
\label{eq:CktCrktSktgg}
\\[2mm]
\cC{k_{\qb}t_q}{}\cC{rk_{\qb}t_q}{}\cS{k_{\qb}t_q}{(0,0)}(\mom{}) \aand=
(8\pi\as\mu^{2\eps})^2 
\frac{2}{s_{kt}s_{\wha{kt}\ha{r}}}\bT_r^2\TR
\nn \\ &&\times
\Bigg(
\frac{\tzz{\ha{r}}{\wha{kt}}}{\tzz{\wha{kt}}{\ha{r}}}
+\tzz{k}{t}\tzz{t}{k}
\frac{s_{\ha{r}\kTt{k,t}}^2}{\kTt{k,t}^2 s_{\wha{kt}\ha{r}}}\Bigg)
\SME{m}{0}{\momt{(\wha{kt},kt)}{}}\,.
\label{eq:CktCrktSktqq}
\eeeq
The momentum fractions and transverse momentum in
\eqnss{eq:CktSkt}{eq:CktCrktSktqq} are defined as usual by
\eqns{eq:zt2}{eq:kTtir}. In \eqn{eq:CktSkt} we use the notation
\beq
\calS_{\ha{j}\ha{l}}^{\mu\nu}(\wha{kt})
=4\frac{\hat{p}_j^{\mu}\hat{p}_l^{\nu}}{s_{\wha{kt}\ha{j}}s_{\wha{kt}\ha{l}}}\,.
\label{eq:Sjlrsmunu}
\eeq
The hatted momenta are given by \eqn{eq:PS_CktCktr_h}.

\subtitle{Momentum mapping and phase space factorization}

The $m$ momenta 
$\momt{(\wha{kt},kt)}{} \equiv \{\ti{p}_1,\ldots,\ti{p}_{m+2}\}$
(momenta $p_k$ and $p_t$ are absent) that appear in the matrix elements
on the right hand sides of \eqnss{eq:CktSkt}{eq:CktCrktSktqq} are
obtained by applying the soft momentum mapping of \eqn{eq:smap} to the
hatted set of momenta of \eqn{eq:PS_CktCktr_h},
\beq
\mom{} \cmap{kt} \momh{(kt)}{+1} \smap{\wha{kt}} \momt{(\wha{kt},kt)}{}\,.
\label{eq:PS_CktSkt}
\eeq
We have
\beq
\ti{p}_{n}^{\mu} = 
\Lambda^{\mu}_{\nu}[Q,(Q-p_{\wha{kt}})/\lambda_{\wha{kt}}] 
(\ha{p}_{n}^{\nu}/\lambda_{\wha{kt}})\,.
\label{eq:PS_CktSkt_t}
\eeq
The hatted momenta are also used to define the momentum fractions
$\tzz{\ha{j}}{\ha{l}}$ in \eqnss{eq:CktSkt}{eq:CktCrktSktqq} and
are given in \eqn{eq:PS_CktCktr_h}.

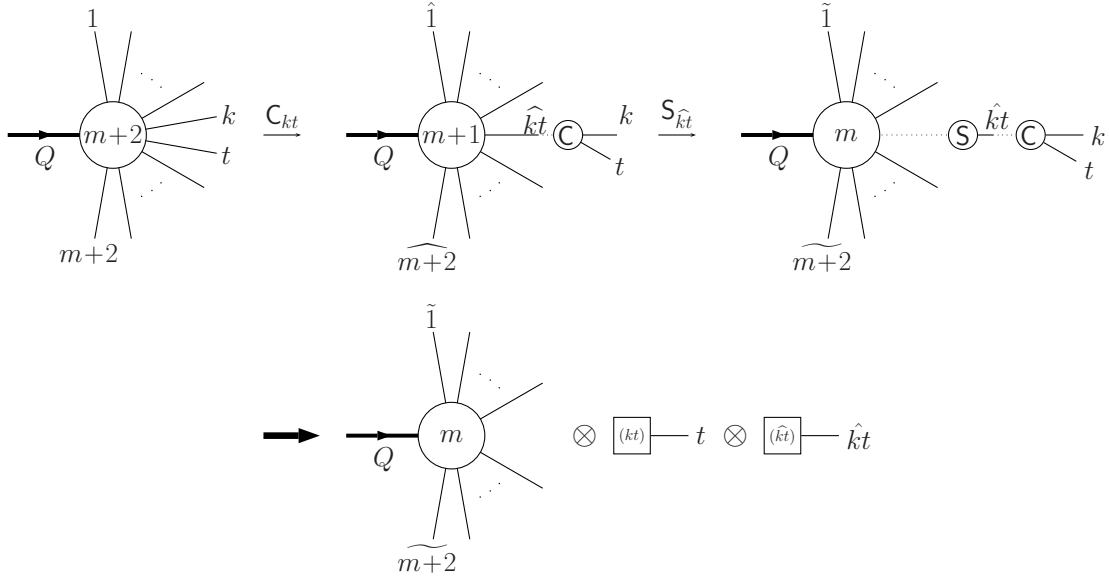
\begin{figure}
\begin{center}
\begin{pspicture}(0,0)(15,4)
\scalebox{0.5}{%

\psline[linewidth=3pt,arrowinset=0]{->}(0.2,4)(1.4,4)\uput{0.3}[d](1.2,4){\LARGE $Q$}
\psline[linewidth=3pt](1.2,4)(2.2,4)

\SpecialCoor
\psline[origin={-3,-4}](0.8;100)(2.8;100)\uput{2.9}[100]{0}(3,4){\LARGE $1$}
\psline[origin={-3,-4}](0.8;80)(2.8;80)
\psdots*[origin={-3,-4},dotscale=0.3](1.8;65)(1.8;55)(1.8;45)
\psline[origin={-3,-4}](0.8;30)(2.8;30)
\psline[origin={-3,-4}](0.8;10)(2.8;10)\uput{2.9}[10]{0}(3,4){\LARGE $k$}
\psline[origin={-3,-4}](0.8;-10)(2.8;-10)\uput{2.9}[-10]{0}(3,4){\LARGE $t$}
\psline[origin={-3,-4}](0.8;-30)(2.8;-30)
\psdots*[origin={-3,-4},dotscale=0.3](1.8;-65)(1.8;-55)(1.8;-45)
\psline[origin={-3,-4}](0.8;-80)(2.8;-80)
\psline[origin={-3,-4}](0.8;-100)(2.8;-100)\uput{2.9}[-100]{0}(3,4){\LARGE $m\!+\!2$}

\pscircle[fillstyle=solid,fillcolor=white](3,4){0.9}\rput(3,4){\LARGE $m\!+\!2$}

\psline{->}(7,4)(8,4)\uput[u](7.5,4){{\LARGE $\mathsf{C}_{kt}$}}

\psline[linewidth=3pt,arrowinset=0]{->}(9.2,4)(10.4,4)\uput{0.3}[d](10.2,4){\LARGE $Q$}
\psline[linewidth=3pt](10.2,4)(11.2,4)

\SpecialCoor
\psline[origin={-12,-4}](0.8;100)(2.8;100)\uput{2.9}[100]{0}(12,4){\LARGE $\hat{1}$}
\psline[origin={-12,-4}](0.8;80)(2.8;80)
\psdots*[origin={-12,-4},dotscale=0.3](1.8;65)(1.8;55)(1.8;45)
\psline[origin={-12,-4}](0.8;30)(2.8;30)
\psline[origin={-12,-4}](0.8;0)(2.3;0)\uput{1.9}[10]{0}(12,4){\LARGE $\widehat{kt}$}
\psline[origin={-12,-4}](0.8;-30)(2.8;-30)
\psdots*[origin={-12,-4},dotscale=0.3](1.8;-65)(1.8;-55)(1.8;-45)
\psline[origin={-12,-4}](0.8;-80)(2.8;-80)
\psline[origin={-12,-4}](0.8;-100)(2.8;-100)\uput{2.9}[-100]{0}(12,4){\LARGE $\widehat{m\!+\!2}$}

\pscircle[fillstyle=solid,fillcolor=white](12,4){0.9}\rput(12,4){\LARGE $m\!+\!1$}

\psline[origin={-12,-4},linestyle=dotted](2.3;0)(2.8;0)

\psline[origin={-12,-4}](3.4;0)(4.4;0)\uput{1.4}[0]{0}(15.05,4.54){\LARGE $k$}
\psline[origin={-15.1,-4}](0.29;-30)(1.3;-30)\uput{1.4}[-30]{0}(15.1,4){\LARGE $t$}

\pscircle[origin={-12,-4},fillstyle=solid,fillcolor=white](3.1;0){0.4}
\rput(15.1,4){\LARGE $\mathsf{C}\,$}

\psline{->}(17.5,4)(18.5,4)\uput[u](18,4){{\LARGE $\mathsf{S}_{\widehat{kt}}$}}

\psline[linewidth=3pt,arrowinset=0]{->}(19.7,4)(20.9,4)\uput{0.3}[d](20.7,4){\LARGE $Q$}
\psline[linewidth=3pt](20.7,4)(21.7,4)

\SpecialCoor
\psline[origin={-22.5,-4}](0.8;100)(2.8;100)\uput{2.9}[100]{0}(22.5,4){\LARGE $\tilde{1}$}
\psline[origin={-22.5,-4}](0.8;80)(2.8;80)
\psdots*[origin={-22.5,-4},dotscale=0.3](1.8;65)(1.8;55)(1.8;45)
\psline[origin={-22.5,-4}](0.8;30)(2.8;30)
\psline[origin={-22.5,-4}](0.8;-30)(2.8;-30)
\psdots*[origin={-22.5,-4},dotscale=0.3](1.8;-65)(1.8;-55)(1.8;-45)
\psline[origin={-22.5,-4}](0.8;-80)(2.8;-80)
\psline[origin={-22.5,-4}](0.8;-100)(2.8;-100)\uput{2.9}[-100]{0}(22.5,4){\LARGE $\widetilde{m\!+\!2}$}

\psline[origin={-22.5,-4},linestyle=dotted](0.8;0)(2.8;0)

\pscircle[fillstyle=solid,fillcolor=white](22.5,4){0.9}\rput(22.5,4){\LARGE $m$}

\psline[origin={-22.5,-4}](3.4;0)(3.9;0)\uput{0.6}[29]{0}(25.6,4){\LARGE $\hat{kt}$}
\psline[origin={-22.5,-4},linestyle=dotted](3.9;0)(4.4;0)

\pscircle[origin={-22.5,-4},fillstyle=solid,fillcolor=white](3.1;0){0.4}
\rput(25.6,4){\LARGE $\mathsf{S}$}

\psline[origin={-27.4,-4}](0.4;0)(1.4;0)\uput{1.6}[0]{0}(27.4,4){\LARGE $k$}
\psline[origin={-27.4,-4}](0.4;-30)(1.4;-30)\uput{1.6}[-30]{0}(27.4,4){\LARGE $t$}

\pscircle[origin={-27.4,-4},fillstyle=solid,fillcolor=white](0;0){0.4}
\rput(27.4,4){\LARGE $\mathsf{C}\,$}


\psline[linewidth=5pt,arrowinset=0]{->}(7,-4)(8.5,-4)

\psline[origin={10.5,8},linewidth=3pt,arrowinset=0]{->}(19.7,4)(20.9,4)\uput{0.3}[d](10.2,-4){\LARGE $Q$}
\psline[origin={10.5,8},linewidth=3pt](20.7,4)(21.7,4)

\SpecialCoor
\psline[origin={-12,4}](0.8;100)(2.8;100)\uput{2.9}[100]{0}(12,-4){\LARGE $\tilde{1}$}
\psline[origin={-12,4}](0.8;80)(2.8;80)
\psdots*[origin={-12,4},dotscale=0.3](1.8;65)(1.8;55)(1.8;45)
\psline[origin={-12,4}](0.8;30)(2.8;30)
\psline[origin={-12,4}](0.8;-30)(2.8;-30)
\psdots*[origin={-12,4},dotscale=0.3](1.8;-65)(1.8;-55)(1.8;-45)
\psline[origin={-12,4}](0.8;-80)(2.8;-80)
\psline[origin={-12,4}](0.8;-100)(2.8;-100)\uput{2.9}[-100]{0}(12,-4){\LARGE $\widetilde{m\!+\!2}$}

\pscircle[fillstyle=solid,fillcolor=white](12,-4){0.9}\rput(12,-4){\LARGE $m$}

\uput{3.7}[0]{0}(11.5,-4){\Huge $\otimes$}
\psline[origin={-11.5,4}](5.8;0)(6.8;0)
\psframe[origin={-11.5,4},fillstyle=solid,fillcolor=white]%
(4.8,-0.5)(5.8,0.5)
\rput(16.8,-4){$(kt)$}
\uput{7}[0]{0}(11.5,-4){\LARGE $t$}

\uput{3.7}[0]{0}(15.5,-4){\Huge $\otimes$}
\psline[origin={-15.5,4}](5.8;0)(6.8;0)
\psframe[origin={-15.5,4},fillstyle=solid,fillcolor=white]%
(4.8,-0.5)(5.8,0.5)
\rput(20.8,-4){$(\widehat{kt})$}
\uput{7}[0]{0}(15.5,-4){\LARGE $\hat{kt}$}
}
\end{pspicture}
\end{center}
~\vskip 26mm
\caption{Graphical representation of the collinear momentum mapping
followed by a soft one and the implied factorization of the phase space.}
\label{fig:CktSkt}
\end{figure}

The phase space factorization inherits the `product' structure of the
momentum mapping and we get
\beq
\PS{m+2}{}(\mom{};Q) = \PS{m}{}(\momt{(\wha{kt},kt)}{};Q)
[\rd p_{1;m}^{(\wha{kt})}(\ha{p}_{kt};Q)]
[\rd p_{1;m+1}^{(kt)}(p_k,\ha{p}_{kt};Q)]\,.
\label{eq:PSfact_CktSkt}
\eeq
The one-parton factorized phase space measures
$[\rd p_{1;m+1}^{(kt)}(p_k,\ha{p}_{kt};Q)]$ and
$[\rd p_{1;m}^{(\wha{kt})}(\ha{p}_{kt};Q)]$ are those in
\eqns{eq:dp_Cir}{eq:dp_Sr}.  
The graphical representation of this iterated momentum mapping and the
corresponding factorization of the phase space is shown in \fig{fig:CktSkt}.


\subsection{Iterated soft counterterms}
\label{ssec:StA2}

\subsubsection{Iterated soft--triple-collinear-type counterterms}
\label{sssec:StCirt}

\subtitle{Counterterms}

The first three subtraction terms on the right hand side of
\eqn{eq:StA2} are defined using the same momentum mapping, therefore
these terms are presented together. They read
\beeq
\cS{t}{}\cC{irt}{(0,0)}(\mom{}) \aand=
(8\pi\as\mu^{2\eps})^2
P^{\rm (S)}_{f_i f_r f_t}
(\tzz{i}{rt},\tzz{r}{it},\tzz{t}{ir},s_{ir},s_{it},s_{rt};\eps)
\nn \\&&\times
\frac{1}{s_{\ha{i}\ha{r}}}
\bra{m}{(0)}{(\momt{(\ha{i}\ha{r},t)}{})}
\hP_{f_i f_r}^{(0)}
(\tzz{\ha{i}}{\ha{r}},\tzz{\ha{r}}{\ha{i}},\kTt{\ha{i},\ha{r}};\eps)
\ket{m}{(0)}{(\momt{(\ha{i}\ha{r},t)}{})}\,,
\label{eq:StCirt}
\\
\cS{t}{}\cSCS{ir;t}{(0,0)}(\mom{}) \aand=
-(8\pi\as\mu^{2\eps})^2
\sum_{j}\sum_{l\ne j}\frac12 \calS_{jl}(t)
\nn\\&&\times
\frac{1}{s_{\ha{i}\ha{r}}}
\bra{m}{(0)}{(\momt{(\ha{i}\ha{r},t)}{})}
\bT_j \bT_l 
\hP_{f_i f_r}^{(0)}
(\tzz{\ha{i}}{\ha{r}},\tzz{\ha{r}}{\ha{i}},\kTt{\ha{i},\ha{r}};\eps)
\ket{m}{(0)}{(\momt{(\ha{i}\ha{r},t)}{})}\,,
\label{eq:StCSirt}
\eeeq
\beeq
\cS{t}{}\cC{irt}{}\cSCS{ir;t}{(0,0)}(\mom{}) \aand=
(8\pi\as\mu^{2\eps})^2
\frac{2}{s_{(ir)t}}\frac{1-\tzz{t}{ir}}{\tzz{t}{ir}} \,\bT^2_{ir}
\nn\\&&\times
\frac{1}{s_{\ha{i}\ha{r}}}
\bra{m}{(0)}{(\momt{(\ha{i}\ha{r},t)}{})}
\hP_{f_i f_r}^{(0)}
(\tzz{\ha{i}}{\ha{r}},\tzz{\ha{r}}{\ha{i}},\kTt{\ha{i},\ha{r}};\eps)
\ket{m}{(0)}{(\momt{(\ha{i}\ha{r},t)}{})}\,.
\label{eq:StCirtCSirt}
\eeeq
The momentum fractions $\tzz{j}{l}$ and $\tzz{j}{kl}$ are given
respectively by \eqns{eq:zt2}{eq:zt3} while the transverse momentum
$\kTt{\ha{i},\ha{r}}$ is defined in \eqn{eq:kTtir}.  The soft
functions $P^{\rm (S)}_{f_i f_r f_s}$ appearing in \eqn{eq:StCirt} were
introduced in \Ref{Somogyi:2005xz}. 
As discussed below \eqn{eq:CirjsCSirs}, whenever $j$ or $l$ equals
$(ir)$ the eikonal factor appearing in \eqn{eq:StCSirt} evaluates to
the expression given in \eqn{eq:Sirls}.

\subtitle{Momentum mapping and phase space factorization}

\begin{figure}
\begin{center}
\begin{pspicture}(0,0)(15,4)
\scalebox{0.5}{%

\psline[linewidth=3pt,arrowinset=0]{->}(0.2,4)(1.4,4)\uput{0.3}[d](1.2,4){\LARGE $Q$}
\psline[linewidth=3pt](1.2,4)(2.2,4)

\SpecialCoor
\psline[origin={-3,-4}](0.8;100)(2.8;100)\uput{2.9}[100]{0}(3,4){\LARGE $1$}
\psline[origin={-3,-4}](0.8;80)(2.8;80)
\psdots*[origin={-3,-4},dotscale=0.3](1.8;65)(1.8;55)(1.8;45)
\psline[origin={-3,-4}](0.8;30)(2.8;30)
\psline[origin={-3,-4}](0.8;15)(2.8;15)\uput{2.9}[15]{0}(3,4){\LARGE $i$}
\psline[origin={-3,-4}](0.8;0)(2.8;0)\uput{2.9}[0]{0}(3,4){\LARGE $r$}
\psline[origin={-3,-4}](0.8;-15)(2.8;-15)\uput{2.9}[-15]{0}(3,4){\LARGE $t$}
\psline[origin={-3,-4}](0.8;-30)(2.8;-30)
\psdots*[origin={-3,-4},dotscale=0.3](1.8;-65)(1.8;-55)(1.8;-45)
\psline[origin={-3,-4}](0.8;-80)(2.8;-80)
\psline[origin={-3,-4}](0.8;-100)(2.8;-100)\uput{2.9}[-100]{0}(3,4){\LARGE $m\!+\!2$}

\pscircle[fillstyle=solid,fillcolor=white](3,4){0.9}\rput(3,4){\LARGE $m\!+\!2$}

\psline{->}(7,4)(8,4)\uput[u](7.5,4){{\LARGE $\mathsf{S}_{t}$}}

\psline[linewidth=3pt,arrowinset=0]{->}(9.2,4)(10.4,4)\uput{0.3}[d](10.2,4){\LARGE $Q$}
\psline[linewidth=3pt](10.2,4)(11.2,4)

\SpecialCoor
\psline[origin={-12,-4}](0.8;100)(2.8;100)\uput{2.9}[100]{0}(12,4){\LARGE $\hat{1}$}
\psline[origin={-12,-4}](0.8;80)(2.8;80)
\psdots*[origin={-12,-4},dotscale=0.3](1.8;65)(1.8;55)(1.8;45)
\psline[origin={-12,-4}](0.8;30)(2.8;30)
\psline[origin={-12,-4}](0.8;15)(2.8;15)\uput{2.9}[15]{0}(12,4){\LARGE $\hat{i}$}
\psline[origin={-12,-4}](0.8;0)(2.8;0)\uput{2.9}[0]{0}(12,4){\LARGE $\hat{r}$}
\psline[origin={-12,-4}](0.8;-30)(2.8;-30)
\psdots*[origin={-12,-4},dotscale=0.3](1.8;-65)(1.8;-55)(1.8;-45)
\psline[origin={-12,-4}](0.8;-80)(2.8;-80)
\psline[origin={-12,-4}](0.8;-100)(2.8;-100)\uput{2.9}[-100]{0}(12,4){\LARGE $\widehat{m\!+\!2}$}

\pscircle[fillstyle=solid,fillcolor=white](12,4){0.9}\rput(12,4){\LARGE $m\!+\!1$}

\psline[origin={-12,-4},linestyle=dotted](0.8;-15)(2.8;-15)

\psline[origin={-12,-4}](3.4;-15)(4.4;-15)\uput{4.5}[-15]{0}(12,4){\LARGE $t$}

\pscircle[origin={-12,-4},fillstyle=solid,fillcolor=white](3.1;-15){0.4}
\rput(14.99,3.20){\LARGE $\mathsf{S}$}

\psline{->}(17.5,4)(18.5,4)\uput[u](18,4){{\LARGE $\mathsf{C}_{\hat{i}\hat{r}}$}}

\psline[linewidth=3pt,arrowinset=0]{->}(19.7,4)(20.9,4)\uput{0.3}[d](20.7,4){\LARGE $Q$}
\psline[linewidth=3pt](20.7,4)(21.7,4)

\SpecialCoor
\psline[origin={-22.5,-4}](0.8;100)(2.8;100)\uput{2.9}[100]{0}(22.5,4){\LARGE $\tilde{1}$}
\psline[origin={-22.5,-4}](0.8;80)(2.8;80)
\psdots*[origin={-22.5,-4},dotscale=0.3](1.8;65)(1.8;55)(1.8;45)
\psline[origin={-22.5,-4}](0.8;30)(2.8;30)
\psline[origin={-22.5,-4}](0.8;10)(2.3;10)\uput{1.9}[20]{0}(22.5,4){\LARGE $\widetilde{ir}$}
\psline[origin={-22.5,-4}](0.8;-30)(2.8;-30)
\psdots*[origin={-22.5,-4},dotscale=0.3](1.8;-65)(1.8;-55)(1.8;-45)
\psline[origin={-22.5,-4}](0.8;-80)(2.8;-80)
\psline[origin={-22.5,-4}](0.8;-100)(2.8;-100)\uput{2.9}[-100]{0}(22.5,4){\LARGE $\widetilde{m\!+\!2}$}

\psline[origin={-22.5,-4},linestyle=dotted](2.3;10)(2.8;10)
\psline[origin={-22.5,-4},linestyle=dotted](0.8;-10)(2.8;-10)

\pscircle[fillstyle=solid,fillcolor=white](22.5,4){0.9}\rput(22.5,4){\LARGE $m$}

\psline[origin={-22.5,-4}](3.4;10)(4.4;10)\uput{1.4}[10]{0}(25.55,4.54){\LARGE $\hat{i}$}
\psline[origin={-25.55,-4.54}](0.29;-20)(1.3;-20)\uput{1.4}[-20]{0}(25.55,4.54){\LARGE $\hat{r}$}

\pscircle[origin={-22.5,-4},fillstyle=solid,fillcolor=white](3.1;10){0.4}
\rput(25.55,4.54){\LARGE $\mathsf{C}\,$}

\psline[origin={-22.5,-4}](3.4;-10)(4.4;-10)\uput{1.4}[-10]{0}(25.55,3.46){\LARGE $t$}

\pscircle[origin={-22.5,-4},fillstyle=solid,fillcolor=white](3.1;-10){0.4}
\rput(25.55,3.46){\LARGE $\mathsf{S}$}


\psline[linewidth=5pt,arrowinset=0]{->}(7,-4)(8.5,-4)

\psline[origin={10.5,8},linewidth=3pt,arrowinset=0]{->}(19.7,4)(20.9,4)\uput{0.3}[d](10.2,-4){\LARGE $Q$}
\psline[origin={10.5,8},linewidth=3pt](20.7,4)(21.7,4)

\SpecialCoor
\psline[origin={-12,4}](0.8;100)(2.8;100)\uput{2.9}[100]{0}(12,-4){\LARGE $\tilde{1}$}
\psline[origin={-12,4}](0.8;80)(2.8;80)
\psdots*[origin={-12,4},dotscale=0.3](1.8;65)(1.8;55)(1.8;45)
\psline[origin={-12,4}](0.8;30)(2.8;30)
\psline[origin={-12,4}](0.8;0)(2.8;0)\uput{2.9}[0]{0}(12,-4){\LARGE $\widetilde{ir}$}
\psline[origin={-12,4}](0.8;-30)(2.8;-30)
\psdots*[origin={-12,4},dotscale=0.3](1.8;-65)(1.8;-55)(1.8;-45)
\psline[origin={-12,4}](0.8;-80)(2.8;-80)
\psline[origin={-12,4}](0.8;-100)(2.8;-100)\uput{2.9}[-100]{0}(12,-4){\LARGE $\widetilde{m\!+\!2}$}

\pscircle[fillstyle=solid,fillcolor=white](12,-4){0.9}\rput(12,-4){\LARGE $m$}

\uput{3.7}[0]{0}(12,-4){\Huge $\otimes$}
\psline[origin={-12,4}](5.8;0)(6.8;0)
\psframe[origin={-12,4},fillstyle=solid,fillcolor=white]%
(4.8,-0.5)(5.8,0.5)
\rput(17.3,-4){$(t)$}
\uput{7}[0]{0}(12,-4){\LARGE $t$}

\uput{3.7}[0]{0}(16,-4){\Huge $\otimes$}
\psline[origin={-16,4}](5.8;0)(6.8;0)
\psframe[origin={-16,4},fillstyle=solid,fillcolor=white]%
(4.8,-0.5)(5.8,0.5)
\rput(21.3,-4){$(\hat{i}\hat{r})$}
\uput{7}[0]{0}(16,-4){\LARGE $\hat{r}$}
}
\end{pspicture}
\end{center}
~\vskip 26mm
\caption{Graphical representation of the soft momentum mapping followed
by a collinear one and the implied factorization of the phase space.}
\label{fig:StCir}
\end{figure}
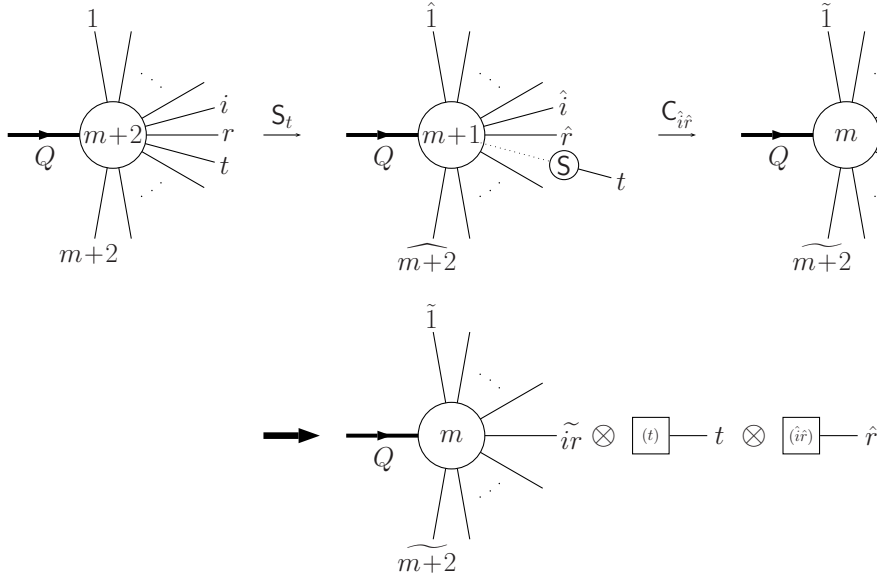

The $m$ momenta $\momt{(\ha{i}\ha{r},t)}{} \equiv
\{\ti{p}_1,\ldots,\ti{p}_{ir},\ldots,\ti{p}_{m+2}\}$ ($p_t$ is absent)
entering the matrix elements on the right hand sides of
\eqnss{eq:StCirt}{eq:StCirtCSirt} are defined by the successive
soft and collinear mappings,
\beq
\mom{} \smap{t} \momh{(t)}{+1} \cmap{\ha{i}\ha{r}}
\momt{(\ha{i}\ha{r},t)}{}\,,
\label{eq:PS_StCirt}
\eeq
where
\beq
\ti{p}_{ir}^{\mu} = \frac{1}{1-\alpha_{\ha{i}\ha{r}}}
(\ha{p}_i^{\mu} + \ha{p}_r^{\mu} - \alpha_{\ha{i}\ha{r}} Q^{\mu})\,,
\qquad
\ti{p}_n^{\mu} = \frac{1}{1-\alpha_{\ha{i}\ha{r}}} \ha{p}_n^{\mu}\,,
\qquad n\ne i,r
\label{eq:PS_StCirt_t}
\eeq
and the hatted momenta, that also appear in the definitions of
$\tzz{\ha{i}}{\ha{r}}$, $\tzz{\ha{i}}{\ha{r}}$ and
$\kTt{\ha{i},\ha{r}}$ in \eqnss{eq:StCirt}{eq:StCirtCSirt}, are
\beq
\ha{p}_n^{\mu} =
\Lambda^{\mu}_{\nu}[Q,(Q-p_t)/\lambda_t] (p_n^{\nu}/\lambda_t)\,,
\qquad n\ne t\,.
\label{eq:PS_StCirt_h}
\eeq
The expressions for $\alpha_{\ha{i}\ha{r}}$ in \eqn{eq:PS_StCirt_t} and
$\lambda_t$ in \eqn{eq:PS_StCirt_h} are given in
\eqns{eq:alphair}{eq:lambdar} respectively.  This momentum mapping
leads to an exact phase space factorization of the following form
\beq
\PS{m+2}{}(\mom{};Q) = \PS{m}{}(\momt{(\ha{i}\ha{r},t)}{};Q)
[\rd p_{1;m}^{(\ha{i}\ha{r})}(\ha{p}_{r},\ti{p}_{ir};Q)]
[\rd p_{1;m+1}^{(t)}(p_t;Q)]\,.
\label{eq:PSfact_StCirt}
\eeq
The collinear and soft one-parton factorized phase space measures 
$[\rd p_{1;m}^{(\ha{i}\ha{r})}(\ha{p}_{r},\ti{p}_{ir};Q)]$ and
$[\rd p_{1;m+1}^{(t)}(p_t;Q)]$ are given respectively in
\eqns{eq:dp_Cir}{eq:dp_Sr}.
The graphical representation of this iterated momentum mapping and the
corresponding factorization of the phase space is shown in \fig{fig:StCir}.

\subsubsection{Iterated soft--double-soft-type counterterms}
\label{sssec:StSrt}

\subtitle{Counterterms}

The remaining four iterated soft counterterms in \eqn{eq:StA2} are all
defined using the same momentum mapping so we discuss them together. We
have 
\beeq
\cS{t}{}\cC{irt}{}\cS{rt}{(0,0)}(\mom{}) \aand=
(8\pi\as\mu^{2\eps})^2
\Bigg[
  \CA \frac{2}{s_{\ha{i}\ha{r}}}
\frac{\tzz{\ha{i}}{\ha{r}}}{\tzz{\ha{r}}{\ha{i}}}
\Bigg(
  \frac{s_{ir}}{s_{it}s_{rt}}
+ \frac{1}{s_{rt}}\frac{\tzz{r}{it}}{\tzz{t}{ir}}
-\frac{1}{s_{it}}\frac{\tzz{i}{rt}}{\tzz{t}{ir}}
\Bigg)
\nn\\ &&\qquad\qquad\qquad
+ \bT_i^2\frac{2}{s_{it}}\frac{\tzz{i}{t}}{\tzz{t}{i}}
  \frac{2}{s_{\ha{i}\ha{r}}}\frac{\tzz{\ha{i}}{\ha{r}}}{\tzz{\ha{r}}{\ha{i}}}
\Bigg]
\bT_i^2\SME{m}{0}{\momt{(\ha{r},t)}{}}\,,
\label{eq:StCirtSrt}
\\[2mm]
\cS{t}{}\cSCS{ir;t}{}\cS{rt}{(0,0)}(\mom{}) \aand=
-(8\pi\as\mu^{2\eps})^2 
\frac{2}{s_{\hat{i}\hat{r}}}
\frac{\tzz{\hat{i}}{\hat{r}}}{\tzz{\hat{r}}{\hat{i}}}\bT_{i}^2
\sum_{j}\sum_{l\ne j}
\frac12 \calS_{jl}(t)\SME{m;(j,l)}{0}{\momt{(\ha{r},t)}{}}\,,\quad~
\label{eq:StCSirtSrt}
\\[2mm]
\cS{t}{}\cC{irt}{}\cSCS{ir;t}{}\cS{rt}{(0,0)}(\mom{}) \aand=
(8\pi\as\mu^{2\eps})^2 
\,\bT_i^2
\,\frac{2}{s_{\ha{i}\ha{r}}}\frac{\tzz{\ha{i}}{\ha{r}}}{\tzz{\ha{r}}{\ha{i}}}
\frac{2}{s_{(ir)t}}\frac{1-\tzz{t}{ir}}{\tzz{t}{ir}}
\bT_i^2\,\SME{m}{0}{\momt{(\ha{r},t)}{}}\,,
\label{eq:StCirtCSirtSrt}
\\
\cS{t}{}\cS{rt}{(0,0)}(\mom{}) \aand= (8\pi\as\mu^{2\eps})^2 \label{eq:StSrt}
\Bigg[\frac18 \sum_{i,j,k,l}\calS_{\hat{i}\hat{k}}(\hat{r})\calS_{jl}(t)
\SME{m;(i,k)(j,l)}{0}{\momt{(\ha{r},t)}{}}
\\ &&
- \frac14 \CA \sum_{i,k}\calS_{\hat{i}\hat{k}}(\hat{r})
  \Big(\calS_{ir}(t) + \calS_{kr}(t) - \calS_{ik}(t)\Big)
  \SME{m;(i,k)}{0}{\momt{(\ha{r},t)}{}}
\Bigg]\,.
\nn
\eeeq
No new notation (besides the set of momenta $\momt{(\ha{r},t)}{}$ to
be defined below) has been introduced in \eqnss{eq:StCirtSrt}{eq:StSrt},
thus we limit the discussion of these terms to two comments. Firstly,
note that in the equations above both $r$ and $t$ are gluons. Secondly,
we emphasize that the discussion below \eqn{eq:CirjsCSirs} also applies
to \eqn{eq:StCSirtSrt}.

\subtitle{Momentum mapping and phase space factorization}

\begin{figure}
\begin{center}
\begin{pspicture}(0,0)(15,4)
\scalebox{0.5}{%

\psline[linewidth=3pt,arrowinset=0]{->}(0.2,4)(1.4,4)\uput{0.3}[d](1.2,4){\LARGE $Q$}
\psline[linewidth=3pt](1.2,4)(2.2,4)

\SpecialCoor
\psline[origin={-3,-4}](0.8;100)(2.8;100)\uput{2.9}[100]{0}(3,4){\LARGE $1$}
\psline[origin={-3,-4}](0.8;80)(2.8;80)
\psdots*[origin={-3,-4},dotscale=0.3](1.8;65)(1.8;55)(1.8;45)
\psline[origin={-3,-4}](0.8;30)(2.8;30)
\psline[origin={-3,-4}](0.8;10)(2.8;10)\uput{2.9}[10]{0}(3,4){\LARGE $r$}
\psline[origin={-3,-4}](0.8;-10)(2.8;-10)\uput{2.9}[-10]{0}(3,4){\LARGE $t$}
\psline[origin={-3,-4}](0.8;-30)(2.8;-30)
\psdots*[origin={-3,-4},dotscale=0.3](1.8;-65)(1.8;-55)(1.8;-45)
\psline[origin={-3,-4}](0.8;-80)(2.8;-80)
\psline[origin={-3,-4}](0.8;-100)(2.8;-100)\uput{2.9}[-100]{0}(3,4){\LARGE $m\!+\!2$}

\pscircle[fillstyle=solid,fillcolor=white](3,4){0.9}\rput(3,4){\LARGE $m\!+\!2$}

\psline{->}(7,4)(8,4)\uput[u](7.5,4){{\LARGE $\mathsf{S}_{t}$}}

\psline[linewidth=3pt,arrowinset=0]{->}(9.2,4)(10.4,4)\uput{0.3}[d](10.2,4){\LARGE $Q$}
\psline[linewidth=3pt](10.2,4)(11.2,4)

\SpecialCoor
\psline[origin={-12,-4}](0.8;100)(2.8;100)\uput{2.9}[100]{0}(12,4){\LARGE $\hat{1}$}
\psline[origin={-12,-4}](0.8;80)(2.8;80)
\psdots*[origin={-12,-4},dotscale=0.3](1.8;65)(1.8;55)(1.8;45)
\psline[origin={-12,-4}](0.8;30)(2.8;30)
\psline[origin={-12,-4}](0.8;10)(2.8;10)\uput{2.9}[10]{0}(12,4){\LARGE $\hat{r}$}
\psline[origin={-12,-4}](0.8;-30)(2.8;-30)
\psdots*[origin={-12,-4},dotscale=0.3](1.8;-65)(1.8;-55)(1.8;-45)
\psline[origin={-12,-4}](0.8;-80)(2.8;-80)
\psline[origin={-12,-4}](0.8;-100)(2.8;-100)\uput{2.9}[-100]{0}(12,4){\LARGE $\widehat{m\!+\!2}$}

\pscircle[fillstyle=solid,fillcolor=white](12,4){0.9}\rput(12,4){\LARGE $m\!+\!1$}

\psline[origin={-12,-4},linestyle=dotted](0.8;-10)(2.8;-10)

\psline[origin={-12,-4}](3.4;-10)(4.4;-10)\uput{4.5}[-10]{0}(12,4){\LARGE $t$}

\pscircle[origin={-12,-4},fillstyle=solid,fillcolor=white](3.1;-10){0.4}
\rput(15.05,3.46){\LARGE $\mathsf{S}$}

\psline{->}(17.5,4)(18.5,4)\uput[u](18,4){{\LARGE $\mathsf{S}_{\hat{r}}$}}

\psline[linewidth=3pt,arrowinset=0]{->}(19.7,4)(20.9,4)\uput{0.3}[d](20.7,4){\LARGE $Q$}
\psline[linewidth=3pt](20.7,4)(21.7,4)

\SpecialCoor
\psline[origin={-22.5,-4}](0.8;100)(2.8;100)\uput{2.9}[100]{0}(22.5,4){\LARGE $\tilde{1}$}
\psline[origin={-22.5,-4}](0.8;80)(2.8;80)
\psdots*[origin={-22.5,-4},dotscale=0.3](1.8;65)(1.8;55)(1.8;45)
\psline[origin={-22.5,-4}](0.8;30)(2.8;30)
\psline[origin={-22.5,-4}](0.8;-30)(2.8;-30)
\psdots*[origin={-22.5,-4},dotscale=0.3](1.8;-65)(1.8;-55)(1.8;-45)
\psline[origin={-22.5,-4}](0.8;-80)(2.8;-80)
\psline[origin={-22.5,-4}](0.8;-100)(2.8;-100)\uput{2.9}[-100]{0}(22.5,4){\LARGE $\widetilde{m\!+\!2}$}

\psline[origin={-22.5,-4},linestyle=dotted](0.8;10)(2.8;10)
\psline[origin={-22.5,-4},linestyle=dotted](0.8;-10)(2.8;-10)

\pscircle[fillstyle=solid,fillcolor=white](22.5,4){0.9}\rput(22.5,4){\LARGE $m$}

\psline[origin={-22.5,-4}](3.4;10)(4.4;10)\uput{1.4}[10]{0}(25.55,4.54){\LARGE $\hat{r}$}

\pscircle[origin={-22.5,-4},fillstyle=solid,fillcolor=white](3.1;10){0.4}
\rput(25.55,4.54){\LARGE $\mathsf{S}$}

\psline[origin={-22.5,-4}](3.4;-10)(4.4;-10)\uput{1.4}[-10]{0}(25.55,3.46){\LARGE $t$}

\pscircle[origin={-22.5,-4},fillstyle=solid,fillcolor=white](3.1;-10){0.4}
\rput(25.55,3.46){\LARGE $\mathsf{S}$}


\psline[linewidth=5pt,arrowinset=0]{->}(7,-4)(8.5,-4)

\psline[origin={10.5,8},linewidth=3pt,arrowinset=0]{->}(19.7,4)(20.9,4)\uput{0.3}[d](10.2,-4){\LARGE $Q$}
\psline[origin={10.5,8},linewidth=3pt](20.7,4)(21.7,4)

\SpecialCoor
\psline[origin={-12,4}](0.8;100)(2.8;100)\uput{2.9}[100]{0}(12,-4){\LARGE $\tilde{1}$}
\psline[origin={-12,4}](0.8;80)(2.8;80)
\psdots*[origin={-12,4},dotscale=0.3](1.8;65)(1.8;55)(1.8;45)
\psline[origin={-12,4}](0.8;30)(2.8;30)
\psline[origin={-12,4}](0.8;-30)(2.8;-30)
\psdots*[origin={-12,4},dotscale=0.3](1.8;-65)(1.8;-55)(1.8;-45)
\psline[origin={-12,4}](0.8;-80)(2.8;-80)
\psline[origin={-12,4}](0.8;-100)(2.8;-100)\uput{2.9}[-100]{0}(12,-4){\LARGE $\widetilde{m\!+\!2}$}

\pscircle[fillstyle=solid,fillcolor=white](12,-4){0.9}\rput(12,-4){\LARGE $m$}

\uput{3.7}[0]{0}(11.5,-4){\Huge $\otimes$}
\psline[origin={-11.5,4}](5.8;0)(6.8;0)
\psframe[origin={-11.5,4},fillstyle=solid,fillcolor=white]%
(4.8,-0.5)(5.8,0.5)
\rput(16.8,-4){$(t)$}
\uput{7}[0]{0}(11.5,-4){\LARGE $t$}

\uput{3.7}[0]{0}(15.5,-4){\Huge $\otimes$}
\psline[origin={-15.5,4}](5.8;0)(6.8;0)
\psframe[origin={-15.5,4},fillstyle=solid,fillcolor=white]%
(4.8,-0.5)(5.8,0.5)
\rput(20.8,-4){$(\hat{r})$}
\uput{7.0}[0]{0}(15.5,-4){\LARGE $\hat{r}$}
}
\end{pspicture}
\end{center}
~\vskip 26mm
\caption{Graphical representation of the iterated soft momentum
mapping and the implied factorization of the phase space.}
\label{fig:StSr}
\end{figure}
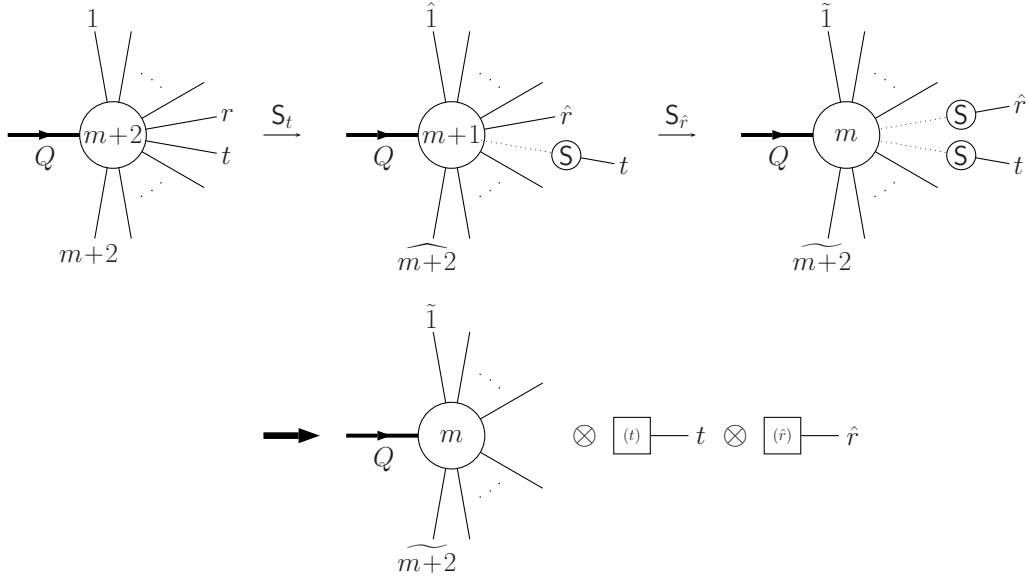

The $m$ momenta $\momt{(\ha{r},t)}{} \equiv
\{\ti{p}_1,\ldots,\ti{p}_{m+2}\}$ (both $r$ and $t$ are omitted) that
appear in the matrix elements on the right hand sides of
\eqnss{eq:StCirtSrt}{eq:StSrt} are obtained by two successive soft-type
mappings (defined in \eqn{eq:PS_Sr}),
\beq
\mom{} \smap{t} \momh{(t)}{+1} \smap{\ha{r}} \momt{(\ha{r},t)}{}\,,
\label{eq:PS_StSrt}
\eeq
where
\beq
\ti{p}_n^{\mu} = \Lambda^{\mu}_{\nu}[Q,(Q-\ha{p}_{r})/\lambda_{\ha{r}}] 
(\ha{p}_n^{\nu}/\lambda_{\ha{r}})\,,\qquad n\ne r\,.
\label{eq:PS_StSrt_t}
\eeq
The hatted momenta are given in \eqn{eq:PS_StCirt_h}.  Naturally, the
iterative structure of the momentum mapping is inherited by the phase
space factorization
\beq
\PS{m+2}{}(\mom{};Q)= \PS{m}{}(\momt{(\ha{r},t)}{};Q)
[\rd p_{1;m}^{(\ha{r})}(\ha{p}_{r};Q)]
[\rd p_{1;m+1}^{(t)}(p_t;Q)]\,,
\label{eq:PSfact_StSrt}
\eeq
where the soft one-parton factorized phase space measures
$[\rd p_{1;m}^{(\ha{r})}(\ha{p}_{r};Q)]$ and
$[\rd p_{1;m+1}^{(t)}\!(p_t;Q)]$ are both given by \eqn{eq:dp_Sr}.
The graphical representation of this iterated momentum mapping and the
corresponding factorization of the phase space is shown in \fig{fig:StSr}.


\subsection{Iterated soft-collinear counterterms}
\label{ssec:StCktA2}

\subsubsection{Iterated soft-collinear--triple-collinear-type counterterms}

\subtitle{Counterterms}

The first two terms on the right hand side of \eqn{eq:CktStA2} turn out
to be defined using the same momentum mapping after a trivial
reindexing. Let us define
\beeq
\cC{it}{}\cS{t}{}\cC{irt}{(0,0)} \aand=
(8\pi\as\mu^{2\eps})^2
\frac{2}{s_{it}}
\frac{\tzz{i}{t}}{\tzz{t}{i}}\,\bT_{it}^2
\nn\\ &&\times
\frac{1}{s_{\hat{i}\hat{r}}}
\bra{m}{(0)}{(\momt{(\ha{i}\ha{r},t)}{})}
\hP_{f_i f_r}^{(0)}
(\tzz{\ha{i}}{\ha{r}},\tzz{\ha{r}}{\ha{i}},\kTt{\ha{i},\hat{r}};\eps)
\ket{m}{(0)}{(\momt{(\ha{i}\ha{r},t)}{})}\,,
\label{eq:CitStCirt}
\\
\cC{kt}{}\cS{t}{}\cSCS{ir;t}{(0,0)}(\mom{}) \aand= (8\pi\as\mu^{2\eps})^2
\frac{2}{s_{kt}}
\frac{\tzz{k}{t}}{\tzz{t}{k}}\bT_i^2
\nn\\ &&\times
\frac{1}{s_{\hat{i}\hat{r}}}
\bra{m}{(0)}{(\momt{(\ha{i}\ha{r},t)}{})}
\hP_{f_i f_r}^{(0)}
(\tzz{\hat{i}}{\hat{r}},\tzz{\hat{r}}{\hat{i}},
\kTt{\hat{i},\hat{r}};\eps)
\ket{m}{(0)}{(\momt{(\ha{i}\ha{r},t)}{})}\,.
\label{eq:CktStCSirt}
\eeeq
Notice how a different indexing of
$\cC{it}{}\cS{t}{}\cC{irt}{(0,0)}(\mom{})$ is given above as compared
to that appearing in \eqn{eq:CktStA2}. As already stated, this is
convenient, because with this indexing exactly the same set of tilded
momenta, $\momt{(\ha{i}\ha{r},t)}{}$, appear in the matrix elements on
the right hand sides of both \eqns{eq:CitStCirt}{eq:CktStCSirt}. The
momentum fractions and transverse momenta are defined as usual via
\eqns{eq:zt2}{eq:kTtir}.

\subtitle{Momentum mapping and phase space factorization}

Exactly the same momentum mapping is used as for the iterated
soft--triple-collinear-type counterterms, \sect{sssec:StCirt}.

\subsubsection{Iterated soft-collinear--double-soft-type counterterms}

\subtitle{Counterterms}

The remaining five terms on the right hand side of \eqn{eq:CktStA2} are
again defined using the same momentum mapping after reindexing some
terms. We have
\beeq
\cC{kt}{}\cS{t}{}\cSCS{ir;t}{}\cS{rt}{(0,0)}(\mom{}) \aand=
(8\pi\as\mu^{2\eps})^2
\frac{2}{s_{\hat{i}\hat{r}}}
\frac{\tzz{\hat{i}}{\hat{r}}}{\tzz{\hat{r}}{\hat{i}}}\,\bT_{ir}^2\,
\frac{2}{s_{kt}}\frac{\tzz{k}{t}}{\tzz{t}{k}}\,\bT_k^2\,
\SME{m}{0}{\momt{(\hat{r},t)}{}}\,,
\label{eq:CktStCSirtSrt}
\\[2mm]
\cC{kt}{}\cS{t}{}\cC{krt}{}\cS{rt}{(0,0)}(\mom{}) \aand=
(8\pi\as\mu^{2\eps})^2
\frac{2}{s_{\hat{k}\hat{r}}}
\frac{\tzz{\hat{k}}{\hat{r}}}{\tzz{\hat{r}}{\hat{k}}}\,\bT_{kr}^2\,
\frac{2}{s_{kt}}\frac{\tzz{k}{t}}{\tzz{t}{k}}\,\bT_k^2\,
\SME{m}{0}{\momt{(\hat{r},t)}{}}\,,
\label{eq:CktStCkrtSrt}
\\[2mm]
\cC{rt}{}\cS{t}{}\cC{krt}{}\cS{rt}{(0,0)}(\mom{}) \aand=
(8\pi\as\mu^{2\eps})^2
\frac{2}{s_{\hat{k}\hat{r}}}
\frac{\tzz{\hat{k}}{\hat{r}}}{\tzz{\hat{r}}{\hat{k}}}\,\bT_{kr}^2\,
\frac{2}{s_{rt}}\frac{\tzz{r}{t}}{\tzz{t}{r}}\,\CA\,
\SME{m}{0}{\momt{(\hat{r},t)}{}}\,,
\label{eq:CrtStCkrtSrt}
\\[2mm]
\cC{kt}{}\cS{t}{}\cS{rt}{(0,0)}(\mom{}) \aand=
-(8\pi\as\mu^{2\eps})^2
\sum_{j}\sum_{l\ne j}\frac12 \calS_{\hat{j}\hat{l}}(\hat{r})
\frac{2}{s_{kt}}\frac{\tzz{k}{t}}{\tzz{t}{k}}\bT_{k}^2
\SME{m;(j,l)}{0}{\momt{(\hat{r},t)}{}}\,,
\nn\\
\label{eq:CktStSrt}
\eeeq
\beeq
\cC{rt}{}\cS{t}{}\cS{rt}{(0,0)}(\mom{}) \aand=
-(8\pi\as\mu^{2\eps})^2
\sum_{j}\sum_{l\ne j}\frac12 \calS_{\hat{j}\hat{l}}(\hat{r})
\frac{2}{s_{rt}}\frac{\tzz{r}{t}}{\tzz{t}{r}}\CA
\SME{m;(j,l)}{0}{\momt{(\hat{r},t)}{}}\,.
\nn\\
\label{eq:CrtStSrt}
\eeeq
Here $\cC{rt}{}\cS{t}{}\cC{krt}{}\cS{rt}{(0,0)}(\mom{})$ and 
$\cC{rt}{}\cS{t}{}\cS{rt}{(0,0)}(\mom{})$ are presented with a
different indexing then in \eqn{eq:CktStA2} so that all matrix elements
on the right hand sides of \eqnss{eq:CktStCSirtSrt}{eq:CrtStSrt} appear
with the same set of tilded momenta, $\momt{(\hat{r},t)}{}$. The momentum
fractions have been defined in \eqn{eq:zt2}, the eikonal factor in
\eqn{eq:Sikr}.  

\subtitle{Momentum mapping and phase space factorization}

The momentum mapping used is identical with the mapping defined for the
iterated soft--double-soft-type conterterms. This mapping and the
corresponding phase space factorization is presented in \sect{sssec:StSrt}.

\section{Cancellation of kinematical singularities}
\label{sec:cancellation}

In \sect{sec:RR_A1}, we have shown that the subtraction terms collected
in \eqn{eq:A1} correctly regularize the kinematical singularities in
the squared matrix element in the singly-unresolved regions of the phase
space. Similarly, the subtraction terms collected in \eqn{eq:RR_A2}
correctly regularize the kinematical singularities in the
doubly-unresolved regions of the phase space. The purpose of the iterated
counterterms is two-fold. These should cancel the
kinematical singularities of the singly-unresolved counterterms in the
doubly-unresolved regions of the phase space and conversely, they should
cancel the kinematical singularities of the doubly-unresolved
counterterms in the singly-unresolved regions of the phase space. The
structure of \eqn{eq:RR_A12} follows that of the candidate subtraction
term $\bA{12}\M{m+2}{(0)}$ found in \Ref{Somogyi:2005xz}, where it was
shown that at the level of the factorization formulae the combination
$(\bA{2}+\bA{1}-\bA{12})\M{m+2}{(0)}$ indeed regularizes the squared
matrix element in all relevant unresolved regions of the phase space.
We show the structure of the cancellations graphically in
\fig{fig:RR-A2-A1+A12}.
\definecolor{mygrey}{rgb}{0.64,0.66,0.62}
\definecolor{mygreyd}{rgb}{0.40,0.40,0.40}
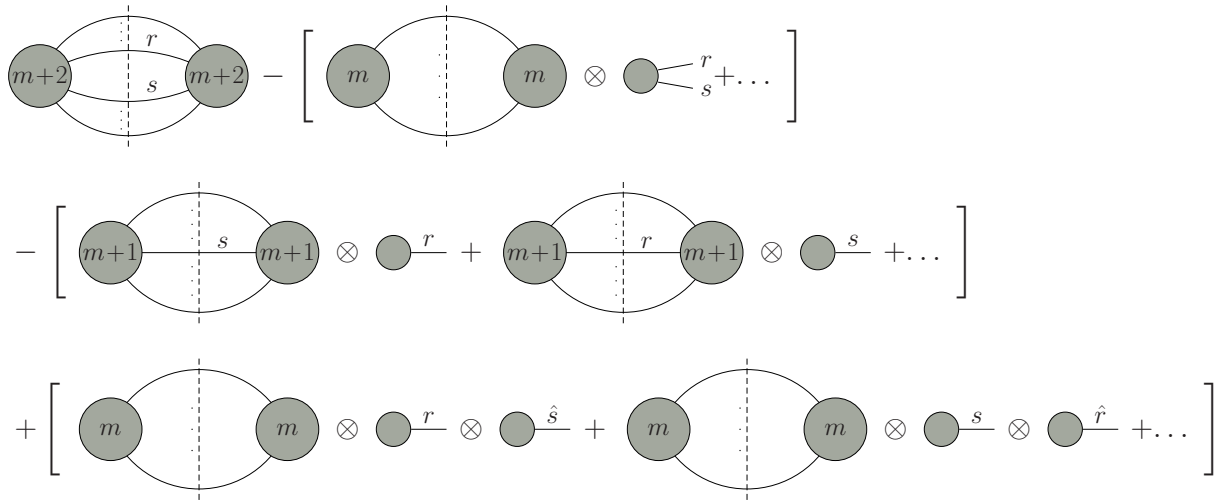
\begin{figure}
\begin{center}
\begin{pspicture}(0,0)(16,4)
\scalebox{0.47}{%


\SpecialCoor
\psarc(3.5,0){4.72}{(2.5,4)}{(-2.5,4)}
\psarc(3.5,3){2.69}{(2.5,1)}{(-2.5,1)}
\psarc(3.5,8){4.72}{(-2.5,-4)}{(2.5,-4)}
\psarc(3.5,5){2.69}{(-2.5,-1)}{(2.5,-1)}
\psline[linestyle=dashed](3.5,2)(3.5,6)
\psdots*[dotscale=0.3](3.3,5)(3.3,5.25)(3.3,5.5)
\psdots*[dotscale=0.3](3.3,3)(3.3,2.75)(3.3,2.5)
\pscircle[fillstyle=solid,fillcolor=mygrey](1,4){0.9}\rput(1,4){\LARGE $m\!+\!2$}
\pscircle[fillstyle=solid,fillcolor=mygrey](6,4){0.9}\rput(6,4){\LARGE $m\!+\!2$}
\rput[l](4,5){\LARGE $r$}
\rput[l](4,3.7){\LARGE $s$}


\rput(1,-1){\Huge $- \,\,\Bigg[$}

\psarc[origin={-2,5}](3.5,3){2.69}{(2.5,1)}{(-2.5,1)}
\psarc[origin={-2,5}](3.5,5){2.69}{(-2.5,-1)}{(2.5,-1)}
\psline[origin={-2,5}](1,4)(6,4)
\psline[origin={-2,5},linestyle=dashed](3.5,2)(3.5,6)
\psdots*[origin={-2,5},dotscale=0.3](3.3,5.2)(3.3,4.8)(3.3,4.4)
\psdots*[origin={-2,5},dotscale=0.3](3.3,3.6)(3.3,3.2)(3.3,2.8)
\rput[l](6,-0.7){\LARGE $s$}
\pscircle[origin={-2,5},fillstyle=solid,fillcolor=mygrey](1,4){0.9}\rput(3,-1){\LARGE $m\!+\!1$}
\pscircle[origin={-2,5},fillstyle=solid,fillcolor=mygrey](6,4){0.9}\rput(8,-1){\LARGE $m\!+\!1$}

\rput(9.7,-1){\Huge $\otimes$}
\pscircle[fillstyle=solid,fillcolor=mygrey](11,-1){0.5}
\psline(11.5,-1)(12.5,-1)\uput[u](12,-1){\LARGE $r$}

\rput(13.2,-1){\Huge $\!+\!$}

\psarc[origin={-14,5}](3.5,3){2.69}{(2.5,1)}{(-2.5,1)}
\psarc[origin={-14,5}](3.5,5){2.69}{(-2.5,-1)}{(2.5,-1)}
\psline[origin={-14,5}](1,4)(6,4)
\psline[origin={-14,5},linestyle=dashed](3.5,2)(3.5,6)
\psdots*[origin={-14,5},dotscale=0.3](3.3,5.2)(3.3,4.8)(3.3,4.4)
\psdots*[origin={-14,5},dotscale=0.3](3.3,3.6)(3.3,3.2)(3.3,2.8)
\rput[l](18,-0.7){\LARGE $r$}
\pscircle[origin={-14,5},fillstyle=solid,fillcolor=mygrey](1,4){0.9}\rput(15,-1){\LARGE $m\!+\!1$}
\pscircle[origin={-14,5},fillstyle=solid,fillcolor=mygrey](6,4){0.9}\rput(20,-1){\LARGE $m\!+\!1$}

\rput(21.7,-1){\Huge $\otimes$}
\pscircle[fillstyle=solid,fillcolor=mygrey](23,-1){0.5}
\psline(23.5,-1)(24.5,-1)\uput[u](24,-1){\LARGE $s$}

\rput[l](25,-1){\Huge $\!+\! \ldots\,\,\Bigg]$}


\rput(8,4){\Huge $- \,\,\Bigg[$}

\psarc[origin={-9,0}](3.5,3){2.69}{(2.5,1)}{(-2.5,1)}
\psarc[origin={-9,0}](3.5,5){2.69}{(-2.5,-1)}{(2.5,-1)}
\psline[origin={-9,0},linestyle=dashed](3.5,2)(3.5,6)
\psdots*[origin={-9,0},dotscale=0.3](3.3,4.6)(3.3,4)(3.3,3.4)
\pscircle[origin={-9,0},fillstyle=solid,fillcolor=mygrey](1,4){0.9}\rput(10,4){\LARGE $m$}
\pscircle[origin={-9,0},fillstyle=solid,fillcolor=mygrey](6,4){0.9}\rput(15,4){\LARGE $m$}

\rput(16.7,4){\Huge $\otimes$}
\psline[origin={-17.5,-4}](1;10)(2;10)\uput{2.2}[10]{0}(17.5,4){\LARGE $r$}
\psline[origin={-17.5,-4}](1;-10)(2;-10)\uput{2.2}[-10]{0}(17.5,4){\LARGE $s$}
\pscircle[fillstyle=solid,fillcolor=mygrey](18,4){0.5}

\rput[l](20.2,4){\Huge $\!+\! \ldots\,\,\Bigg]$}


\rput(1,-6){\Huge $\!+\! \,\,\Bigg[$}

\psarc[origin={-2,10}](3.5,3){2.69}{(2.5,1)}{(-2.5,1)}
\psarc[origin={-2,10}](3.5,5){2.69}{(-2.5,-1)}{(2.5,-1)}
\psline[origin={-2,10},linestyle=dashed](3.5,2)(3.5,6)
\psdots*[origin={-2,10},dotscale=0.3](3.3,4.6)(3.3,4)(3.3,3.4)
\pscircle[origin={-2,10},fillstyle=solid,fillcolor=mygrey](1,4){0.9}\rput(3,-6){\LARGE $m$}
\pscircle[origin={-2,10},fillstyle=solid,fillcolor=mygrey](6,4){0.9}\rput(8,-6){\LARGE $m$}

\rput(9.7,-6){\Huge $\otimes$}
\pscircle[fillstyle=solid,fillcolor=mygrey](11,-6){0.5}
\psline(11.5,-6)(12.5,-6)\uput[u](12,-6){\LARGE $r$}

\rput(13.2,-6){\Huge $\otimes$}
\pscircle[fillstyle=solid,fillcolor=mygrey](14.5,-6){0.5}
\psline(15,-6)(16,-6)\uput[u](15.5,-6){\LARGE $\hat{s}$}

\rput(16.7,-6){\Huge $\!+\!$}

\psarc[origin={-17.5,10}](3.5,3){2.69}{(2.5,1)}{(-2.5,1)}
\psarc[origin={-17.5,10}](3.5,5){2.69}{(-2.5,-1)}{(2.5,-1)}
\psline[origin={-17.5,10},linestyle=dashed](3.5,2)(3.5,6)
\psdots*[origin={-17.5,10},dotscale=0.3](3.3,4.6)(3.3,4)(3.3,3.4)
\pscircle[origin={-17.5,10},fillstyle=solid,fillcolor=mygrey](1,4){0.9}\rput(18.5,-6){\LARGE $m$}
\pscircle[origin={-17.5,10},fillstyle=solid,fillcolor=mygrey](6,4){0.9}\rput(23.5,-6){\LARGE $m$}

\rput(25.2,-6){\Huge $\otimes$}
\pscircle[fillstyle=solid,fillcolor=mygrey](26.5,-6){0.5}
\psline(27,-6)(28,-6)\uput[u](27.5,-6){\LARGE $s$}

\rput(28.7,-6){\Huge $\otimes$}
\pscircle[fillstyle=solid,fillcolor=mygrey](30,-6){0.5}
\psline(30.5,-6)(31.5,-6)\uput[u](31,-6){\LARGE $\hat{r}$}

\rput[l](32,-6){\Huge $\!+\! \ldots\,\,\Bigg]$}
}
\end{pspicture}
\end{center}
~\vskip 26mm
\caption{Graphical representation of the squared matrix element and its
factorization formulae in the singly- and doubly-unresolved (soft and/or
collinear) limits.}
\label{fig:RR-A2-A1+A12}
\end{figure}

The first picture corresponds to the squared matrix element of the
$m+2$ final-state partons, while the following terms in the squared
brackets correspond to the terms that build $\bA{2}\M{m+2}{(0)}$,
$\bA{1}\M{m+2}{(0)}$ and $\bA{12}\M{m+2}{(0)}$, respectively. The
factorized one- and two-parton factors correspond to Altarelli-Parisi
kernels, with azimuthal correlations included, or eikonal factors, with
colour-correlations included, or the combinations of these. In the
first bracket we find all those terms that regularize the kinematical
singularities of the squared matrix element in the doubly-unresolved
regions of the phase space. In these phase space regions the terms in
the second bracket, corresponding to $\bA{1}\M{m+2}{(0)}$, are
regularized by terms in the third bracket, corresponding to
$\bA{12}\M{m+2}{(0)}$. In particular, if partons $r$ and $s$ become
unresolved, than the two terms shown in the third line regularize those
two in the second line.

In the singly-unresolved regions, the second bracket contains all terms,
necessary to regularize the squared matrix element. Note however, that
these terms also contain spurious singularities. For instance, when $r$
is unresolved, then the first term in the second bracket regularizes the
squared matrix element, while the second becomes singular. This term will
be regularized by the second term in the third bracket. At the same time,
the terms in the first bracket will lead to a strongly-ordered
factorization formula, that will be regularized by the first term in the
third bracket. If $s$ becomes unresolved, then the role of the two terms
in the third bracket interchanges: the first term will regularize the
spurious singularity in the first term of the second bracket, while the
second will regularize the term in the first bracket.

In defining the full subtraction
$(\bcA{2}{}+\bcA{1}{}-\bcA{12}{})\M{m+2}{(0)}$, we keep the structure
of $(\bA{2}+\bA{1}-\bA{12})\M{m+2}{(0)}$ and replace the momenta in the
squared matrix elements of $m$ or $m+1$ final-state partons in
\fig{fig:RR-A2-A1+A12} with tilded momenta, defined by different types
of momentum mappings for the various terms. In order that the
cancellations described in the previous two paragraphs take place it is
crucial that these momentum mappings obey the following three conditions:
\begin{enumerate}
\itemsep -2pt
\item
The mappings used for defining the doubly-unresolved subtraction terms
should be such that in the singly-unresolved regions of the phase space,
the mapped momenta tend to the same limit as those of the iterated
mappings, used for defining the terms in $\bcA{12}{}\M{m+2}{(0)}$.
\item
In those regions of the phase space, where only momentum $p_s^\mu$
becomes unresolved, the singly-collinear and soft momentum mappings,
used in the definition of those singly-unresolved subtraction terms, in
which momentum $p_r^\mu$ is factorized, tend to the same limit as the
iterated mappings used in the definition of those terms in
$\bcA{12}{}\M{m+2}{(0)}$, in which the first mapping factorizes
$p_r^\mu$ and the second $\hat{p}_s^\mu$.
\item
In the doubly-unresolved regions of the phase space, the
singly-collinear and soft momentum mappings, used in the definition of
the $\bcA{1}{}\M{m+2}{(0)}$ term, tend to the same limit as the
iterated mappings used in the definition of $\bcA{12}{}\M{m+2}{(0)}$. 
\end{enumerate}
The first of these conditions is necessary in order that the required
cancellations between $\bcA{2}{}\M{m+2}{(0)}$ and
$\bcA{12}{}\M{m+2}{(0)}$ take place. The second requirement ensures
that the cancellations between the first terms in the second and third
bracket of \fig{fig:RR-A2-A1+A12} happens when momentum $p_s^\mu$ becomes
unresolved. Finally, the third condition is needed for cancelling all
kinematical singularities of the $\bcA{1}{}\M{m+2}{(0)}$ terms by the
corresponding terms in $\bcA{12}{}\M{m+2}{(0)}$ in the doubly-unresolved
regions of the phase space. It is not difficult to check that the
above requirements are fulfilled for all mappings. In particular, the
third condition follows from the construction of the iterative mappings,
namely these are successive applications of the singly-collinear and/or
soft mappings. Here we consider two illustrative examples of the first
two conditions.

\begin{figure}
\begin{center}
\begin{pspicture}(0,0)(16,4)

\scalebox{0.5}{%

\psline[linewidth=3pt,arrowinset=0]{->}(0.2,4)(1.4,4)
\uput{0.3}[d](1.2,4){\LARGE $Q$}
\psline[linewidth=3pt](1.2,4)(2.2,4)

\SpecialCoor
\psline[origin={-3,-4}](0.8;100)(2.8;100)\uput{2.9}[100]{0}(3,4){\LARGE $\tilde{1}$}
\psline[origin={-3,-4}](0.8;80)(2.8;80)
\psdots*[origin={-3,-4},dotscale=0.3](1.8;65)(1.8;55)(1.8;45)
\psline[origin={-3,-4}](0.8;30)(2.8;30)
\psline[origin={-3,-4}](0.8;0)(2.3;0)\uput{1.6}[12]{0}(3,4){\LARGE $\widetilde{ktr}$}
\psline[origin={-3,-4}](0.8;-30)(2.8;-30)
\psdots*[origin={-3,-4},dotscale=0.3](1.8;-65)(1.8;-55)(1.8;-45)
\psline[origin={-3,-4}](0.8;-80)(2.8;-80)
\psline[origin={-3,-4}](0.8;-100)(2.8;-100)\uput{2.9}[-100]{0}(3,4){\LARGE $\widetilde{m\!+\!2}$}

\pscircle[fillstyle=solid,fillcolor=white](3,4){0.9}\rput(3,4){\LARGE $m$}

\psline[origin={-3,-4},linestyle=dotted](2.3;0)(2.8;0)
\psline[origin={-6.1,-4}](0.3;0)(1.3;0)\uput{1.4}[0]{0}(6.1,4){\LARGE $k$}
\psline[origin={-6.1,-4}](0.3;-30)(1.3;-30)\uput{1.4}[-30]{0}(6.1,4){\LARGE $t$}
\psline[origin={-6.1,-4}](0.3;-60)(1.3;-60)\uput{1.4}[-60]{0}(6.1,4){\LARGE $r$}

\pscircle[origin={-3,-4},fillstyle=solid,fillcolor=white](3.1;0){0.4}
\rput(6.1,4){\LARGE $\mathsf{C}\,$}

\uput{0}[0]{0}(8.3,4){\LARGE ,}

\psline[linewidth=3pt,arrowinset=0]{->}(9.2,4)(10.4,4)\uput{0.3}[d](10.2,4){\LARGE $Q$}
\psline[linewidth=3pt](10.2,4)(11.2,4)
\SpecialCoor
\psline[origin={-12,-4}](0.8;100)(2.8;100)\uput{2.9}[100]{0}(12,4){\LARGE $\tilde{1}$}
\psline[origin={-12,-4}](0.8;80)(2.8;80)
\psdots*[origin={-12,-4},dotscale=0.3](1.8;65)(1.8;55)(1.8;45)
\psline[origin={-12,-4}](0.8;30)(2.8;30)
\psline[origin={-12,-4}](0.8;0)(2.3;0)\uput{1.6}[14]{0}(12,4){\LARGE $\widetilde{ktr}$}
\psline[origin={-12,-4}](0.8;-30)(2.8;-30)
\psdots*[origin={-12,-4},dotscale=0.3](1.8;-65)(1.8;-55)(1.8;-45)
\psline[origin={-12,-4}](0.8;-80)(2.8;-80)
\psline[origin={-12,-4}](0.8;-100)(2.8;-100)\uput{2.9}[-100]{0}(12,4){\LARGE $\widetilde{m\!+\!2}$}

\psline[origin={-12,-4},linestyle=dotted](2.3;0)(2.8;0)

\pscircle[fillstyle=solid,fillcolor=white](12,4){0.9}\rput(12,4){\LARGE $m$}

\psline[origin={-12,-4}](3.4;0)(3.9;0)\uput{0.6}[29]{0}(15.1,4){\LARGE $\hat{kt}$}
\psline[origin={-12,-4},linestyle=dotted](3.9;0)(4.4;0)
\psline[origin={-15.1,-4}](0.29;-30)(1.3;-30)\uput{1.4}[-30]{0}(15.1,4){\LARGE $\hat{r}$}

\pscircle[origin={-12,-4},fillstyle=solid,fillcolor=white](3.1;0){0.4}
\rput(15.1,4){\LARGE $\mathsf{C}\,$}
\psline[origin={-16.9,-4}](0.4;0)(1.4;0)\uput{1.6}[0]{0}(16.9,4){\LARGE $k$}
\psline[origin={-16.9,-4}](0.4;-30)(1.4;-30)\uput{1.6}[-30]{0}(16.9,4){\LARGE $t$}

\pscircle[origin={-16.9,-4},fillstyle=solid,fillcolor=white](0;0){0.4}
\rput(16.9,4){\LARGE $\mathsf{C}\,$}


\psline{->}(19.7,4)(21.8,4)\uput[u](20.8,3){{\LARGE $p_k || p_t$}}

\psline[linewidth=3pt,arrowinset=0]{->}(22.7,4)(23.9,4)\uput{0.3}[d](23.7,4){\LARGE $Q$}
\psline[linewidth=3pt](23.7,4)(24.7,4)

\SpecialCoor
\psline[origin={-25.5,-4}](0.8;100)(2.8;100)\uput{2.9}[100]{0}(25.5,4){\LARGE $\tilde{1}$}
\psline[origin={-25.5,-4}](0.8;80)(2.8;80)
\psdots*[origin={-25.5,-4},dotscale=0.3](1.8;65)(1.8;55)(1.8;45)
\psline[origin={-25.5,-4}](0.8;30)(2.8;30)
\psline[origin={-25.5,-4}](0.8;0)(2.3;0)\uput{1.6}[14]{0}(25.5,4){\LARGE
$\widetilde{(kt)r}$}
\psline[origin={-25.5,-4}](0.8;-30)(2.8;-30)
\psdots*[origin={-25.5,-4},dotscale=0.3](1.8;-65)(1.8;-55)(1.8;-45)
\psline[origin={-25.5,-4}](0.8;-80)(2.8;-80)
\psline[origin={-25.5,-4}](0.8;-100)(2.8;-100)\uput{2.9}[-100]{0}(25.5,4){\LARGE $\widetilde{m\!+\!2}$}

\psline[origin={-25.5,-4},linestyle=dotted](2.3;0)(3.2;0)
\pscircle[fillstyle=solid,fillcolor=white](25.5,4){0.9}\rput(25.5,4){\LARGE $m$}

\psline[origin={-25.9,-4}](3.4;0)(4.9;0)\uput{0.6}[29]{0}(29,4){\LARGE $(kt)$}
\psline[origin={-29,-4}](0.29;-30)(1.3;-30)\uput{1.4}[-30]{0}(29,4){\LARGE $r$}

\pscircle[origin={-25.9,-4},fillstyle=solid,fillcolor=white](3.1;0){0.4}
\rput(29,4){\LARGE $\mathsf{C}\,$}
\psline[origin={-30.8,-4}](0;10)(1.4;10)\uput{1.6}[10]{0}(30.8,4){\LARGE $k$}
\psline[origin={-30.8,-4}](0;-10)(1.4;-10)\uput{1.6}[-10]{0}(30.8,4){\LARGE $t$}
}
\end{pspicture}
\end{center}
\vskip-8mm
\caption{Graphical representation of the collinear limit of the
triply-collinear and and iterated singly-collinear mappings. $(kt)$ means
the momentum $p_k^\mu+p_t^\mu$ in the collinear direction in the
collinear limit.}
\label{fig:Cktlimit}
\end{figure}
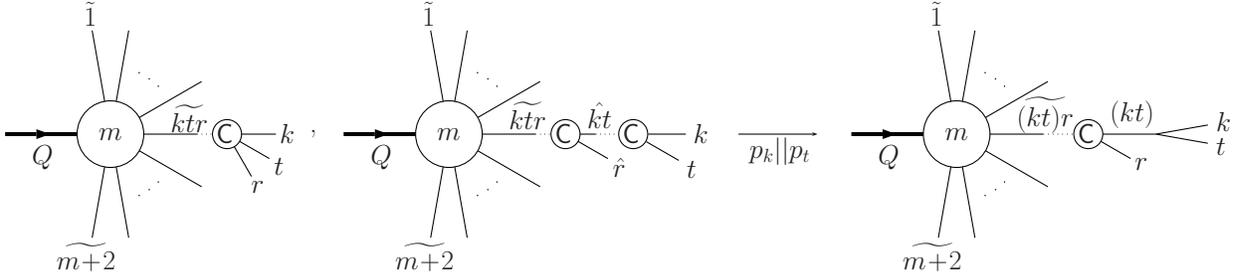

In the case of the first condition, the least trivial is that the
singly-collinear limit of the triply-collinear mapping is the same
as that of two successive collinear mappings, defined by
\eqnss{eq:PS_CktCktr}{eq:PS_CktCktr_t}, as shown graphically in
\fig{fig:Cktlimit}. In the limit when momenta $p_k^\mu$ and $p_t^\mu$ are
collinear, using \eqns{eq:zt3}{eq:kTtris} we find 
\beq
\tzz{k}{ts} \arrowlimit{p_k || p_t} z_k z_{(kt),s}\,,
\qquad
\tzz{t}{ks} \arrowlimit{p_k || p_t} z_t z_{(kt),s}\,,
\qquad
\tzz{s}{kt} \arrowlimit{p_k || p_t} z_{s,(kt)}
\label{eq:Cktst3}
\eeq
and
\beq
\kTt{k,ts}^{\mu} \arrowlimit{p_k || p_t} z_k \kTt{(kt),s}^{\mu}\,,
\qquad
\kTt{t,ks}^{\mu} \arrowlimit{p_k || p_t} z_t \kTt{(kt),s}^{\mu}\,,
\qquad
\kTt{s,kt}^{\mu} \arrowlimit{p_k || p_t} \kTt{s,(kt)}^{\mu}\,,
\label{eq:CktkTt3}
\eeq
i.e., both the momentum fractions and the transverse momenta tend to the
limit of the corresponding variables of the iterative collinear mapping.
As a result, the kinematics defined by the two mappings tends to the same
limit and the cancellation of the singularities takes place.

For the second requirement, let us consider for example, the soft limit
$p_s^\mu \to 0$.  The singly-unresolved counterterms depend on
$\ti{p}_s^\mu$, while the iterated terms depend on $\hat{p}_s^\mu$.
The singly-unresolved mappings, that lead to these different momenta in
the two cases, are linear in $p_s^\mu$, therefore, $\hat{p}_s^\mu \to 0$
in the soft limit, when $p_s^\mu \to 0$.  It follows that
$\lambda_{\hat{s}}  \to 1$ when $p_s^\mu \to 0$ so
in the soft limit the hatted momenta tend to those with tilde. In
particular, $\hat{p}_s^\mu \to \ti{p}_s^\mu$ if $p_s^\mu \to 0$, so
the kinematics become identical and the cancellation takes place.
The graphical representations of the soft limits of the two mappings
are shown in \fig{fig:Sslimit}.
\begin{figure}
\begin{center}
\begin{pspicture}(0,0)(15,4)
\scalebox{0.5}{%

\psline[linewidth=3pt,arrowinset=0]{->}(0.2,4)(1.4,4)\uput{0.3}[d](1.2,4){\LARGE $Q$}
\psline[linewidth=3pt](1.2,4)(2.2,4)

\SpecialCoor
\psline[origin={-3,-4}](0.8;100)(2.8;100)\uput{2.9}[100]{0}(3,4){\LARGE $\ti{1}$}
\psline[origin={-3,-4}](0.8;80)(2.8;80)
\psdots*[origin={-3,-4},dotscale=0.3](1.8;65)(1.8;55)(1.8;45)
\psline[origin={-3,-4}](0.8;30)(2.8;30)
\psline[origin={-3,-4}](0.8;10)(2.3;10)\uput{1.9}[20]{0}(3,4){\LARGE $\wti{ir}$}
\psline[origin={-3,-4}](0.8;-10)(2.8;-10)\uput{3.0}[-10]{0}(3,4){\LARGE $\wti{s}$}
\psline[origin={-3,-4}](0.8;-30)(2.8;-30)
\psdots*[origin={-3,-4},dotscale=0.3](1.8;-65)(1.8;-55)(1.8;-45)
\psline[origin={-3,-4}](0.8;-80)(2.8;-80)
\psline[origin={-3,-4}](0.8;-100)(2.8;-100)\uput{2.9}[-100]{0}(3,4){\LARGE $\wti{m\!+\!2}$}

\pscircle[fillstyle=solid,fillcolor=white](3,4){0.9}\rput(3,4){\LARGE $m\!+\!1$}

\psline[origin={-3,-4},linestyle=dotted](2.3;10)(2.8;10)
\psline[origin={-3,-4},linestyle=dotted](2.3;-10)(2.8;-10)

\psline[origin={-3,-4}](3.4;10)(4.4;10)\uput{1.4}[10]{0}(6.05,4.54){\LARGE $i$}
\psline[origin={-6.05,-4.54}](0.29;-20)(1.3;-20)\uput{1.4}[-20]{0}(6.05,4.54){\LARGE $r$}

\pscircle[origin={-3,-4},fillstyle=solid,fillcolor=white](3.1;10){0.4}
\rput(6.05,4.54){\LARGE $\mathsf{C}\,$}

\uput{0}[0]{0}(8.3,4){\LARGE ,}

\psline[linewidth=3pt,arrowinset=0]{->}(9.2,4)(10.4,4)\uput{0.3}[d](10.2,4){\LARGE $Q$}
\psline[linewidth=3pt](10.2,4)(11.2,4)

\SpecialCoor
\psline[origin={-12,-4}](0.8;100)(2.8;100)\uput{2.9}[100]{0}(12,4){\LARGE $\tilde{1}$}
\psline[origin={-12,-4}](0.8;80)(2.8;80)
\psdots*[origin={-12,-4},dotscale=0.3](1.8;65)(1.8;55)(1.8;45)
\psline[origin={-12,-4}](0.8;30)(2.8;30)
\psline[origin={-12,-4}](0.8;10)(2.3;10)\uput{1.9}[20]{0}(12,4){\LARGE $\widetilde{ir}$}
\psline[origin={-12,-4}](0.8;-30)(2.8;-30)
\psdots*[origin={-12,-4},dotscale=0.3](1.8;-65)(1.8;-55)(1.8;-45)
\psline[origin={-12,-4}](0.8;-80)(2.8;-80)
\psline[origin={-12,-4}](0.8;-100)(2.8;-100)\uput{2.9}[-100]{0}(12,4){\LARGE $\widetilde{m\!+\!2}$}

\psline[origin={-12,-4},linestyle=dotted](2.3;10)(2.8;10)
\psline[origin={-12,-4},linestyle=dotted](0.8;-10)(2.8;-10)

\pscircle[fillstyle=solid,fillcolor=white](12,4){0.9}\rput(12,4){\LARGE $m$}

\psline[origin={-12,-4}](3.4;10)(4.4;10)\uput{1.4}[10]{0}(15.05,4.54){\LARGE $i$}
\psline[origin={-15.05,-4.54}](0.29;-20)(1.3;-20)\uput{1.4}[-20]{0}(15.05,4.54){\LARGE $r$}

\pscircle[origin={-12,-4},fillstyle=solid,fillcolor=white](3.1;10){0.4}
\rput(15.05,4.54){\LARGE $\mathsf{C}\,$}

\psline[origin={-12,-4}](3.4;-10)(4.4;-10)\uput{1.4}[-10]{0}(15.05,3.46){\LARGE $\hat{s}$}

\pscircle[origin={-12,-4},fillstyle=solid,fillcolor=white](3.1;-10){0.4}
\rput(15.05,3.46){\LARGE $\mathsf{S}$}

\psline[linewidth=3pt,arrowinset=0]{->}(9.2,4)(10.4,4)\uput{0.3}[d](10.2,4){\LARGE $Q$}
\psline[linewidth=3pt](10.2,4)(11.2,4)

\psline{->}(17.5,4)(19.5,4)\uput[u](18.5,3){{\LARGE $p_s \to 0$}}

\psline[linewidth=3pt,arrowinset=0]{->}(20.2,4)(21.4,4)\uput{0.3}[d](21.2,4){\LARGE $Q$}
\psline[linewidth=3pt](21.2,4)(22.2,4)

\SpecialCoor
\psline[origin={-23,-4}](0.8;100)(2.8;100)\uput{2.9}[100]{0}(23,4){\LARGE $\tilde{1}$}
\psline[origin={-23,-4}](0.8;80)(2.8;80)
\psdots*[origin={-23,-4},dotscale=0.3](1.8;65)(1.8;55)(1.8;45)
\psline[origin={-23,-4}](0.8;30)(2.8;30)
\psline[origin={-23,-4}](0.8;10)(2.3;10)\uput{1.9}[20]{0}(23,4){\LARGE $\widetilde{ir}$}
\psline[origin={-23,-4}](0.8;-30)(2.8;-30)
\psdots*[origin={-23,-4},dotscale=0.3](1.8;-65)(1.8;-55)(1.8;-45)
\psline[origin={-23,-4}](0.8;-80)(2.8;-80)
\psline[origin={-23,-4}](0.8;-100)(2.8;-100)\uput{2.9}[-100]{0}(23,4){\LARGE $\widetilde{m\!+\!2}$}

\psline[origin={-23,-4},linestyle=dotted](2.3;10)(2.8;10)
\psline[origin={-23,-4},linestyle=dotted](0.8;-10)(2.8;-10)

\pscircle[fillstyle=solid,fillcolor=white](23,4){0.9}\rput(23,4){\LARGE $m$}

\psline[origin={-23,-4}](3.4;10)(4.4;10)\uput{1.4}[10]{0}(26.05,4.54){\LARGE $i$}
\psline[origin={-26.05,-4.54}](0.29;-20)(1.3;-20)\uput{1.4}[-20]{0}(26.05,4.54){\LARGE $r$}

\pscircle[origin={-23,-4},fillstyle=solid,fillcolor=white](3.1;10){0.4}
\rput(26.05,4.54){\LARGE $\mathsf{C}\,$}

\psline[origin={-23,-4}](2.8;-10)(3.8;-10)\uput{0.8}[-10]{0}(26.05,3.46)
{\LARGE $\hat{s} \to \wti{s}$}
}
\end{pspicture}
\end{center}
~\vskip -20mm
\end{figure}
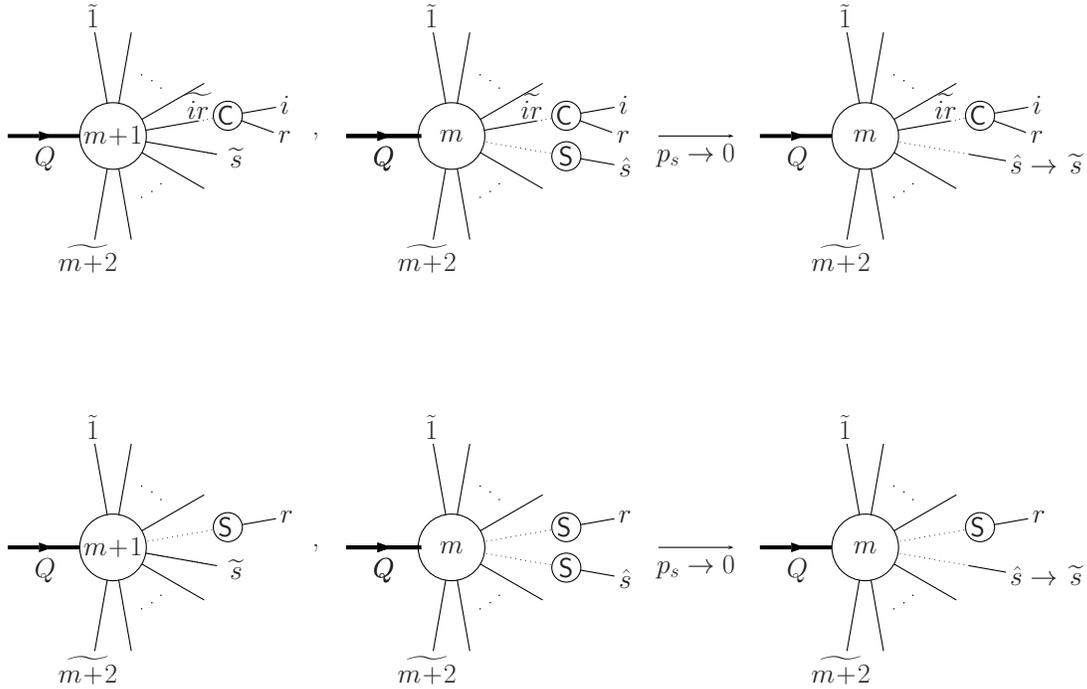
\begin{figure}
\begin{center}
\begin{pspicture}(0,0)(15,4)
\scalebox{0.5}{%

\psline[linewidth=3pt,arrowinset=0]{->}(0.2,4)(1.4,4)\uput{0.3}[d](1.2,4){\LARGE $Q$}
\psline[linewidth=3pt](1.2,4)(2.2,4)

\SpecialCoor
\psline[origin={-3,-4}](0.8;100)(2.8;100)\uput{2.9}[100]{0}(3,4){\LARGE $\ti{1}$}
\psline[origin={-3,-4}](0.8;80)(2.8;80)
\psdots*[origin={-3,-4},dotscale=0.3](1.8;65)(1.8;55)(1.8;45)
\psline[origin={-3,-4}](0.8;30)(2.8;30)
\psline[origin={-3,-4},linestyle=dotted](0.8;10)(2.8;10)
\psline[origin={-3,-4},linestyle=dotted](0.8;-10)(2.8;-10)
\psline[origin={-3,-4}](0.8;-10)(2.8;-10)\uput{3.0}[-10]{0}(3,4){\LARGE $\wti{s}$}
\psline[origin={-3,-4}](0.8;-30)(2.8;-30)
\psdots*[origin={-3,-4},dotscale=0.3](1.8;-65)(1.8;-55)(1.8;-45)
\psline[origin={-3,-4}](0.8;-80)(2.8;-80)
\psline[origin={-3,-4}](0.8;-100)(2.8;-100)\uput{2.9}[-100]{0}(3,4){\LARGE $\wti{m\!+\!2}$}

\pscircle[fillstyle=solid,fillcolor=white](3,4){0.9}\rput(3,4){\LARGE $m\!+\!1$}

\psline[origin={-3,-4}](3.4;10)(4.4;10)\uput{1.4}[10]{0}(6.05,4.54){\LARGE $r$}

\pscircle[origin={-3,-4},fillstyle=solid,fillcolor=white](3.1;10){0.4}
\rput(6.05,4.54){\LARGE $\mathsf{S}\,$}

\uput{0}[0]{0}(8.3,4){\LARGE ,}

\psline[linewidth=3pt,arrowinset=0]{->}(9.2,4)(10.4,4)\uput{0.3}[d](10.2,4){\LARGE $Q$}
\psline[linewidth=3pt](10.2,4)(11.2,4)

\SpecialCoor
\psline[origin={-12,-4}](0.8;100)(2.8;100)\uput{2.9}[100]{0}(12,4){\LARGE $\tilde{1}$}
\psline[origin={-12,-4}](0.8;80)(2.8;80)
\psdots*[origin={-12,-4},dotscale=0.3](1.8;65)(1.8;55)(1.8;45)
\psline[origin={-12,-4}](0.8;30)(2.8;30)
\psline[origin={-12,-4},linestyle=dotted](0.8;10)(2.8;10)
\psline[origin={-12,-4},linestyle=dotted](0.8;-10)(2.8;-10)
\psline[origin={-12,-4}](0.8;-30)(2.8;-30)
\psdots*[origin={-12,-4},dotscale=0.3](1.8;-65)(1.8;-55)(1.8;-45)
\psline[origin={-12,-4}](0.8;-80)(2.8;-80)
\psline[origin={-12,-4}](0.8;-100)(2.8;-100)\uput{2.9}[-100]{0}(12,4){\LARGE $\widetilde{m\!+\!2}$}

\pscircle[fillstyle=solid,fillcolor=white](12,4){0.9}\rput(12,4){\LARGE $m$}

\psline[origin={-12,-4}](3.4;10)(4.4;10)\uput{1.4}[10]{0}(15.05,4.54){\LARGE $r$}
\pscircle[origin={-12,-4},fillstyle=solid,fillcolor=white](3.1;10){0.4}
\rput(15.05,4.54){\LARGE $\mathsf{S}\,$}

\psline[origin={-12,-4}](3.4;-10)(4.4;-10)\uput{1.4}[-10]{0}(15.05,3.46){\LARGE $\hat{s}$}

\pscircle[origin={-12,-4},fillstyle=solid,fillcolor=white](3.1;-10){0.4}
\rput(15.05,3.46){\LARGE $\mathsf{S}$}

\psline[linewidth=3pt,arrowinset=0]{->}(9.2,4)(10.4,4)\uput{0.3}[d](10.2,4){\LARGE $Q$}
\psline[linewidth=3pt](10.2,4)(11.2,4)

\psline{->}(17.5,4)(19.5,4)\uput[u](18.5,3){{\LARGE $p_s \to 0$}}

\psline[linewidth=3pt,arrowinset=0]{->}(20.2,4)(21.4,4)\uput{0.3}[d](21.2,4){\LARGE $Q$}
\psline[linewidth=3pt](21.2,4)(22.2,4)

\SpecialCoor
\psline[origin={-23,-4}](0.8;100)(2.8;100)\uput{2.9}[100]{0}(23,4){\LARGE $\tilde{1}$}
\psline[origin={-23,-4}](0.8;80)(2.8;80)
\psdots*[origin={-23,-4},dotscale=0.3](1.8;65)(1.8;55)(1.8;45)
\psline[origin={-23,-4}](0.8;30)(2.8;30)
\psline[origin={-23,-4},linestyle=dotted](0.8;10)(2.8;10)
\psline[origin={-23,-4},linestyle=dotted](0.8;-10)(2.8;-10)
\psline[origin={-23,-4}](0.8;-30)(2.8;-30)
\psdots*[origin={-23,-4},dotscale=0.3](1.8;-65)(1.8;-55)(1.8;-45)
\psline[origin={-23,-4}](0.8;-80)(2.8;-80)
\psline[origin={-23,-4}](0.8;-100)(2.8;-100)\uput{2.9}[-100]{0}(23,4){\LARGE $\widetilde{m\!+\!2}$}

\pscircle[fillstyle=solid,fillcolor=white](23,4){0.9}\rput(23,4){\LARGE $m$}

\psline[origin={-23,-4}](3.4;10)(4.4;10)\uput{1.4}[10]{0}(26.05,4.54){\LARGE $r$}

\pscircle[origin={-23,-4},fillstyle=solid,fillcolor=white](3.1;10){0.4}
\rput(26.05,4.54){\LARGE $\mathsf{S}\,$}

\psline[origin={-23,-4}](2.8;-10)(3.8;-10)\uput{0.8}[-10]{0}(26.05,3.46)
{\LARGE $\hat{s} \to \wti{s}$}
}
\end{pspicture}
\end{center}
~\vskip -5mm
\caption{Graphical representation of the soft limit of the
singly-unresolved and iterated momentum mappings.}
\label{fig:Sslimit}
\end{figure}

\section{Numerical results}
\label{sec.results}

In Sections \ref{sec:RR_A1}, \ref{sec:RR_A2} and \ref{sec:RR_A12}
we have defined explicitly all subtraction terms in
\eqnss{eq:dsigRRA2}{eq:dsigRRA12} that are necessary to make
$\dsig{RR}_{m+2}$ integrable in $d = 4$ dimensions. In order to further
demonstrate that the subtraction terms indeed regularize the cross
section for doubly-real emission, we consider the non-trivial example
of the contribution of the $e^+ e^- \to q \qb g g g$ subprocess to
the moments of three-jet event-shape variables thrust ($T$) and
$C$-parameter, when  the jet function is a functional
\beq
J_n(p_1,\ldots,p_n;O) = \delta(O-O_3(p_1,\ldots,p_n))\:,
\eeq
with $O_3(p_1,\ldots,p_n)$ being the value of either $\tau \equiv 1-T$
or $C$ for a given event $(p_1,\ldots,p_n)$. The $e^+ e^- \to q \qb g g
g$ subprocess gives rise to the most general kinds of kinematical
singularities and colour structures. The only additional complication
present in the four-quark subprocess $e^+e^- \to q \qb Q \Qb g$ is the
identical flavour contribution that does not require any addition to
the subtraction scheme.  

Starting from randomly chosen phase space points and approaching the
various singly- and doubly-unresolved regions of the phase space in
successive steps, we have checked numerically that the sum of the
subtraction terms has the same limits (up to integrable square-root
singularities) as the squared matrix element itself.

The perturbative expansion to the $n^{\rm th}$ moment of a three-jet
observable at a fixed scale $\mu = Q$ and NNLO accuracy can be
parametrized as 
\beeq
\la O^n \ra \aand\equiv
\int\!\rd O\,O^n\,\frac{1}{\sigma_0}\frac{\dsig{}}{\rd O} =
\nn\\&&=
  \left(\frac{\as(Q)}{2\pi}\right)   A_{O}^{(n)}
+ \left(\frac{\as(Q)}{2\pi}\right)^2 B_{O}^{(n)}
+ \left(\frac{\as(Q)}{2\pi}\right)^3 C_{O}^{(n)}\:,
\eeeq
where according to \eqn{eq:sigmaNNLOfin}, the NNLO correction is a sum
of three contributions
\beq
C_{O}^{(n)} =
C_{O;5}^{(n)} + C_{O;4}^{(n)} + C_{O;3}^{(n)}\:.
\eeq
Carrying out the phase space integrations in \eqn{eq:sigmaNNLOm+2}, we
computed the five-parton contribution $C_{O;5}^{(n)}(O)$ as defined in
this article. The predictions for the first three moments of $\tau$ and
the $C$-parameter, obtained using about ten million Monte Carlo events,
are presented in \tab{tab:C}. These numbers are unphysical, and given
only to demonstrate that the $(m+2)$-parton NNLO cross section defined
in this paper is finite. In particular, the relatively small negative
values simply indicate that the subtractions altogether subtract
slightly more then the full doubly-real cross section. If needed, the
colour decomposition is straightforward.
\begin{table}
\begin{center}
\caption{The moments $C_{\tau;5}^{(n)}$ and $C_{C;5}^{(n)}$.}
\label{tab:C}
\begin{tabular}{|c|c|c|}
\hline
\hline
 & & \\[-4mm]
 $n$ & $C_{\tau;5}^{(n)}$ & $C_{C;5}^{(n)}$ \\
 & & \\[-4mm]
\hline
 & & \\[-4mm]
  1  & -\res{9.27}{0.33}{ } & -\res{3.44}{0.13}{ 2} \\
  2  & -\res{0.31}{0.04}{ } & -\res{1.41}{0.03}{ 2} \\
  3  & -\res{0.20}{0.01}{ } & -\res{0.63}{0.18}{  } \\
\hline
\hline
\end{tabular} 
\end{center}
\end{table}

%
%

\section{Conclusions}

In this paper we have set up a subtraction scheme for computing NNLO
corrections to QCD jet cross sections to processes without coloured partons
in the initial state. The scheme is completely general in the sense
that any number of massless coloured final-state partons (massive
vector bosons are assumed to decay into massless fermions) are allowed
provided the necessary squared matrix elements are known. It is also
general in the sense that it is algorithmic in a straightforward manner,
therefore, the generalization to N$^n$LO accuracy does not require new
concepts. Each step of the computation can in principle be extended to
any order in perturbation theory, which is useful in setting up parton
shower algorithms that can be matched to fixed-order approximations
naturally.

Three types of corrections contribute to the NNLO corrections: the
doubly-real, the real-virtual and the doubly-virtual ones. Here we have
constructed the subtraction terms for the doubly-real emissions; those
to the real-virtual corrections will be presented in a companion paper.
By rendering these two contributions finite in $d = 4$ dimensions, the
KLN theorem ensures that for infrared safe observables adding the
subtractions above to the doubly-virtual correction, that becomes also
finite in $d = 4$ dimensions.  

The subtraction terms for the doubly-real corrections presented here
are local in $d = 4 - 2 \eps$ dimensions and include complete colour and
azimuthal correlations. The expressions were derived by extending the
various singly- and doubly-unresolved limits of QCD squared matrix
elements over the whole phase space, which was achieved by introducing
momentum mappings which define exactly factorized phase-space measures.
Although the number of subtraction terms is rather large,
the implementation of the scheme is not so complicated, because only
five different types of phase-space mappings have to be defined, all
other mappings being obtained by employing those five basic ones
iteratively.

In order to demonstrate that the subtracted cross section is indeed
integrable, we have computed the corresponding contributions to the
first three moments of two three-jet event-shape observables, the
thrust and the $C$-parameter.

\section*{Acknowledgements}
We are grateful to Z. Nagy for his comments on choosing the transverse
momenta and for the hospitality of the CERN Theory Division, where this
work was completed.
This research was supported in part by the Hungarian Scientific Research
Fund grant OTKA K-60432.

%
%


\end{document}